\documentclass[11pt,a4paper]{article}
\pdfoutput=1
\usepackage{jheppub}
\usepackage[utf8]{inputenc}

\usepackage{multirow}
\usepackage{amsmath,amssymb,amsfonts}
\usepackage{calc}
\usepackage{graphicx}
\usepackage{color}
\usepackage{bbm}
\usepackage{url}
\usepackage{slashed}
\usepackage{subfigure}
\usepackage{color}
\usepackage[table,xcdraw,usenames,dvipsnames]{xcolor}
\usepackage{float} 
\usepackage{epsfig}
\usepackage{graphicx}
\usepackage{euscript}
\usepackage{slashed}
\usepackage{epstopdf}
\usepackage{mathbbol}
\usepackage{mathrsfs}
\usepackage[normalem]{ulem}
\usepackage{pifont}
\usepackage{bbold}
\usepackage{adjustbox}
\usepackage{float}
\allowdisplaybreaks
\usepackage{soul,xcolor}
\setstcolor{red}

\usepackage{bbding}

\usepackage{verbatim}
\usepackage{graphicx}
\usepackage{latexsym}
\usepackage{stackengine}
\stackMath
\newcommand\tsup[2][2]{%
 \def\useanchorwidth{T}%
  \ifnum#1>1%
    \stackon[-.5pt]{\tsup[\numexpr#1-1\relax]{#2}}{\scriptscriptstyle\sim}%
  \else%
    \stackon[.5pt]{#2}{\scriptscriptstyle\sim}%
  \fi%
}

\newcommand{\xmark}{\ding{55}}

\newcommand{\be}{\begin{equation}}
\newcommand{\ee}{\end{equation}}
\newcommand{\bea}{\begin{align}}
\newcommand{\eea}{\end{align}}

\newcommand{\rb}{\bar{r}}

\newcommand{\cH}{\mathcal{H}} 
\newcommand{\cO}{\mathcal{O}} 
\newcommand{\cP}{\mathcal{P}}

\newcommand{\td}{\tilde{\delta}}

\definecolor{darkgreen}{rgb}{0.2,0.6,0}
\definecolor{lightblue}{rgb}{0,0.5,0.8}
\definecolor{lightred}{rgb}{0.8,0.2,0.2}
\definecolor{darkorange}{rgb}{1,0.549,0}
\definecolor{brown}{rgb}{0.609, 0.164, 0.164}

\allowdisplaybreaks

\title{Zooming into the horizon region \\ of black hole-type objects}
\author[a]{Jesse Daas,} \emailAdd{j.daas@science.ru.nl}
\affiliation[a]{Institute for Mathematics, Astrophysics and Particle Physics (IMAPP),
Radboud University, Heyendaalseweg 135, 6525 AJ Nijmegen, The Netherlands}

\author[a]{Cristobal Laporte,} \emailAdd{cristobal.laportemunoz@ru.nl}

\author[a]{Frank Saueressig,} \emailAdd{f.saueressig@science.ru.nl}

\author[a]{Tim van Dijk} \emailAdd{tim.vandijk@ru.nl}

\abstract{
A universal prediction of quantum gravity is that the dynamics of general relativity is augmented by interactions that are of higher order in the spacetime curvature. Numerical explorations indicate that such terms may have a drastic impact on black hole-type solutions by modifying the geometry close to the would-be event horizon in a substantial way. In this work, we perform the first systematic investigation of this blow-up mechanism within general relativity supplemented by quadratic gravity terms, the Goroff-Sagnotti counterterm, the combination of the two, and Einstein-Cubic Gravity. By studying linear perturbations of the Schwarzschild solution close to the Schwarzschild radius, we discover the following picture: the higher-derivative terms giving rise to extra degrees of freedom play a distinguished role. Once couplings associated with these terms enter the solution in the asymptotically flat region, a blow-up mechanism removes the event horizon and one deals with either a naked singularity or a wormhole. We believe that this finding is highly relevant when constraining the coefficients appearing in the Wilsonian description of gravity by observations.
}

\begin{document}
\maketitle
\section{Introduction and Motivation}
\label{Sect.Intro}
Shadow observations made by the Event Horizon Telescope (EHT) \cite{EventHorizonTelescope:2019dse,EventHorizonTelescope:2022wkp} and the direct detection of gravitational waves emitted from the merger and subsequent ringdown of binary black hole systems \cite{LIGOScientific:2016aoc} have put a spotlight on black hole research \cite{Cardoso:2019rvt,Buoninfante:2024oxl}. Currently, all these tests, probing the strong gravity regime, are in agreement with the predictions made by classical general relativity. At the same time, it is expected that quantum effects will modify the dynamics of general relativity in one way or another. This invites the question whether the objects featuring in these observations are indeed the black holes occurring in general relativity and whether current observations can be used to rule out some modifications of the gravitational dynamics. Our work initiates a novel path towards answering this question in a systematic way. 

According to general relativity, black holes are very simple objects. Uniqueness theorems \cite{heusler1996black} ensure that the stationary solutions are characterized completely by their mass, angular momentum, and charge. Arguably, the most fascinating property of the underlying spacetimes is the presence of an event horizon. Loosely speaking, once an object has crossed this horizon, it is no longer able to communicate with outside observers. Classically, the horizon acts as a one-way street: everything that crosses the horizon will not be able to escape again. This feature is so prominent that it has been advocated as the defining property of a black hole \cite{Hawking:2014tga,Curiel:2018cbt}. From general relativity we are so used to the omnipresence of event horizons that this feature is often taken for granted even when the dynamics is altered, e.g., by going to an effective field theory description. The importance of an event horizon for gravitational physics can hardly be understated, as it comes with a number of important consequences. Firstly, it is at the foundation of the cosmic censorship hypothesis \cite{Penrose:1969pc} (also see \cite{Penrose:1999vj,Landsman:2021mjt} for more recent accounts and further references) and screens the curvature singularity lurking in the interior of the black hole from an outside observer. Secondly, it is at the heart of black hole thermodynamics and the prediction that black holes emit Hawking radiation \cite{Hawking:1975vcx}. The horizon also plays a crucial role in setting up boundary conditions when studying the dynamics of black hole accretion (relevant for interpreting shadow images), gravitational wave emission related to the ringdown signal, and when performing a stability analysis of the static spacetime with respect to the time-evolution of small perturbations. This provides a strong impetus for understanding the existence of event horizons on general grounds.

The second characteristic feature of black holes in general relativity is the curvature singularity hidden inside the event horizon. The associated breakdown of the laws of physics at the singularity together with the omnipresence of singularities in general relativity is a strong pointer that a satisfactory description of the gravitational dynamics requires modifying general relativity in one way or another. A prominent path along these lines is to add matter degrees of freedom. This leads to scalar-tensor theories \cite{Bronnikov:2005gm,Bronnikov:2006fu,Sotiriou:2011dz,Herdeiro:2015waa,Karakasis:2023hni}, Horndeski gravity and its extensions \cite{Hui:2012qt,Anabalon:2013oea,Babichev:2017guv}, and non-linear electrodynamics \cite{Ayon-Beato:1998hmi,Bronnikov:2000vy,Dymnikova:2004zc,Bronnikov:2017sgg}. Along a more conservative path, one may supplementing general relativity by higher-order terms in the spacetime curvature which may either be local or non-local.  The occurrence of non-local terms is well-motivated from the effective field theory formulation of general relativity where such contributions already appear at one-loop level \cite{Donoghue:2017pgk}. More general, non-local extensions have played an essential role in constructing regular black hole solutions \cite{Bambi:2023try}. For instance, non-local, ghost-free gravity \cite{BasiBeneito:2022wux} argues that the curvature singularity of the Schwarzschild geometry is not compatible with the non-local equations of motion \cite{Edholm:2016hbt,Koshelev:2018hpt,Buoninfante:2018xif,Koshelev:2024wfk}. Along a similar line, it has been argued based on resummation arguments that the dynamics arising from an infinite tower of curvature terms may also lead to black hole geometries void of curvature singularities  \cite{Bueno:2024dgm,Giacchini:2024exc,DiFilippo:2024mwm,Konoplya:2024hfg,Konoplya:2024kih}. Furthermore, quantum gravity has motivated a number of black hole-type objects including fuzzballs \cite{Bena:2022rna}, Planck stars \cite{Rovelli:2014cta,Saueressig:2015xua}, boson stars \cite{Liebling:2012fv}, and asymptotically safe black holes \cite{Bonanno:2000ep,Eichhorn:2022bgu,Platania:2023srt}.

The dynamics studied in our work investigates local modifications of the gravitational dynamics, characterized by the feature that the number of derivatives appearing in the equations of motion is finite. This already leads 
to a rich phase space of black-hole type objects. Besides (modified) black holes solutions exhibiting event horizons, this space also contains wormholes and naked singularities \cite{Lu:2015cqa,Lu:2015psa,Pravda:2016fue,Bonanno:2019rsq,Silveravalle:2022wij,Daas:2022iid}. This raises the interesting question whether one can identify generic features in the dynamics which determine the structure of the phase space and, in particular, ensure the presence of an event horizon in the geometry. In order to address this question we perform a comparative study within the realm of theories where general relativity is supplemented by local higher-derivative terms. Specifically, we cover the cases of quadratic gravity \cite{Stelle:1976gc,Stelle:1977ry}, Einstein-Cubic Gravity \cite{Bueno:2016xff}, general relativity supplemented by the Goroff-Sagnotti counterterm \cite{Goroff:1985sz,Goroff:1985th}, and the combination of quadratic gravity with the Goroff-Sagnotti term. Following the discussion \cite{Daas:2024pxs}, terms are readily grouped into two categories according to how they manifest themselves in the asymptotically flat regime. In the case of quadratic gravity, the construction of linear perturbations on a flat background gives rise to Yukawa-type interactions associated with the massive degrees of freedom of the theory. These also come with new free parameters, enhancing the space of asymptotically flat solutions as compared to the one available in general relativity. In contrast, Einstein-Cubic Gravity and the Goroff-Sagnotti term do not modify the flat space analysis and just give rise to power-law corrections to the metric functions without giving rise to new, undetermined parameters. Throughout this work, we adopt the viewpoint that the extra degrees of freedom associated with the higher-derivative terms are a genuine feature of the theory. In this sense, we deviate from the effective field theory viewpoint where these effects are excluded by definition.\footnote{Since our analysis deals with classical equations of motion only, we will not discuss the physics implications of these extra degrees of freedom at the level of the quantum theory and refer to \cite{Platania:2020knd,Buoninfante:2025klm} for recent discussions on this point.} The implications of our analysis for an effective field theory setting are readily deduced by switching off the parameter space associated with the additional degrees of freedom. 

At this stage, the following clarification is in order. The analysis performed in this work is orthogonal to the stability analysis \cite{Whitt:1985ki,Deffayet:2023wdg} which studies the time-evolution of small perturbations in the corresponding black hole background. Our work focuses on purely static solutions only. Starting from initial conditions imposed in the asymptotically flat region, we construct the spacetime geometry by ``shooting inward'' towards the center of the geometry. Our interest is in how small perturbations (either given by Yukawa-type corrections or power-law corrections to the Schwarzschild geometry) modify the geometry in the strong gravity regime where one expects the event horizon. In particular, we seek to answer the question whether a small perturbation of the static geometry placed in the asymptotically flat region remains small or destabilizes the horizon. The result has a clear impact on the stability analysis related to the time-evolution of small perturbations since the absence of the event horizon modifies the boundary conditions in the dynamical setting.

The rest of this work is organized as follows. We start with a mini-review on black hole-type objects appearing in quadratic gravity, Einstein-Cubic Gravity, and general relativity supplemented by the Goroff-Sagnotti term in Sect.\ \ref{Sect.review}. This introduces the spacetime geometries investigated in this work. Subsequently, we introduce our analytic framework for determining the fate of the event horizon in Sect.\ \ref{Sect.main}. In Sect.\ \ref{Sect.main2} we work out the general formalism for the black hole-type geometries appearing in general relativity (Sect.\ \ref{Sect.main.1}), Einstein-Cubic Gravity (Sect.\ \ref{Sect.main.2}), quadratic gravity (Sect.\ \ref{Sect.main.3}), general relativity supplemented by the Goroff-Sagnotti counterterm (Sect.\ \ref{Sect.main.4}), and quadratic gravity supplemented by the Goroff-Sagnotti counterterm (Sect.\ \ref{Sect.main.5}). The results of this comperative study are summarized in Table \ref{Table.3}. As part of our conclusions, Sect.\ \ref{Sect.conc} formulates our general conjecture that 
the blow-up mechanism, leading to a removal of the event horizon, is triggered by the presence of extra degrees of freedom visible in a perturbative analysis around flat space. Some technical details are included in two appendices and the ancillary Mathematica notebook. In particular, App.\ \ref{App.A} shows that our conclusions are robust when changing coordinate systems by demonstrating that Schwarzschild- and conformal-to-Kundt coordinates (representing coordinate systems which are singular and regular at an event horizon) give the same answers. Finally, lengthy formulas appearing in the analysis of the quadratic gravity case 
are collected in App.\ \ref{App.B}.

\section{Black holes beyond general relativity - a mini-review}
\label{Sect.review}
Throughout this work, we will focus on static, spherically symmetric spacetimes. The line-elements used to capture these geometries in various coordinate systems are introduced in Sect.\ \ref{Sect.2.1}. The dynamical frameworks determining the geometries are introduced in Sect.\ \ref{Sect.2.2}. This subsection also contains a brief summary of black hole-type solutions found in these settings.
%

\subsection{Spherically symmetric, static spacetimes}
\label{Sect.2.1}
Static, spherically symmetric geometries are determined by two metric functions of a single variable. Using Schwarzschild coordinates $x^\mu = \{t,r,\theta,\phi\}$, their line-element takes the form
\be\label{met1}
ds^2 = -h(r) dt^2 + f(r)^{-1} dr^2 + r^2 \left( d\theta^2 + \sin^2\theta \, d\phi^2 \right) \, . 
\ee
An event horizon corresponds to a point $r_h$ where $f(r_h) = h(r_h) = 0$ while the characteristic feature of a wormhole with a throat at $r_h$ is a point where $f(r_t) = 0$ with $h(r_t) > 0$. Throughout this work, we will demand that the geometries are asymptotically flat, i.e., they approach the Minkowski metric in the limit $r \rightarrow \infty$.

The Schwarzschild solution (SS), giving the spherically symmetric black hole solutions in general relativity, is given by
\be\label{SSsol}
h^{\rm SS}(r) = f^{\rm SS}(r) = 1 - \frac{2 GM}{r} \, . 
\ee
Here $M$ is the asymptotic mass and $G$ denotes Newton's constant. The geometry is asymptotically flat and exhibits an unstable, circular photon orbit at $r= 3GM$. Moreover, there is an event horizon at $r=2GM$ which screens the curvature singularity at $r=0$. The line-element degenerates at the horizon, so that this point corresponds to a coordinate singularity when using Schwarzschild coordinates. 


The Schwarzschild coordinates used in eq.\ \eqref{met1} are not the only way to express the underlying geometry. Depending on the problem at hand, it may be more convenient to move to Eddington-Finkelstein, Kruskal, isotropic, or harmonic coordinates. In the context of solving equations of motion arising in higher-derivative gravity theories, the transformation to conformal-to-Kundt-coordinates has turned out to be extremely powerful \cite{Podolsk__2018,Daas:2023axu}. In this case, the static, spherically symmetric geometry takes the form
\be\label{metctoK}
ds^2 = \Omega^2(\rb) \left[ d\theta^2 + \sin^2\theta d\phi^2 - 2 dv d\rb + \cH(\rb) dv^2 \right] \, . 
\ee
It was shown in \cite{Pravda:2016fue} that all static, spherically symmetric metrics can be cast into conformal-to-Kundt form \eqref{metctoK}. The change of coordinates connecting \eqref{met1} and \eqref{metctoK} is
\be\label{coordchange}
r = \Omega(\rb) \, , \qquad t = v - \int \frac{d\rb}{\cH(\rb)} \, . 
\ee
This entails that the domain of geometrical interest is $\Omega \in [0, \infty[$. This domain may be covered by either a compact or non-compact interval in the new coordinate $\rb$. Moreover, the condition that the change of coordinates is well-defined requires that $\Omega^\prime$ obeys a suitable monotonicity condition on this interval. The metric functions appearing in \eqref{met1} and \eqref{metctoK} are related by
\be\label{metricfctsmap}
h = - \Omega^2 \, \cH \, , \qquad f = - \left( \frac{\Omega^\prime}{\Omega} \right)^2 \, \cH \, . 
\ee
Hence an event horizon, situated at a finite value $r$, is mapped to a root of $\cH$. The Schwarzschild solution written in conformal-to-Kundt coordinates is given by
\be\label{SS-ctK}
\Omega^{\rm SS}(\rb) = - \frac{1}{\rb} \, , \qquad \cH^{\rm SS}(\rb) = - \rb^2 \left(1 + 2 G M \rb \right) \, . 
\ee
One easily verifies that the event horizon is situated at $\rb = - (2GM)^{-1}$.

The main advantage of conformal-to-Kundt coordinates is that the metric \eqref{metctoK} has no explicit $\rb$-dependence. This ensures that the dynamical systems determining the metric functions $\Omega, \cH$ are autonomous and typically much simpler than the ones obtained in Schwarzschild coordinates \cite{Daas:2023axu}. Since the equations of motion determining the metric functions are known to be notoriously difficult to solve due to their non-linearity, this is a very welcome feature which makes the construction and analysis much more tractable.

\subsection{Modified gravitational dynamics }
\label{Sect.2.2}
We proceed by introducing our models giving rise to the modified gravitational dynamics. For the sake of systematics, we limit the discussion to cases where general relativity is supplemented by local higher-derivative terms. Concretely, we consider Einstein-Cubic Gravity (ECG), quadratic gravity (QG), the addition of the Goroff-Sagnotti (GS) counterterm, and the combined contributions of Quadratic Gravity and the Goroff-Sagnotti (QG+GS). The corresponding actions are conveniently written in a unified way
\be\label{actionans}
S[g] = \frac{1}{16\pi G}\int d^4x\sqrt{-g}\big[R + \mathcal{P} ]\,.
\ee
Here $g$ is the determinant of the spacetime metric $g_{\mu\nu}$, $R$ is the Ricci scalar, and we will ignore the contribution of the cosmological constant throughout the analysis. The higher-derivative contributions to the dynamics are captured by $\cP$ and read
\begin{subequations}\label{def:models}
\begin{equation}\label{def:QG}
\begin{split}
    \cP^{\rm QG} = & \, G \Big[ \beta R^2 - \alpha C_{\mu\nu\rho\sigma} C^{\mu\nu\rho\sigma}   \Big]\, ,
    \end{split}
\end{equation}
\begin{equation}\label{def:ECG}
\begin{split}
    \cP^{\rm EQG} = & \, - G^2 \lambda \Big[12 R_{\mu \enspace \nu}^{\enspace \rho \enspace \sigma}R_{\rho\enspace\sigma}^{\enspace\gamma\enspace\delta}R_{\gamma\enspace\delta}^{\enspace\mu\enspace\nu} + R_{\mu\nu}^{\phantom{\alpha\beta}\rho\sigma}R_{\rho\sigma}^{\phantom{\alpha\beta}\gamma\delta}R_{\gamma\delta}^{\phantom{\alpha\beta}\mu\nu} 
    \\ & \qquad \qquad - 12 R_{\mu\nu\rho\sigma}R^{\mu\rho}R^{\nu\sigma} + 8 R_{\mu}^{\phantom{\alpha}\nu}R_{\nu}^{\phantom{\alpha}\rho}R_{\rho}^{\phantom{\alpha}\mu}\Big]\, , 
    \end{split}
\end{equation}
\begin{equation}\label{def:GS}
\begin{split}
    \cP^{\rm GS} = & \, G^2 \lambda \Big[ C_{\mu\nu}^{\phantom{\alpha\beta}\rho\sigma}C_{\rho\sigma}^{\phantom{\alpha\beta}\gamma\delta}C_{\gamma\delta}^{\phantom{\alpha\beta}\mu\nu}  \Big]\, .
    \end{split}
\end{equation}
\end{subequations}
Here $R_{\mu\nu\rho\sigma}$ and $R_{\mu\nu} \equiv R^\alpha{}_{\mu\alpha\nu}$ are the Riemann and Ricci tensor, respectively, and $C_{\mu\nu\rho\sigma}$ is the Weyl tensor. The parameters $\alpha,\beta,\lambda$ have been made dimensionless by extracting suitable powers of Newton's coupling. The equations of motion arising from these actions are readily obtained using computer algebra software as, e.g., the xAct package \cite{xActwebpage} for Mathematica. The resulting expressions are quite cumbersome and little illuminating. Therefore, we refrain from giving explicit expressions at this point and refer the interested reader to the ancillary notebook for details.

As highlighted in \cite{Daas:2024pxs}, the higher-derivative terms introduced in eq.\ \eqref{def:GS} are on different conceptual footing. The quadratic gravity contributions give rise to additional massive degrees of freedom \cite{Stelle:1977ry}, where
\be\label{massesQG}
(m_2)^2 = (2 \, G \, \alpha)^{-1} \, , \qquad (m_0)^2 = (6 \, G \, \beta)^{-1} \, ,
\ee
are the masses of the new spin-two and spin-zero degree of freedom, respectively. These modes manifest themselves as Yukawa-type contributions to the gravitational potential. In contrast, the coupling $\lambda$ associated with the cubic curvature contributions is not associated with new degrees of freedom \cite{Bueno:2016xff,Bueno:2016lrh,Daas:2023axu,Daas:2024pxs} and, in general, leads to power-law corrections of the Schwarzschild geometry at higher order in the post-Newtonian expansion.

\subsection{Black hole-type geometries}
\label{Sect.2.3}
\begin{figure}[t!]
\centering
\includegraphics[width=0.9\textwidth]{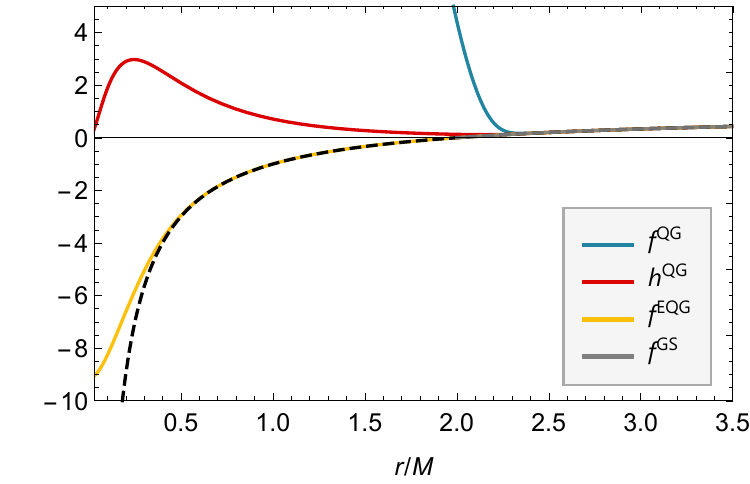}
\caption{\label{Fig.Illustration} Illustration of the metric functions $h(r)$ and $f(r)$ determining static, spherically symmetric geometries \eqref{met1} solving the equations of motion of Einstein Cubic Gravity (ECG) (yellow), general relativity supplemented by the Goroff-Sagnotti (GS) term (gray), and quadratic gravity (QG) (blue and red). The Schwarzschild solution (SS) is added as a black, dashed line for comparison. All solutions essentially follow the Schwarzschild solution up to the event horizon region where $r \approx 2GM$. In the case of quadratic gravity, the solutions exhibit strong dynamics in this regime which eliminates the event horizon. We will refer to this type of dynamics as ``blow-up mechanism''.}
\end{figure}

Equations of motion including the higher-derivative terms \eqref{def:GS} are readily obtained by determining the stationary points of $S[g]$ using a suitable computer algebra software, and subsequently restricting the general equations to spacetime geometries of the form \eqref{met1}. Solutions of these equations have been analyzed based on either local expansions in terms of power series or numerical integration. The goal of this subsection is to briefly review the resulting solution spaces spanned by asymptotically flat geometries. Fig.\ \ref{Fig.Illustration} displays a selection of the solutions discussed below for illustrative purposes. In order to lighten our notation, we will set $G=1$ in the sequel.

\subsubsection{Quadratic Gravity}
Black hole-type solutions of quadratic gravity have already been studied in much detail \cite{Lu:2015cqa,Lu:2015psa,Podolsky:2019gro,Huang:2022urr}, and the structure of the solution space has been given explicitly in \cite{Silveravalle:2022wij,Daas:2022iid}. The Schwarzschild solution is also a solution in this case. Linearizing the equations of motion around flat space, the form of the perturbations can be found analytically
\be\label{lin:QG}
\begin{split}
h(r) = & \, 1 + C_t - \frac{2  M}{r} + 2 S_2^- \frac{e^{-m_2 r}}{r} + S_0^- \, \frac{e^{-m_0 r}}{r} + 2 S_2^+ \frac{e^{m_2 r}}{r} + S_0^+ \, \frac{e^{m_0 r}}{r} \, , \\
f(r) = & \, 1 - \frac{2  M}{r} + S_2^- \frac{e^{-m_2 r}}{r} \left(1+m_2 \, r \right)- S_0^- \, \frac{e^{-m_0 r}}{r} \left(1+m_0 \, r \right)  \, \\ 
& \qquad \qquad \; \; + S_2^+ \frac{e^{m_2 r}}{r} \left(1+m_2 \, r \right)- S_0^+ \, \frac{e^{-m_0 r}}{r} \left(1+m_0 \, r \right) \, . 
\end{split}
\ee
For fixed, non-zero masses $m_0$, $m_2$ the solution space is spanned by six free parameters $\{M, C_t, S_2^-, S_0^-, S_2^+, S_0^+ \}$. These coefficients have a different status: $C_t$ is a redundant parameter which can be set to zero by a redefinition of the time coordinate $t$. Moreover, non-zero values of $S_0^+$ and $S_2^+$ lead to solutions which deviate exponentially from the Schwarzschild solution once $r$ is larger than the masses $m_0$ or $m_2$. Thus asymptotically flat solutions are determined by three free parameters $M$, $S_2^-$, and $S_0^-$. The new parameters determine the strength of the (exponentially decaying) Yukawa-type contributions. Solutions with $S_2^-$ and $S_0^-$ being non-zero thereby satisfy $f(r) \not = h(r)$.

Integrating the equations of motion inwards, the space of solutions can be characterized as follows. The Schwarzschild solution is the only solution which exhibits an event horizon and exists for all values of $M$. For $M$ below a critical value, there is a second branch of black holes exhibiting an event horizon \cite{Lu:2015cqa,Lu:2015psa}. Staying above this critical value, and turning on the new parameters $S_2^-$ or $S_0^-$ removes the event horizon and one deals with naked singularities or wormhole solutions. 
\subsubsection{Einstein-Cubic Gravity}
Motivated by generalized quasi-topological theories of gravity \cite{Hennigar:2017ego,Bueno:2019ltp,Bueno:2019ycr} in four spacetime-dimensions, the specific combination of higher-derivative terms given in $\cP^{\rm ECG}$ has been proposed in \cite{,Bueno:2016xff} and the resulting black hole solutions have subsequently been explored in \cite{Bueno:2016lrh, Hennigar:2016gkm,Hennigar:2018hza}. The specific combination of terms appearing in eq.\ \eqref{def:ECG} ensures that the resulting equations of motion have consistent solutions where $f(r) = h(r)$. This property simplifies the construction of the corresponding solution space in a significant way.

We may then expand the metric function $f(r)$ in the asymptotically flat regime,
\begin{equation}
    f(r) = 1 - \frac{2M}{r} + \sum_{i = 2}^\infty \frac{a_i}{r^i} \, . 
\end{equation}
Substituting this ansatz into the equation of motion and solving for the coefficients $a_i$ yields
\begin{equation}\label{eq: ECG asymptotic polynomial}
    f(r)=1-\frac{2M}{r} - \frac{16  M^2 \left(27 r - 46 M \right) \lambda}{r^7}  -\frac{1536  M^3 \left(1134 r^2-4617 M r + 4669 M^2 \right) \lambda^2}{r^{13}} + \mathcal{O}(\lambda^3).
\end{equation}
In agreement with the general expectation \cite{Holdom:2002xy}, the asymptotic analysis for large values $r$ shows that the leading corrections to the Schwarzschild geometry \eqref{SSsol} are proportional to $\lambda M^2$ and occur at order $1/r^6$. 

Based on the structure of the equations of motion, one can conclude through dimensional analysis that the coefficients appearing at each order in $\lambda$ can contain a finite number of terms coming with different inverse powers of $r$. As a consequence, it is possible to recast the result in terms of a series expansion at an arbitrary reference point $r_0$
\be
f(r) = \sum_{i = 0}^\infty \frac{a_i}{r^i} = \sum_{i = 0}^\infty \frac{a_i}{(r_0 + (r - r_0))^i} \, . 
\ee
The first terms in this expansion read
\begin{equation}\label{eq: ECG polynomial corrections from AF solution}
\begin{split}
    f(r) 
    &= \bigg(f_0^{\rm SS} + \tfrac{16M^2\,(46M - 27r_0)}{r_0^7}\,\lambda - \tfrac{1536M^3\,(4669M^2 - 4617r_0\,M + 1134r_0^2)}{r_0^{13}}\,\lambda^2 + \mathcal{O}(\lambda^3)\bigg) \\
    &\, + \bigg(f_1^{\rm SS} -\tfrac{32M^2\,(161M - 81r_0)}{r_0^8}\,\lambda + \tfrac{1536M^3\,(60697M^2 - 55404r_0\,M + 12474r_0^2)}{r_0^{14}}\,\lambda^2 + \mathcal{O}(\lambda^3)\bigg)\,(r - r_0)\\
    &\, + \mathcal{O}(r-r_0)^2\,.
\end{split}
\end{equation}
This intermediate result will be useful when assessing the stability of the solution \eqref{eq: ECG asymptotic polynomial} in the vicinity of an event horizon in Sect.\ \ref{Sect.main.2}.

Following a strategy similar to the one leading to the asymptotic solutions in quadratic gravity, one may also expand the metric about a background geometry and study the equations of motion for the linearized perturbations. For a Schwarzschild background, this has been implemented in \cite{Bueno:2016lrh}. The perturbation then satisfies a linear second-order differential equation. Besides the particular solution to this equation, \cite{Bueno:2016lrh} also constructed the homogeneous solutions of this system. These solutions exhibit exponential growth or decay as $r \rightarrow \infty$.  For $r \rightarrow \infty$, the decaying branch is given by the approximate solution 
\begin{equation}\label{eq: ECG exponential correction}
    f(r) \approx 1 - \frac{2M}{r} + B\, \mbox{exp}\left(-\frac{r^{5/2}}{15\sqrt{2M \lambda }}\right)\,.
\end{equation}
The exponential corrections are non-analytic in $M$ and the coupling $\lambda$. Comparing to \eqref{lin:QG}, it is also evident that the exponential decay is faster than the one associated with the Yukawa corrections \eqref{lin:QG}. This nurtured the interpretation that the exponential corrections are not associated with new degrees of freedom but constitute ``pseudo-modes''. The exponential tail also disappears from the expansion \eqref{eq: ECG asymptotic polynomial}, since the term is non-analytic at the expansion point \cite{Saueressig:2021wam}.

Global solutions are readily obtained by solving the equations of motion numerically. These solutions exhibit an event horizon, and the metric function $f(r)$ remains finite and smooth at the core. Despite $f(r)$ being regular everywhere, the core at $r=0$ is still a singularity, as the Kretschmann scalar $K \equiv R_{\mu\nu\rho\sigma}R^{\mu\nu\rho\sigma}$ diverges as $r^{-4}$. This represents an improvement over the singularity of the Schwarzschild black hole, where the Kretschmann invariant scales as $r^{-6}$ \cite{Bueno:2016lrh}. Although the first evidence of traversable wormholes connecting two asymptotically AdS spacetimes has been found \cite{Lu:2024prz}, this spacetime exhibits a geometric deficit. Furthermore, no naked singularities have been found in the solution space.

\subsubsection{Solutions including the Goroff-Sagnotti term}
The space of static, spherically symmetric solutions in the presence of the GS-term has been investigated in \cite{Alvarez:2023gfg,Daas:2023axu}, but the structure of the solution space is not completely understood yet. The asymptotic analysis in the weak field regime shows that asymptotically flat solutions are completely fixed by their asymptotic mass $M$ once $\lambda$ is given. The leading corrections are polynomial in $1/r$ and appear at 6$^{\rm th}$ order in the Post-Newtonian expansion \cite{Anselmi:2013wha,deRham:2020ejn} and are given by 
\be\label{GS:asmptcorrections}
\begin{split}
h(r) = & 1 -\frac{2 M}{r} + \frac{8 M^2 \left(3 r - 4 M\right) \lambda}{r^7} + \mathcal{O}\left(\frac{\lambda^2\, M^3}{r^{11}} \right) \, ,  \\
f(\rb) = & 1 - \frac{2 M}{r} + \frac{8 M^2\left(9 r - 16 M\right) \lambda}{r^7} + \mathcal{O}\left(\frac{\lambda^2\, M^3}{r^{11}} \right) \, .  
\end{split}
\ee
Already the leading corrections to the SS-geometry break the degeneracy between the metric functions so that $f(r) \not = h(r)$ if $\lambda$ is non-vanishing. As long as $\lambda > 0$ is sufficiently small the asymptotic solution can uniquely be extended up to an event horizon by numerical integration, giving rise to a black hole solution. The power series \eqref{GS:asmptcorrections} remains an excellent approximation throughout the entire exterior of the spacetime \cite{Daas:2023axu}.  

\subsubsection{Quadratic Gravity supplemented by the Goroff-Sagnotti term} 
To the best of our knowledge, this work is the first case where this particular combination of higher-derivative terms is studied in detail.

\section{Stability of the event horizon - general framework}
\label{Sect.main}
The event horizon is a critical feature when it comes to understanding black hole-type objects beyond general relativity. At the level of static spacetimes obtained from general relativity this feature is essentially omnipresent and thus often taken for granted. Once higher-order derivative terms are added to the dynamics, this feature becomes non-trivial though. The reason is that the modified equations of motions generically possess fixed singularities situated at the points where the metric functions vanish. This property also persists in coordinate systems which are regular at a spacetime horizon. A particularly simple illustration of this feature is provided by the equations of motion arising from the GS-system \eqref{def:GS} written in terms of conformal-to-Kundt coordinates \cite{Daas:2023axu}.

The pole in the equations of motion may render the near-horizon region highly dynamical in the sense that the metric functions change rapidly in a very small coordinate interval. For solutions starting in the asymptotically flat region and extended towards the center of the geometry, this effect can give rise to several distinct physics scenarios: 
\begin{itemize}
\item[a)] The dynamics associated with the pole drives the solution into the singular point. In this case, the solution may not continue past the pole and terminates at the horizon.  
\item[b)] The dynamics drives the solution into the pole. The solution is special in the following sense though. The pole, associated with the vanishing of the denominator, is compensated by a root of sufficiently high order appearing in the numerator. In this case, the solution passes through the singular point and continues into the spacetime region inside the horizon.
\item[c)] The pole is avoided dynamically. The horizon configuration is repulsive in the sense that solutions are driven away from the pole, thereby preventing the occurrence of the event horizon. This mechanism allows to continue the solutions beyond the point where the Schwarzschild solution with the same asymptotic mass would have its event horizon.
\end{itemize}
The options a) and b) create an event horizon while option c) results in a horizonless object which may be given by a wormhole or a naked singularity.

Clearly, it is desirable to understand these scenarios in more detail. The numerical integration of the equations of motion thereby comes with the drawback that it may not be able to reliably identify the correct scenario since numerical errors in the solution may easily convert option b) into option a). Hence an analytic understanding of the local behavior close to the pole is highly desirable. In the remainder of this section, we describe the generic strategy for such an analysis. In Sect.\ \ref{Sect.main2}, this framework will then be applied to the dynamics specified in \eqref{def:models}. Throughout the exposition, we will work with Schwarzschild coordinates \eqref{SSsol}, using the metric functions $h(r)$ and $f(r)$. The same strategy is readily extended to other coordinate systems as well (cf.\ App.\ \ref{App.A}). 

We are interested in the local stability properties of the metric functions. We then choose an arbitrary reference point $r_0$. Assuming that the metric functions are regular at this point, they can be represented in terms of a power series
\be\label{powerseries}
h(r) = \sum_{n=0}^\infty \, h_n (r-r_0)^n \, , \qquad \qquad 
f(r) = \sum_{n=0}^\infty \, f_n (r-r_0)^n \, . 
\ee
For the Schwarzschild solution \eqref{SSsol}, the expansion coefficients read
\be\label{coeff.SS}
h_0^{\rm SS} = 1 - \frac{2M}{r_0} \, , \qquad \quad
h_n^{\rm SS} = 2 M \left( - \frac{1}{r_0} \right)^{n+1} \, , \; n \ge 1 \, , 
\ee
with $f_n^{\rm SS} = h_n^{\rm SS}$. Since these coefficients arise from expanding a geometric series, it is straightforward to verify that the series \eqref{powerseries} converges for $r \in ]0, 2 r_0 [$. 

Substituting the expansions \eqref{powerseries} into the equations of motion, the Frobenius method allows to determine the expansion coefficients in terms of a low number $N^{\rm free}$ of undetermined parameters
\be\label{a-relevant}
a_i \in \{ h_n, f_n \} \, , \qquad i = 0,\cdots, N^{\rm free}-1 \, . 
\ee
Canonically, one chooses coefficients $h_n$ and $f_n$ with the lowest possible index value. The precise choice depends on the underlying dynamical system though. Note that the local expansion is agnostic about the condition that the resulting solution is asymptotically flat. Hence $N^{\rm free}$ can be larger than the sets of free parameters discussed in Sect.\ \ref{Sect.main2}.

The next step perturbs the coefficients $a_i$ around their value for the Schwarzschild geometry
\begin{equation}\label{aidef}
    a_i = a_i^{\rm SS} + \delta_i\, , \qquad i = 0,\cdots, N^{\rm free}-1 \, , 
\end{equation}
where the perturbations $\delta_i$ are taken to be small. The relations obtained from the Frobenius analysis propagate the perturbations of the free parameters into perturbations of the other expansion coefficients. Working at linear order in $\delta_i$, the resulting relations are of the schematic form
\be\label{ND-expressions}
\begin{split}
h_n = & \, h_n^{\rm SS} +\frac{1}{D}\sum_{i=0}^{N^{\rm free}-1} N_{n,i}^h \, \delta_i + \mathcal{O}(\delta_i^2)\,, \qquad
    f_n =  \, f_n^{\rm SS} +\frac{1}{D}\sum_{i=0}^{N^{\rm free}-1} N_{n,i}^f \,\delta_i + \mathcal{O}(\delta_i^2)\, . 
\end{split}
\ee
For the constraint coefficients, the expressions $\{N_{n,i}^h, N_{n,i}^f, D\}$ depend on $\{h_n^{\rm SS}, f_n^{\rm SS} \}$ and the reference point $r_0$. Compatibility with \eqref{aidef} implies that for the free coefficients the denominator equals one, while the numerators are given by Kronecker symbols.

In the next step, one investigates the dependence of the denominator $D$ on $r_0 \in (0, \infty )$. Points associated with a strong dynamics correspond to values $r_0^{\rm sing}$ where the denominators vanish. In this case, one then distinguishes two cases:  the corresponding numerator $\{ N_{n,i}^h, N^f_{n,i}\}$ may come with a zero of the same order. In this case, the combination $N/D$ remains bounded. Conversely, if the root in the denominator is not compensated by a root in the numerator, the coefficient $N/D$ will blow up close to $r_0^{\rm sing}$. In this case, small deviations $\delta_i$ have a large impact on the coefficients $\{h_n, f_n\}$. So even as the perturbations $\delta_i$ can be arbitrarily small, by choosing an expansion around a point $r_0$ close enough to $r_0^{\rm sing}$, some of the coefficients $\{h_n, f_n\}$ deviate substantially from their Schwarzschild values. This is the origin of the instability underlying the options a)-c) stated above. 
\section{Event horizons in the presence of modified dynamics}
\label{Sect.main2}
We proceed to the central analysis of our work, investigating the position and stability of an event horizon when small perturbations about the Schwarzschild background are included. We start with a detailed exposition for the case of general relativity and Einstein-Cubic Gravity in Sects.\ \ref{Sect.main.1} and \ref{Sect.main.2} before proceeding to the case of quadratic gravity in Sect.\ \ref{Sect.main.3}. The cases including the Goroff-Sagnotti term are analyzed in Sects.\ \ref{Sect.main.4} and \ref{Sect.main.5}.
\subsection{General Relativity}
\label{Sect.main.1}
We start by illustrating the stability analysis of the event horizon within general relativity. In this case, Birkhoff's theorem ensures that spherically symmetric solutions of the vacuum field equation $R_{\mu\nu} = 0$ must be static and asymptotically flat \cite{Wald:1984rg}. The only solutions are then the Schwarzschild black holes characterized by their asymptotic mass $M$ and an event horizon situated at $r=2M$. 

The local analysis close to the event horizon then proceeds as follows. Substituting the ansatz \eqref{met1} into the vacuum Einstein's equations gives rise to two independent differential equations for the metric functions $f(r)$ and $h(r)$,
\be\label{diff-GR}
\frac{1 - f(r) - r f^\prime(r)}{r^2} = 0 \, , \qquad 
\frac{(1 - f(r)) h(r) - r f(r) h^\prime(r)}{r^2 h(r)} = 0 \, . 
\ee
Here we deliberately retained the denominators which could, potentially, give rise to an instability at the horizon. Substituting the expansion \eqref{powerseries} and carrying out the Frobenius analysis at the generic expansion point $r_0$ gives rise to two free parameters
\be\label{free-parameters-SS}
a_i = \{f_0, h_0 \} \, , \qquad i = 0,1 \, . 
\ee
These are associated with the mass $M$ of the solution and the possibility of redefining the time-coordinate by a multiplicative factor, $t \mapsto c t$. The Frobenius analysis then expresses the higher-order coefficients in terms of the $a_i$'s. At the lowest order, the relations read
\be
\begin{split}
f_1 = \frac{1-a_0}{r_0} \, , \qquad h_1 = -\frac{1-a_0}{r_0}  \, \frac{a_1}{a_0} \, . 
\end{split}
\ee

In the next step, we perturb the free parameters around the Schwarzschild solution,
\be\label{pert-SS2}
a_i = a_i^{\rm SS} + \delta_i \, .
\ee
The equation fixing $f_1$ does not give rise to a pole and therefore will not put constraints on the deformation parameters $\delta_i$. Perturbing the equation for $h_1$ gives
\be\label{SS-constraint}
h_1 = h_1^{\rm SS} + \frac{2 M \delta_0 - r_0 \delta_1}{r_0 (r_0 - 2M)} + \cO(\delta^2) \, . 
\ee
Formally, we can now move the expansion point close to the horizon of the unperturbed solution, setting $r_0 = 2 M + \epsilon$. Here $\epsilon$ serves as a control parameter fixing the order of the pole and the perturbations. 
Substituting the expression for $r_0$ into \eqref{SS-constraint} and expanding in $\epsilon$ highlights the $1/\epsilon$-pole in the expansion
\be\label{h1-def}
h_1 = h_1^{\rm SS} + \frac{1}{\epsilon} \left(\delta_0 - \delta_1 \right) - \frac{\delta_0}{2M}  + \cO(\epsilon,\delta^2) \, . 
\ee
For solutions avoiding the instability, the numerator on top of the pole has to vanish. This is satisfied if the perturbations in \eqref{pert-SS2} satisfy
\be\label{pole-avoidance-condition-GR}
\delta \equiv \delta_0 = \delta_1 \, . 
\ee
The same condition is obtained when considering perturbations of the numerator and denominators in $h_1$ separately and asking that both expressions have a common root. Remarkably, the condition \eqref{pole-avoidance-condition-GR} \emph{is sufficient to also eliminate the blow-up mechanism in the higher-order coefficients} $\{f_n, h_n\}$, $n \ge 2$ as well. Thus, at this point, we are left with a one-parameter family of deformations which avoids an instability at the horizon.

In order to understand this result, it is instructive to consider a perturbation of the mass parameter $M$ featuring in the Schwarzschild background
\be\label{SS-mass-shift}
M^\prime = M - \frac{1}{2} \, r_0 \, \delta \, . 
\ee
At the level of the free parameters $a_i$, this induces the deformation \eqref{pert-SS2}, provided that the identification \eqref{pole-avoidance-condition-GR} is used. Moreover, the shift in the mass also generates the $\epsilon$-independent deformations appearing in the Frobenius coefficients $f_n, h_n$, $n \ge 1$. This is easily checked explicitly for $h_1^{\rm SS}$ by noting that
\be
h_1^{\rm SS}(M^\prime) = h_1^{\rm SS}(M) - \frac{\delta}{2M} + \cO(\epsilon) \, ,
\ee
and comparing to the expansion \eqref{h1-def}. This holds for all values of $n$. Thus the remaining free parameter $\delta$ can be absorbed into a shift in $M$. 

Hence, we arrive at the following conclusion. Fixing the asymptotic mass of the solution to be $M$ fixes $\delta = 0$. Conversely, it is also clear that solutions with non-zero $\delta$ can pass the pole in the Frobenius coefficients since these just correspond to the Schwarzschild solution with a different asymptotic mass $M^\prime$. Thus the Schwarzschild solution is the only solution that is not subject to the blow-up mechanism at the horizon. This is of course compatible with the well-known properties of spherically symmetric vacuum solutions of general relativity and, in particular, Birkhoff's theorem. 
\subsection{Einstein-Cubic Gravity}
\label{Sect.main.2}
The defining feature of Einstein-Cubic Gravity is that the equations of motion support solutions where the metric functions coincide, $f(r) = h(r)$. As a consequence, it is sufficient to study the stability of $f(r)$ which leads to significant simplifications. Thus, we start with this specific case.
\subsubsection{Analytic perturbations}
We consider the expansion of $f(r)$ at an arbitrary point $r_0$. Stipulating that $f(r)$ is analytic at the expansion point allows to expand the function in a power series
\begin{equation}\label{eq: ECG expansion}
    f(r) = \sum_{n = 0}^\infty f_n\,(r - r_0)^n\,.
\end{equation}
Generically, one has $f_0 \not = 0$. The case $f_0 = 0$ is special in the sense that   $r_0$ corresponds to the position of a horizon. This follows from the observation that $f_0 = 0$ entails that $f(r_0) = 0$.

\begin{table}[t]
\begin{tabular}{ll}
\hline \hline
 $\bigg. p_2^{(0)}$ &   $r_0^2 \, \big(2 M - r_0 (1-f_0) \big)$ \\
 $\bigg. p_2^{(1)}$ &   $4 f_1 \, \big(6 f_0 (1-f_0)+ r_0 f_1 (3 + r_0 f_1) \big)$ \\ \hline
 $\bigg. p_3^{(0)}$ & $- r_0^4 \, \big(2 M - r_0 (1-f_0) \big)^2 $\\
 \multirow{2}{*}{$\bigg. p_3^{(1)}$} & 
 $4 M r_0 \big( 48 f_0 \, (1 - f_0)^2 + 48 f_0 f_1 r_0 (1 - f_0)  - 4 r_0^2 f_1^2 (3 - 6 f_0 + r_0 f_1)\big)$  
 \\[1.2ex]
 & $-4r_0 \, \big( 36 f_0 r_0 \, (1 - f_0)^3  
 + 24 r_0^2 f_0 f_1 \, (1 - f_0)^2 
 -r_0^3 f_1^2 (6 - 9 f_0 + 3 f_0^2 + 2 r_0 f_1 + r_0 f_0 f_1)
 \big)$\\[1.2ex]
\multirow{2}{*}{$\bigg. p_3^{(2)}$} & $16 f_1^2 \, \bigg( 
 108 f_0 \, (1 - f_0)^3 
 + 132 r_0 f_0 f_1 \, (1 - f_0)^2$ \\[1.2ex] & \qquad \quad 
 $- r_0^2 f_1^2 \, \big(3 (1 - f_0) (3 - 20 f_0) + r_0 f_1 (6 - 12 f_0 + r_0 f_1)\big)  
 \bigg)$ \\
\hline \hline
\end{tabular}
\caption{\label{Tab.1} Top: polynomials appearing in the coefficients $f_2$ and $f_3$, eq.\ \eqref{eq: ECG f2}.}
\end{table}
Substituting this ansatz into the equations of motion and performing a Frobenius analysis reveals that the system has two free parameters. These are chosen as $\{a_i\} = \{f_0, f_1\}$, $i=1,2$. In addition, there is another free parameter $M$ which arises as an integration constant in the equations of motion and can be identified with the asymptotic mass of the geometry. The coefficients $f_n(f_0, f_1, M; r_0)$ with $n \ge 2$ are obtained as functions of these parameters and the expansion point $r_0$. For the two lowest coefficients, the explicit expressions are
\be\label{eq: ECG f2}
\begin{split}
f_2 = &\frac{1}{24 \, r_0\,\lambda\,f_0\,(2 - 2f_0 + r_0\,f_1)} \left( p_2^{(0)} + \lambda p_2^{(1)} \right) \, , \\
f_3 = & \frac{1}{864 \, r_0 \, \lambda^2 \, f_0^2 \left(2 - 2f_0 + r_0 f_1\right)^3} \left( p_3^{(0)} + \lambda p_3^{(1)}  + \lambda^2 p_3^{(2)} \right) \, . 
\end{split}
\ee
Anticipating the structures relevant in the latter discussion, we wrote the numerators as power series in the higher-derivative coupling $\lambda$ and listed the explicit form of the prefactors $p_n^{(a)}$ in Table \ref{Tab.1}.

At this point the following observation is in order. We find that
\be 
f_n \propto (\lambda f_0)^{1-n} \, , \quad n \ge 2 \, . 
\ee
 Hence the coefficients $f_n$ may diverge at a horizon unless there is a cancellation in the numerator. This is the manifestation of the (potential) blow-up mechanism introduced in the previous section. Secondly, the appearance of $\lambda$ in the denominator makes a well-defined limit $\lambda \rightarrow 0$ non-trivial. This reflects that the order of the equations of motion is set by the higher-order curvature terms supplementing the Einstein-Hilbert action. A well-defined limit $\lambda \rightarrow 0$ then requires the vanishing of $p_2^{(0)}, p_3^{(0)}$ and $p_3^{(1)}$. This condition fixes $f_0 =f_0^{\rm SS}$ and $f_1 = \{-\frac{4 M}{r_0^2},\frac{2 M}{r_0^2} \}$. The first option is excluded, since it corresponds to a pole generated by the vanishing of $(2-2 f_0 + r_0 f_1)$. Thus imposing regularity of the limit $\lambda \rightarrow 0$ fixes two of the three free parameters to their Schwarzschild values \eqref{coeff.SS} up to deformations proportional to $\lambda$.

We proceed by studying perturbations of the free parameters about their Schwarzschild values, keeping the asymptotic mass $M$ fixed. It is natural to organize the perturbations in terms of a power series in $\lambda$, i.e., 
\be\label{relevant-perturbation}
f_i = f_i^{\rm SS} + \lambda \, \delta_i^{(1)} + \lambda^2 \, \delta_i^{(2)} + \cO(\lambda^3) \, , \qquad i=0,1. 
\ee
Here we already included the result that deformations at zeroth order in $\lambda$ are excluded. In order to keep the resulting expressions compact, we carry out the analysis using a bootstrap strategy and work order by order in the coupling $\lambda$. Substituting the perturbative expansion into \eqref{eq: ECG f2}, the $\lambda$-independent terms are found to be
\be
\begin{split}
f_2 = &  -\frac{160 \, M^3 - 144 \, M^2 \, r_0 +r_0^7 \, \delta^{(1)}_0}{144 M \, r_0^3 \, (r_0 - 2 M)} + \cO(\lambda) \, , \\
f_3 = & - \frac{1}{186624 M^3 r_0^4 (r_0-2 M)^2} \Big( 320512 M^6-152064 M^5 r_0 + (\delta^{(1)}_0)^2 r_0^{14}  \\
& \qquad +32 M^3 r_0^7 (89 \delta^{(1)}_0 +27 \delta^{(1)}_1 r_0) -432 M^2 r_0^8 (3 \delta^{(1)}_0 +\delta^{(1)}_1 r_0) \Big) + \cO(\lambda) \, . 
\end{split}
\ee
Imposing that the limit $\lambda \rightarrow 0$ of these expressions agrees with the Schwarzschild values, i.e.\ 
$
\lim_{\lambda \to 0}f_{2} = f_{2}^{\rm SS}$ and $\lim_{\lambda \to 0}f_{3} = f_{3}^{\rm SS}$,
fixes the expansion coefficients $\delta_i^{(1)}$
\begin{equation}\label{eq: ECG delta condition 1}
    \delta_0^{(1)} = \frac{16M^2\,(46M - 27r_0)}{r_0^7} \,,\quad \delta_1^{(1)} = -\frac{32M^2\,(161M - 81r_0)}{r_0^8} \, . 
\end{equation}
Inserting this result into the expansion of \eqref{eq: ECG f2} and moving to the next order in $\lambda$ then yields
\begin{equation}\label{eq: ECG pole appearence}
\begin{split}
    &f_2 = f_2^{\rm SS} + \frac{1236480M^5 - 1511424\, r_0\,M^4 + 435456r_0^2\,M^3 + r_0^{13}\,\delta_0^{(2)}}{144\, r_0^9\,M\,(r_0 - 2M)}\,\lambda + \mathcal{O}(\lambda^2)\,, \\
    &f_3 = f_3^{\rm SS} - \frac{3956736M^5 - 2032128 \, r_0\,M^4 - r_0^{13}\,(5\delta_0^{(2)} + r_0\,\delta_1^{(2)})}{432 \, r_0^{10}\,M\,(r_0-2M)}\,\lambda + \mathcal{O}(\lambda^2)\,.
\end{split}
\end{equation}
The key feature of the new terms is the vanishing of the denominator at $r_0 = 2M$.\footnote{The position of the pole up to terms of order $\lambda^3$ is given in eq.\ \eqref{pole-corrected}. Given our bootstrap method, the $\mathcal{O}(\lambda)$ corrections to the pole are included in the expansion in $\lambda$. Applying this expansion in the denominator changes nothing to the proceeding analysis though.} This signals a potential instability of the solutions: a small perturbation in the initial perturbations $\delta_i^{(2)}$ can magnify and have a drastic impact on $f_n$ if the expansion point $r_0$ is moved close to this root. This will happen for generic values $\delta_i^{(2)}$. The exception occurs if the numerator also develops a zero at $r_0=2M$. A straightforward computation shows that this occurs for the special values
\begin{equation}\label{eq: ECG delta condition 2}
    \delta_0^{(2)}\big|_{r_0 = 2M} = \frac{87}{16M^8}\,,\quad \delta_1^{(2)}\big|_{r_0 = 2M} = -\frac{645}{32M^9}\,.
\end{equation}
Thus, it is possible for the solution to be close to the Schwarzschild solution near the would-be horizon, if and only if the conditions \eqref{eq: ECG delta condition 1} and \eqref{eq: ECG delta condition 2} are satisfied. 

At this stage, it is interesting to compare these conditions to the polynomial solution \eqref{eq: ECG asymptotic polynomial}. Recasting the resummed expressions 
\eqref{eq: ECG polynomial corrections from AF solution} into the form \eqref{relevant-perturbation} shows that the solution corresponds to
\be\label{poly-sol-coeff}
\begin{array}{ll}
\delta_0^{(1)} = \tfrac{16M^2\,(46M - 27r_0)}{r_0^7} \, , \quad & \delta_0^{(2)} = - \tfrac{1536M^3\,(4669M^2 - 4617r_0\,M + 1134r_0^2)}{r_0^{13}} \, , \\[1.2ex]
\delta_1^{(1)} = -\tfrac{32M^2\,(161M - 81r_0)}{r_0^8} \, , \quad & \delta_1^{(2)} = \tfrac{1536M^3\,(60697M^2 - 55404r_0\,M + 12474r_0^2)}{r_0^{14}} \, . \\
\end{array}
\ee
Taking the limit $r_0 \rightarrow 2M$ of these expressions then establishes that the polynomial corrections to the Schwarzschild solution dictated by Einstein-Cubic Gravity indeed satisfy the stability criteria \eqref{eq: ECG delta condition 1} and \eqref{eq: ECG delta condition 2}. Hence, we expect that asymptotically flat solutions can pass an event horizon without being subject to the blow-up mechanism. This is in agreement with the results obtained by numerical integration, c.f.\ Fig.\ \ref{Fig.Illustration}.

\subsubsection{Non-analytic perturbations}
When setting up the stability analysis in eq.\ \eqref{relevant-perturbation}, we made the assumption that the perturbation is analytic in the coupling $\lambda$. For the exponential terms displayed in eq.\ \eqref{eq: ECG exponential correction} this is not the case though. In order to incorporate these contributions, we perturb the analytic part of the solution by additional parameters $\tilde{\delta}_i$,
\begin{equation}\label{eq: ECG free par assumption exp}
    f_i = f_i^{\rm SS} + \lambda \, \delta_i^{(1)} + \lambda^2 \, \delta_i^{(2)} + \cO(\lambda^3) + \td_i  \, , \qquad i=0,1 \, . 
\end{equation}
Ultimately, the $\delta_i^{(a)}$ will be fixed to the result \eqref{poly-sol-coeff}, but at this point, we retain them in general form. The only assumption entering in \eqref{eq: ECG free par assumption exp} is that the $\td_i$ are clearly distinguishable from the analytic perturbations.

Substituting \eqref{eq: ECG free par assumption exp} into the expressions for $f_2$ and $f_3$ given in eq.\ \eqref{eq: ECG f2} and focusing on the terms linear in $\td_i$ shows that the stability against perturbations is controlled by an expression of the schematic form
\be
f_n =   \frac{1}{D} \left( N_{n,0} \, \td_0 + N_{n,1} \, \td_1 \right) + \ldots \,  ,  
\ee
where the dots represent all the terms analytic in $\lambda$. Thus we encounter the general structure anticipated in eq.\ \eqref{ND-expressions}. While the explicit expression for $N_{n,i}$ and $D$ are lengthy and little illuminating it is straightforward to establish the following properties. The denominator $D$ possesses a root of multiplicity one at 
\begin{equation}\label{pole-corrected}
    r_0^{\rm pole} = 2M + \frac{2}{M^3}\,\lambda - \frac{75}{8M^7}\,\lambda^2 + \mathcal{O}(\lambda^3)\,.
\end{equation}
The value of $r_0^{\rm pole}$ is the point where the function $f(r)$ vanishes, corresponding to the shifted position of the event horizon. Notably, reading off the position of the horizon from the root structure of $D$ is much more efficient than doing so by numerical integration.

Setting the polynomial corrections $\delta_0, \delta_1$ equal to the one obtained from the full asymptotically flat solution \eqref{eq: ECG polynomial corrections from AF solution}, substituting $r_0 = r_0^{\rm pole}$ and expanding in $\lambda$ yields 
\begin{equation}
    N_{n,i} = 0\, , \qquad i = 0,1 \, . 
\end{equation}
Thus the pole induced by the root of $D$ is canceled for the specific polynomial solution. This property is independent of the precise form of the $\td_i$. Thus we conclude that the exponential corrections do not trigger an instability near the horizon. Again, this property is confirmed by solving the exact equations of motion numerically.

\subsection{Quadratic Gravity}
\label{Sect.main.3}
We proceed with investigating the stability of the event horizon in quadratic gravity. Thus, eq.\ \eqref{lin:QG} indicates that, unlike EQG, the metric functions $h(r)$ and $f(r)$ are no longer degenerate. The local analysis of the equations of motion based on the Frobenius method takes this feature into account by allowing for different expansion coefficients in the metric function, c.f.\ eq.\ \eqref{powerseries}. Expanding the solutions at a generic, regular point $r_0$ then shows that there are six free parameters spanning the local solution space. Following the notation in eq.\ \eqref{a-relevant}, we take these parameters as \cite{Perkins:2016imn}
\be\label{free-parameters-QG}
a_i = \{f_0, h_0, f_1, h_1, f_2, h_2 \} \, , \qquad i = 0,\cdots,5 \, . 
\ee
This number agrees with the number of free parameters seen in the linearized solution \eqref{lin:QG}. This also entails that physically distinct, asymptotically flat solutions are found within a subspace spanned by the $a_i$'s only. The higher-order coefficients appearing in the expansion \eqref{powerseries} are then obtained as functions of the free parameters and the expansion point $r_0$. Schematically,
\be\label{Frob.sol.QG}
h_n = h_n(a_i; r_0)  \, , \qquad f_n = f_n(a_i; r_0)  \, , \qquad n \ge 3 \, .  
\ee
The stability of a solution in the vicinity of a horizon will then be inferred based on the relations satisfied by $f_3$, $h_3$, $h_4$, and $f_4$. The explicit form of these relations is rather lengthy and little illuminating. Thus we do not give them at this point and refer to the ancillary notebook. Moreover, we reduce the complexity of the computation by adopting a bootstrap approach, i.e., we determine the constraints following the precise order noted above and then use the results from the previous steps to simplify the next step in the analysis.

Following our general strategy, we perturb the local solutions about the Schwarzschild solution
\be\label{def.QG}
a_i = a_i^{\rm SS} + \, \delta_i \, , \qquad i = 0,\cdots,5 \, .  
\ee
 In contrast to ECG these perturbations may also induce a change in the asymptotic mass of the perturbed solution. We retain this freedom, studying the effect of all perturbations included in \eqref{def.QG}. 

In the next step, we substitute the deformations \eqref{def.QG} into the relations \eqref{Frob.sol.QG}. Furthermore, we choose the expansion point close to the position of the event horizon for the Schwarzschild black hole with the same asymptotic mass, $r_0 = 2M + \epsilon$. Subsequently, we perform a double expansion tracking all terms with negative powers of $\epsilon$ and up to linear order in the perturbations $\delta_i$. Starting with $f_3(a_i; r_0)$ this results in the following, schematic expression
\begin{equation}\label{f3expansion}
    f_3 = f_3^{\rm SS} + \frac{1}{\epsilon^3}\Big[ N_{3,0}^f(\delta_i) + N_{3,1}^f(\delta_i)\, \epsilon + N_{3,2}^f(\delta_i)\, \epsilon^2 \Big] + \mathcal{O}\left(\epsilon^0\right)\,.
\end{equation}
The explicit expressions for $N_{3,j}^f$, $j = 0,1,2$, are listed in the first block of Table \ref{Tab.QG}, provided in App.\ \ref{App.B}. The terms of $\cO(\epsilon^0)$ are finite as $\epsilon = 0$ and thus are irrelevant for the blow-up mechanism. Requiring that the $N_{3,j}^f$ vanish then yields three constraints on the $\delta_i$
\begin{equation}\label{conditions f3}
    \begin{split}
        &\delta_0 = \delta_1\,, \\  
        & \delta_2 = \delta_3\,,\\
        &\delta_4 = \frac{1}{27\beta}\Big(\delta_1 + 4M\,\delta_3 + \big(4M^2 + 27 \beta)\,\delta_5\Big) + \frac{6}{27\alpha}\Big(\delta_1 + M\,\delta_3 - 2M^2 \,\delta_5\Big)\,.
    \end{split}
\end{equation}
The first condition has the profound consequence that it enforces $f(r) = h(r)$ at $r= r^{\rm pole}$ in order to avoid the instability. As a consequence, we must have $f(r=r^{\rm pole}) = h(r=r^{\rm pole}) = 0$ at the horizon. Generically, this may not be the case.

We proceed by analyzing the constraints arising from $h_3$. Expanding in powers of $\epsilon$ yields
\begin{equation}\label{h3expansion}
    h_3 = h_3^{\rm SS} + \frac{1}{\epsilon^3}\Big[ N_{3,0}^h(\delta_i) + N_{3,1}^h(\delta_i)\, \epsilon + N_{3,2}^h(\delta_i)\, \epsilon^2 \Big] + \mathcal{O}\left(\epsilon^0\right)\,.
\end{equation}
The expansion coefficients are given in the second block of Table \ref{Tab.QG}. Substituting the relations \eqref{conditions f3} shows that $N_{3,0}^h(\delta_i)$ and $N_{3,1}^h(\delta_i)$ vanish already and do not give rise to new constraints. Solving $N_{3,2}^h = 0$ gives the new relation
\be\label{conditions h3}
 \delta_5 = -\,\frac{6\beta\big(\delta_1 + M\,\delta_3\big) + \alpha\big(\delta_1 + 4M\,\delta_3\big)}{4M^2(\alpha - 3\beta)}\,.
\ee

In the next step, it is convenient to analyze $h_4$. In this case, the expansion starts with a potential pole of order four in $\epsilon$
\begin{equation}\label{h4expansion}
    h_4 = h_4^{\rm SS} + \frac{1}{\epsilon^4}\Big[ N_{4,0}^h(\delta_i) + N_{4,1}^h(\delta_i)\, \epsilon + N_{4,2}^h(\delta_i)\, \epsilon^2 + N_{4,3}^h(\delta_i)\, \epsilon^3 \Big] + \mathcal{O}\left(\epsilon^0\right)\, ,
\end{equation}
with the expansion coefficients in the third block of Table \ref{Tab.QG}.
Imposing the relations \eqref{conditions f3} and \eqref{conditions h3} the only non-trivial relation results from $N_{4,3}^h$. This leads to the relation
\be\label{conditions h4}
 \delta_3 = - \frac{M^2 - \alpha - 3 \beta}{2M^3 - 2M(\alpha + \beta)}\,\delta_1\,.
\ee

The final part of the analysis considers the expansion of $f_4$, yielding
\begin{equation}\label{f4expansion}
    f_4 = f_4^{\rm SS} + \frac{1}{\epsilon^4}\Big[ N_{4,0}^h(\delta_i) + N_{4,1}^f(\delta_i)\, \epsilon + N_{4,2}^f(\delta_i)\, \epsilon^2 + N_{4,3}^f(\delta_i)\, \epsilon^3 \Big] + \mathcal{O}\left(\epsilon^0\right)\, .
\end{equation}
Investigating the coefficients given in the bottom block of Table \ref{Tab.QG} gives the final constraint
\be\label{conditions f4}
 \delta_1 = 0 \,.
\ee
Combining the relations \eqref{conditions f3}, \eqref{conditions h3}, \eqref{conditions h4}, and \eqref{conditions f4} establishes that the only admissible solution which cancels all poles has
\be\label{analysis-final}
\delta_i = 0 \, , \qquad  i = 0, \cdots, 5 \, . 
\ee
Thus, we conclude: \textit{The only asymptotically flat solution that avoids the instability is the Schwarzschild solution}. This also entails that the Schwarzschild geometry is the only asymptotically flat solution of quadratic gravity (for sufficiently large $M$) that has a horizon. This is in line with the argument provided in \cite{Lu:2015cqa,Lu:2015psa} that this is the unique geometry in the solution space of quadratic gravity that exhibits an event horizon for all values of $M$. Furthermore, the condition $\delta_0 = \delta_1 = 0$ implies that the location of the pole is $r^{\rm pole} = 2M$. Thus, the event horizon must be located exactly at the Schwarzschild radius set by the asymptotic mass of the solution. App.\ \ref{App.A} repeats this analysis using Conformal-to-Kundt coordinates, establishing that the result of this subsection is independent of the coordinate system.

We close with a cautious remark. A key assumption of our analysis is that the perturbed solutions are close to the Schwarzschild geometry in the asymptotically flat region. This entails that there could be other solutions that do avoid the instability, but that are not asymptotically flat, or deviate from the Schwarzschild solution before reaching event horizon scales. The initial assumption that the solution is close to Schwarzschild is a crucial ingredient for the analysis carried out here. Other black hole solutions possessing an event horizon such as the Schwarzschild-Bach black holes obtained in \cite{Podolsky:2019gro} are known to have an event horizon and are not in conflict with the results obtained here. Our conclusions do suggest however that they are either not asymptotically flat, or that the corrections to the Schwarzschild solution become significant already before the event horizon scales are reached.

\subsection{The Goroff-Sagnotti Counterterm}
\label{Sect.main.4}
We proceed with the stability analysis for general relativity supplemented by the GS-term. In this case the action is given by eq.\ \eqref{actionans} with $\cP$ given in eq.\ \eqref{def:GS}. The resulting equations of motion are conveniently constructed using computer algebra software. For spherically symmetric solutions have been investigated in \cite{Alvarez:2023gfg,Daas:2023axu}. In the asymptotic region, solutions can be constructed as a power series in inverse powers of $r$. The first few terms of this expansion read \cite{Daas:2023axu}
\be\label{fhasympexp}
\begin{split} 
h(r) = & \, 1 - \frac{2M}{r} + \frac{24 \lambda M^2}{r^6} - \frac{32 \lambda M^3}{r^7} + \cO\left( \frac{\lambda^2 M^3}{r^{11}} \right) \, , \\
f(r) =  & \, 1 - \frac{2M}{r} + \frac{72 \lambda M^2}{r^6} - \frac{128 \lambda M^3}{r^7} + \cO\left( \frac{\lambda^2 M^3}{r^{11}} \right) \, .
\end{split}
\ee
Owed to the mass-dimension of $\lambda$ the corrections to $h(r), f(r)$ can be constructed perturbatively in a power series in $\lambda$, so that \eqref{fhasympexp} is exact in $r$ up to terms of order $\lambda^2$. Already the leading terms show
that the metric functions are no longer degenerate and one has $h(r) \not = f(r)$. This is the distinguishing feature of the ECG-case. An example solution, obtained by integrating the equations of motion numerically, is shown by the black line in Fig.\ \ref{Fig.2}. This suggests that the addition of $\cP^{\rm GS}$ does not trigger a blow-up mechanism and the horizon remains intact.

This stability analysis of the horizon is conveniently carried out in conformal-to-Kundt-coordinates \eqref{metctoK}, since this drastically simplifies the computation. Owed to the still rather lengthy expressions, we summarize the main findings and refer to the ancillary notebook for the complete analysis. Converting to the new coordinate system, the asymptotic expansion \eqref{fhasympexp} takes the form
\be\label{OHasympexp}
\begin{split} 
\Omega(\rb) = & \, -\frac{1}{\rb} + \frac{24}{7}\, \lambda  M^2 \rb^5  + \cO\left( \lambda^2 M^3 \rb^{10} \right) \, , \\
\mathcal{H}(\rb) =  & \, -\rb^2 - 2M \rb^3 - \frac{216}{7}\, \lambda M^2 \rb^8 - \frac{368}{7}\, \lambda M^3 \rb^9 + \cO\left(\lambda^2 M^3 \rb^{13}\right) \, .
\end{split}
\ee
The Frobenius analysis is then implemented by expanding the metric functions at an arbitrary point $\rb_0$,
\be\label{conf-to-Kundt-exp}
    \Omega(\rb) = \sum^{\infty}_{n=0} \Omega_n \left(\rb-\rb_0\right)^n \, , 
    \qquad \mathcal{H}(\rb) = \sum^{\infty}_{n=0} \mathcal{H}_n \left(\rb - \rb_0\right)^n \, .
\ee
Substituting this ansatz into the equations of motion then shows that the system has five free parameters
\be\label{free-GS} 
a_i = \{\Omega_0, \Omega_1, \cH_0, \cH_1, \cH_2 \} \, , \qquad i = 0,\cdots, 4 \, . 
\ee

For the Schwarzschild solution \eqref{SS-ctK}, the expansion coefficients are  given by
\be\label{conf-SS-1} 
    \Omega^{\rm SS}_n = \left(-\frac{1}{\rb_0}\right)^{n+1} \, , 
\ee 
and
 \be\label{conf-SS-2}     
\begin{array}{ll}
     \mathcal{H}^{\rm SS}_0 = -\rb^{\rm 2}_0 \left(1 + 2 M \rb_0\right) \, , \quad & \qquad \mathcal{H}^{\rm SS}_1 = -2 \rb_0 \left(1 + 3 M \rb_0\right) \, , \\[1.3ex]
    \mathcal{H}^{\rm SS}_2 = -1 -6 M \rb_0 \, , \quad & \qquad \mathcal{H}^{\rm SS}_3 = -2 M \, , 
\end{array}
\ee
with $\cH^{\rm SS}_n = 0$ for $n \ge 4$. Perturbing the coefficients $a_i$ about their Schwarzschild values, 
\be\label{def.QG.Kundt}
a_i = a_i^{\rm SS} + \, \delta_i \, , \qquad i = 0,\cdots,4 \, ,  
\ee
and substituting these expressions into the constrained coefficients gives rise to a first-order pole whose location is slightly shifted away from the Schwarzschild radius. In contrast to the quadratic gravity case, the numerators $N_i(\delta_i)$ are inhomogeneous equations for the $\delta_i$. This reflects the fact that the Schwarzschild metric is no longer a solution of the combined system. The condition to cancel the pole can then be solved order-by-order in $\lambda$ and fixes all the free parameters $\delta_i$ to non-trivial values. Reading off the specific values of $\delta_i$ resulting from \eqref{OHasympexp} and substituting the result into the constraint equations shows that they are satisfied. Hence, it is precisely the deformation \eqref{OHasympexp} which bypasses the pole. In other words  the asymptotically flat solution \eqref{GS:asmptcorrections} is \textit{not} subject to the blow-up mechanism. This remains true even if additional non-analytic terms are included. The conclusion is supported by the presence of an event horizon in the numerical solutions obtained in \cite{Daas:2023axu}.

We close this subsection with an interesting observation. The shift of the position of the pole away from its value for the Schwarzschild solution is given by a power series in $\lambda$. Switching to Schwarzschild coordinates, we find that the pole appearing in the expansion is located at
\be
r_h = 2\,M - \frac{\lambda}{4M^3} + \mathcal{O}\left(\frac{\lambda^2}{M^7}\right)\,.
\ee
This result is completely in line with \cite{Daas:2023axu}, see their equation (15). Thus the location of the pole can then be used to determine the position of the event horizon. 

\subsection{Quadratic Gravity supplemented by the Goroff-Sagnotti counterterm}
\label{Sect.main.5}
%
\begin{figure}[t!]
\includegraphics[width = 0.9\textwidth]{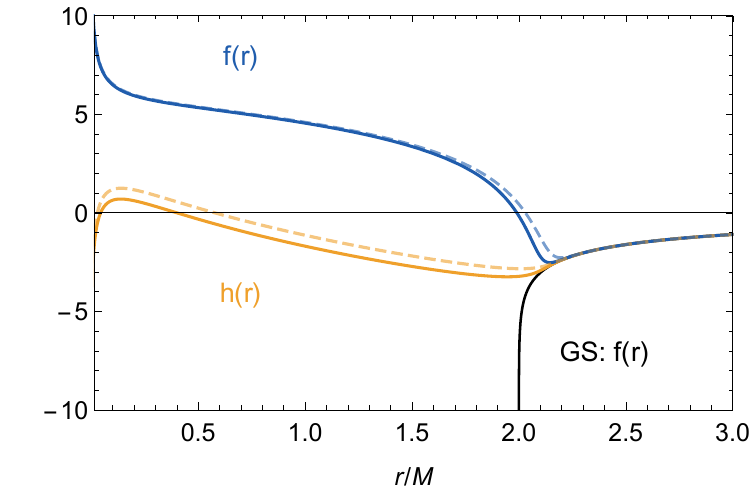}
\caption{\label{Fig.2} Example geometries obtained from solving the equations of motion of the combined quadratic gravity plus Goroff-Sagnotti system numerically.
All metric functions are given on a logarithmic scale so that an horizon where $f(r) = 0$ corresponds to a vertical line. The solution for $f(r)$ the pure GS-system (with $\alpha=\beta=0$)  is represented by the black line (GS) and exhibits an event horizon. A typical solution including the exponential corrections is shown by the blue and orange lines. The initial conditions are obtained from \eqref{AFQGGS} and \eqref{AFQGGSe} where we set $M = 10, \lambda = 0.1, \alpha = \frac{1}{2}, \beta = \frac{1}{3}$, and $S_2^- = S_0^- = 1/10$. In this case, the event horizon is absent and one deals with a naked singularity. The dashed line represents the solution obtained from $M = 10, \lambda = 0.1, \alpha = \frac{1}{2}, \beta = \frac{1}{3}$ with $S_2^- = S_0^- = 0$, indicating that non-zero values for $\alpha$ and $\beta$ in the polynomial part is sufficient to trigger the blow-up mechanism.}
\end{figure}
The analysis of our previous sections revealed that the equations of motion of quadratic gravity give rise to a blow-up mechanism while the addition of the Goroff-Sagnotti term does not lead to an instability. At this stage, it is interesting to combine the two systems and set up a competition between the instabilities encountered in Sect.\ \ref{Sect.main.3} and the stability of the horizon found in Sect.\ \ref{Sect.main.4}. Again, the equations of motion for the combined system are readily found using computer algebra software. Again the analysis can be carried out using either Schwarzschild or conformal-to-Kundt coordinates. In this section, we opt for the former for two reasons: first, it gives more intuition in particular in the asymptotically flat region and, second, the choice simplifies the technical complexity of the analysis.

We start with constructing the solution analytically in the asymptotically flat region. The metric functions of the Schwarzschild geometry inherit polynomial (pol) corrections triggered by the GS counterterm as well as Yukawa-type contributions (exp) originating from the quadratic gravity terms
\be\label{metfctcombined}
h(r) \approx h^{\rm pol}(r) + h^{\rm exp}(r) \, , \qquad f(r) \approx  f^{\rm pol}(r) + f^{\rm exp}(r) \, . 
\ee
The polynomial part extends \eqref{fhasympexp}. Up to order $\cO(r^{-10})$ it is given by
\begin{equation}\label{AFQGGS}
    \begin{split}
        h^{\rm pol}(r) = & \,  1 - \frac{2M}{r} + \frac{24 \lambda M^2}{r^6} - \frac{32 \lambda M^3}{r^7} + \frac{1440 M^2 \alpha \lambda }{r^8} -\frac{32M^3\big(163\,\alpha + 6\,\beta\big)\lambda}{r^9} \\
        &\quad+ \mathcal{O}\left(r^{-10}\right)\,,\\
        f^{\rm pol}(r) = &\, 1 - \frac{2M}{r} + \frac{72 \lambda M^2}{r^6} - \frac{128 \lambda M^3}{r^7} + \frac{5760 \lambda \alpha M^2 }{r^8}
        -\frac{288 M^3 (89\,\alpha - 6\,\beta\big)\lambda}{r^9}\\
        &\quad + \mathcal{O}\left(r^{-10}\right)\,.
    \end{split}
\end{equation}
Importantly, there are terms originating from the interplay of the quadratic gravity and GS-term which appear before the $\lambda^2$-terms enter into the expansion \cite{Daas:2024pxs}. In addition, the solution of the linearized equations of motion contains contributions in the form of exponentials. Since the GS counterterm does not survive the linearization, the solution to these equations is the same as for quadratic gravity giving rise to the Yukawa modes of eq.\ \eqref{lin:QG}. Imposing asymptotic flatness then yields for the exponential corrections
\be\label{AFQGGSe}
\begin{split}
h_{\rm exp}(r) =\, & 2 S_2^- \,\frac{e^{-m_2 r}}{r} + S_0^- \, \frac{e^{-m_0 r}}{r} \, , \\
f_{\rm exp}(r) = \,& S_2^-\, \frac{e^{-m_2 r}}{r} \left(1+m_2 \, r \right)- S_0^- \, \frac{e^{-m_0 r}}{r} \left(1+m_0 \, r \right)  \, . 
\end{split}
\ee

Fig.\ \ref{Fig.2} serves as an illustration of typical solutions obtained from integrating the equations of motion of the combined system numerically. We start by placing initial conditions based on the asymptotic solutions \eqref{AFQGGS} and \eqref{AFQGGSe} the asymptotically flat region and then integrate inward. 
The black line shows the solution of the GS system with $\alpha = \beta = 0$ for reference. Once the Yukawa-type terms \eqref{AFQGGSe} are turned on when setting the initial conditions, the numerical investigation reveals that the solutions undergo a strong dynamics in the region where the GS solution has its event horizon. While a fulfledged numerical investigation of the solution space of this system is beyond the scope of the present work, varying parameters shows that phase space resembles quadratic gravity with the Schwarzschild solution being replaced by its GS counterpart.

The analytic investigation of the dynamics in the vicinity of the event horizon then proceeds analogously to the previous sections. Similarly to the quadratic gravity case, it is most economic to carry out this investigation using Schwarzschild coordinates. Carrying out the Frobenius analysis around an arbitrary point $r_0$, eq.\ \eqref{powerseries}, then shows that there are seven free parameters characterizing the solution locally. These can be taken as
\begin{equation}
    \{a_i\} = \{f_0, h_0, f_1, h_1, f_2, h_2, h_3\} \, , \qquad i = 0,\cdots,6 \, . 
\end{equation}
These parameters are again perturbed around their Schwarzschild values
\be\label{def.QG2}
a_i = a_i^{\rm SS} + \, \delta_i \, , \qquad i = 0,\cdots,6 \, . 
\ee
Substituting the perturbation into the constrained coefficients and analyzing the resulting pole structure immediately shows that the cancellation of the poles in $f_3$ and $h_4$ requires
\begin{equation}\label{constraint GQGS}
    \delta_i = 0\,, \quad i = 0, ..., 4 \, .
\end{equation}
Extending the analysis to $f_4$, gives non-trivial relations for the remaining deformations $\delta_i, i = 5, 6$ as functions of the asymptotic mass $M$ and couplings of the theory, $\alpha, \beta, \lambda$. This is reminiscent of the fact that the Schwarzschild spacetime is not a solution of the system as long as $\lambda \neq 0$. 

In order to interpret this result, we recast the asymptotically flat solution \eqref{AFQGGS} into its expansion around the value of pole. Retaining non-vanishing values for $\alpha, \beta,$ and $\lambda$ shows that the resulting coefficients are not the ones solving the constrained equations. This can be verified along the same lines as the analysis of eq.\ \eqref{poly-sol-coeff}.\footnote{An efficient strategy to arrive at this conclusion is to work on the subspace of theories where $\alpha = \beta = \lambda$. This results in a significant simplification of the constrained equations. At the same time, it is sufficient to confirm our general conclusion.} Thus, the local analysis suggests that there are no asymptotically flat solutions featuring an event horizon in the quadratic gravity plus Goroff-Sagnotti system, as long as dimensionless asymptotic mass is considerably larger than the couplings.

The construction of solutions using numerical methods corroborates this picture. We find several naked singularity and wormhole solutions, an example of which is shown in Fig.\ \ref{Fig.2}. The instability also applies to initial conditions where the exponential correction terms are switched off. This is readily seen from the generic example resulting from the initial conditions $\alpha = \beta = \lambda = 0.1$ and $S_2^- = S_0^- = 0$, which leads to a naked singularity solution. While it is difficult to obtain the full phase space of solutions along this lines, we take this as an additional confirmation that the analytic picture is correct.

\section{Conclusion and outlook}
\label{Sect.conc}
Our comparative study analyzes the existence of event horizons in static, spherically symmetric spacetime geometries which solve the equations of motion of general relativity supplemented by local higher-derivative terms. Specifically, we initiate the systematic exploration of a blow-up mechanism that operates in the vicinity of the Schwarzschild radius and strongly affects the geometry in this region. Our findings are summarized in Table \ref{Table.3}. These suggest to group the higher-derivative terms according to whether or not they give rise to additional massive degrees of freedom visible in the asymptotically flat region. Examples of the former include the higher-derivative terms contained in quadratic gravity. The extensions considered in Einstein-Cubic Gravity and the Goroff-Sagnotti counterterm constitute examples of the latter type. Adopting this classification results in an intriguingly simple picture: the event horizon is removed if the dynamics includes higher-derivative terms that give rise to Yukawa-type contributions in the asymptotically flat region. The analysis of the combined quadratic gravity and Goroff-Sagnotti system furthermore suggests that it is not even necessary to switch on the exponential correction terms to trigger the blow-up mechanism and expel the horizon. We conjecture that this feature is true on general grounds beyond the specific cases studied in this work.
\begin{table}[t!]
\centering
\begin{tabular}{|l||
>{\columncolor[HTML]{FFFFFF}}c |
>{\columncolor[HTML]{FFFFFF}}c |
>{\columncolor[HTML]{FFFFFF}}c ||
>{\columncolor[HTML]{FFFFFF}}c |}
\hline
\cellcolor[HTML]{C0C0C0}Model & \cellcolor[HTML]{C0C0C0}Adds d.o.f.\ & \cellcolor[HTML]{C0C0C0}Exponentials & \cellcolor[HTML]{C0C0C0}Schwarzschild solution & \cellcolor[HTML]{C0C0C0}Blow-up mechanism \\ \hline
QG            &      \checkmark        &                 \checkmark                   &         \checkmark                                            &               \checkmark                        \\ \hline
ECG    &         \xmark  &          \checkmark   & \xmark            &    \xmark    \\ \hline
GS              &         \xmark    & \xmark   & \xmark  &   \xmark                                                \\ \hline
QG+GS                         & \checkmark   & \checkmark   & \xmark   & \checkmark       \\ \hline
\end{tabular}
\caption{\label{Table.3} Overview of the various extensions of general relativity that were discussed in this work. The model is specified in the first column. The key properties of these models including the presence of additional massive degrees of freedom visible in the asymptotically flat region (Adds.\ d.o.f.), the presence of exponentially suppressed corrections (Exponentials), and whether the Schwarzschild metric solves the equations of motion (Schwarzschild solution) are listed in the second, third, and fourth column, respectively. The presence or absence of the blow-up mechanism is given in the last column. Notably, only the column ``Adds.\ d.o.f.'' has a perfect match with the occurrence of the blow-up mechanism. }
\end{table}

Of course, it would be interesting to corroborate (or refute) the systematics discovered in Table \ref{Table.3} at the level of additional examples. We note that, ultimately, the connection between the existence of an event horizon and certain couplings appearing in the gravitational dynamics may restrict these couplings in a novel and highly non-trivial way. Since shadow observations by the Event Horizon Telescope \cite{EventHorizonTelescope:2019dse,EventHorizonTelescope:2022wkp} may be able to discriminate between compact objects with and without an event horizon \cite{Daas:2022iid,Carballo-Rubio:2022aed,Eichhorn:2022oma,Aktar:2024akk} (also see \cite{Ayzenberg:2023hfw} for a summary of opportunities opened up by the Next-Generation Event Horizon Telescope), this mechanism may provide a unique opportunity to constrain the Wilson coefficients appearing in the gravitational dynamics by observations.

As a side result, our analysis revealed an intriguing relation between the pole structure arising when studying perturbations of the Schwarzschild solution using the Frobenius method and the shifted position of an event horizon (provided that the latter exists). Corrections to the position of the horizon can be obtained analytically from the root structure of the denominators $D$, introduced in eq.\ \eqref{ND-expressions}. This analytic approach is much more efficient in finding the position of the event horizon than numerical integration.

\section*{Acknowledgments}
We thank A.\ Bonanno, V.\ Nikoli\'{c}, and A.\ Pos for illuminating discussions. The work of C.L. is funded by Becas Chile, ANID-PCHA/2020-72210073.
\appendix


\section{Horizon stability using Conformal-to-Kundt coordinates}
\label{App.A}
The presentation in Sect.\ \ref{Sect.main} relied on expressing the spacetime metric in Schwarzschild coordinates. This raises the question to which extend the (in-)stability of the horizon depends on the choice of coordinate system. In this appendix we repeat the analysis of the main part of our work using a coordinate system which is manifestly regular at an event horizon. This demonstrates that the results of our analysis are not tied to the use of a specific coordinate system. While our presentation is limited to the case of quadratic gravity, we also confirmed that the robustness of the analysis extends to the other dynamical settings as well. In addition, all our results were corroborated using Eddington-Finkelstein coordinates.

We repeat the analysis of Sect.\ \ref{Sect.main.3} using Conformal-to-Kundt coordinates. In this case the line-element is given by
\be
ds^2 = \Omega^2(\rb) \left[ d\theta^2 + \sin^2\theta d\phi^2 - 2 dv d\rb + \cH(\rb) dv^2 \right] \, . 
\ee
Structurally, these coordinates are similar to Eddington-Finkelstein coordinates in the sense that the line-element remains regular at a horizon where the metric function $\cH$ vanishes. The Schwarzschild solution in conformal-to-Kundt coordinates is
\be\label{SS-CTK}
\Omega^{\rm SS}(\rb) = - \frac{1}{\rb} \, , \qquad \cH^{\rm SS}(\rb) = - \rb^2 \left(1 + 2 M \rb \right) \, , 
\ee
and exhibits an event horizon at $\rb = - (2M)^{-1}$.

Following the general algorithm introduced in Sect.\ \ref{Sect.main}, we expand the metric functions around an arbitrary reference point $\rb_0$ using the expansion \eqref{conf-to-Kundt-exp}. Upon substituting these power series into the equations of motion, the Frobenius analysis reveals seven free parameters, in addition to the arbitrary point $\rb_0$. Thus $N_{\rm free} = 7$ in this case. Following the notation in eq.\ \eqref{a-relevant}, these are taken as
\be\label{App-free-coeffs}
    a_i = \{\Omega_0,\Omega_1,\Omega_2,\mathcal{H}_0,\mathcal{H}_1,\mathcal{H}_2,\mathcal{H}_3\} \, .
\ee
The number of free parameters here aligns with those present in Schwarzschild coordinates, as the line element \eqref{metctoK} retains an additional gauge freedom compared to \eqref{met1}. 

In the next step, we perturb the free coefficients \eqref{App-free-coeffs} around their Schwarzschild values \eqref{conf-SS-1} and \eqref{conf-SS-2}
\be
    a_i = a^{\rm SS}_i + \delta_i \, , \quad i=0, \cdots, 6 \, .
\ee
In contrast to the analysis building on Schwarzschild coordinates, it is then found that all constraints on the perturbations $\delta_i$ arise from the dependent coefficients $\Omega_n$, $n \ge 3$. The analogue of eq.\ \eqref{f3expansion} obtained from $\Omega_3$ is
\be\label{Omega3exp}
    \Omega_3 = \Omega^{\rm SS}_3 + \frac{1}{\epsilon} \, N_{3,0}^\Omega   + \cO(\epsilon^0)\, , 
\ee
where,
\be
\begin{split}
    N_{3,0}^\Omega = -\frac{2}{27 \beta} \Big(&144 M^7 \delta_1 - 36 M \beta \delta_2 + 24 M^6 \delta_3 + 
    6 M^4 (\delta_0 + 2 (\alpha - 3 \beta) \delta_3) + 
    9 \beta \delta_4 \\
    &+ 
    M^3 ( 8 (\alpha - 3 \beta) \delta_5 -3 \delta_2) + 
    3 M^2 (12 \beta \delta_0 + (\alpha - 3 \beta )\delta_6 ) \Big) \, .
\end{split}
\ee
Insisting on the absence of a pole requires setting $N_{3,0}^\Omega = 0$. This leads to a condition among the $\delta_i$'s. We then follow the bootstrap method used in the main part of the work and analyze the constraints resulting from $\Omega_4, \Omega_5, \Omega_6$. These are sufficient to deduce that
\be\label{condition2}
  \delta_i = 0 \, , \qquad  i = 0,\cdots, 6 \, . 
\ee 
An analysis of $\cH_4, \cH_5$, and $\cH_6$ shows that all poles appearing in these expressions are eliminated by the conditions \eqref{condition2} as well. 

Thus we conclude that the Schwarzschild solution is the only geometry that avoids the blow-up mechanism. The same conclusion holds if the analysis is carried out in Eddington-Finkelstein coordinates. In this way, we establish that our conclusions of the main part remain valid once one adopts a coordinate system that is regular at the event horizon. 


\section{Expansion coefficients appearing in Quadratic Gravity}
\label{App.B}
%
\vspace{-0.6cm}
\begin{table}[H]
\centering
\begin{tabular}{ll}
\hline \hline
 $\bigg. N^f_{3,0}$ &  $16 M^4 \left(\alpha - 3 \beta\right) \big(M^3 \, (\alpha - 3\beta) + 9 \alpha \beta\big) \, (\delta_0 - \delta_3)$ \\
 \multirow{2}{*}{$\bigg. N^f_{3,1}$} &   $8 M^3 \Big(M^2 \, (\alpha - 39 \beta) \, (\alpha - 3\beta) \, (\delta_0 - \delta_3) + 18 \alpha \beta \, (\alpha - 12\beta) \, (\delta_0 - \delta_3)\Big)$ \\[1.2ex] & \qquad \quad $-2 M^3 \, (\alpha - 3\beta)^2 \, (\delta_1 - \delta_4) - 18 M \alpha \beta (\alpha - 3\beta) \, (\delta_1 - \delta_4) \Big)$ \\
 \multirow{3}{*}{$\bigg. N^f_{3,2}$} & $4 M^2 \Big(4 M^3 \, (\alpha - 3 \beta) \, \big(\beta \, (15 \delta_4 - 9 \delta_1) + \alpha \, (3 \delta_1 + \delta_4)\big) + 54 \alpha \beta (\alpha + 6 \beta) \, (\delta_0 - \delta_3)$ \\[1.2ex] & \qquad \quad
 $+ M^2 \, (\alpha - 3\beta) \, \big(7 \alpha \delta_0 + 3 \beta \, (41 \delta_0 - 33 \delta_3) - 3 \alpha \, (36 \beta \, (\delta_2 - \delta_5) + \delta_3)\big)$ \\[1.2ex] & \qquad \quad
 $+324 M \alpha \beta^2 \, (\delta_4 - \delta_1) + 16 M^4 \, (\alpha - 3 \beta)^2 \, \delta_5 \Big)$\\ \hline
 $\bigg. N^{h}_{3,0}$ & $16 M^4 \left(\alpha - 3 \beta\right) \, (\delta_0 - \delta_3)$\\
 $\bigg. N^{h}_{3,1}$ & 
 $-16 M^3 \, \big(M \alpha \, (\delta_1 - \delta_4) + 3 \beta \, (3 \delta_0 - M \delta_1 - 3 \delta_3 + M \delta_4)\big)$\\[1.2ex]
$\bigg. N^{h}_{3,2}$ & $8 M^2 \big(2 M^2 \, (\alpha - 3\beta) \, (\delta_2 - \delta_5) + 18 M \beta \, (\delta_1 - \delta_4) - (5 \alpha - 12\beta - 6 M^2) \, (\delta_0 - \delta_3)\big)$ \\ \hline
 $\bigg. N^{h}_{4,0}$ & $16 M^4 \left(\alpha - 3 \beta\right) \, (\delta_0 - \delta_3)$\\
 $\bigg. N^{h}_{4,1}$ & $-16 M^3 \, \big(M \alpha \, (\delta_1 - \delta_4) + 3 \beta \, (3 \delta_0 - M \delta_1 - 3 \delta_3 + M \delta_4)\big)$\\
 $\bigg. N^{h}_{4,2}$ & $8 M^2 \big(2 M^2 \, (\alpha - 3\beta) \, (\delta_2 - \delta_5) + 18 M \beta \, (\delta_1 - \delta_4) - (5 \alpha - 12\beta - 6 M^2) \, (\delta_0 - \delta_3)\big)$ \\
 \multirow{2}{*}{$\bigg. N^h_{4,3}$} & $4 M \Big( 2 M^2 \, \big(3 \, (4 \delta_0 - 3 \delta_3) - 2 \alpha \, (\delta_2 - 5 \delta_5) + 6 \beta \, (\delta_5 - 5 \delta_2)\big) - 2 \delta_0 \, (\alpha - 3 \beta)$ \\[1.2ex] & \qquad \quad
 $+ M \, \big(\alpha \, (3 \delta_1 - 5 \delta_4) + \beta \, (27 \delta_1 - 21 \delta_4)\big) + 12 M^3 \, \delta_4 \Big)$\\ \hline
 $\bigg. N^{f}_{4,0}$ & $-32 M^5 \, (\alpha - 3 \beta) \, (\delta_0 - \delta_3) \, \big(5 M^2 \, (\alpha - 3 \beta) + 54 \alpha \beta\big)$ \\
 \multirow{2}{*}{$\bigg. N^f_{4,1}$} & $-16 M^4 \Big( \big(M^2 \, (5 \alpha - 213 \beta) \, (\alpha - 3\beta) - 18 \alpha \beta \, (11\alpha - 87\beta)\big) \, \big(\delta_0 - \delta_3\big)$ \\[1.2ex] & \qquad \quad
 $-2 M \, \big(\alpha - 3\beta\big) \, \big(\delta_1 - \delta_4\big) \, \big(5 M^2 \, (\alpha - 3\beta) + 54 \alpha\beta\big)\Big)$ \\
 \multirow{4}{*}{$\bigg. N^f_{4,2}$} & $8 M^3 \Big(M^2 \big(9\beta^2 \, (231 \delta_3 - 271 \delta_0) + 12 \alpha \beta \big(83 \delta_0 - 78 \epsilon_0 + 144 \beta \, (\delta_2 - \delta_5)\big)\big)$ \\[1.2ex] & \qquad \quad
 $+4 M^3 \, (\alpha - 3\beta) \big(2 \beta \, (42 \delta_5 - 27 \delta_1) + 5 \alpha \, (3 \delta_1 + \delta_4) + 12 M  (\delta_0 - \delta_3)\big)$ \\[1.2ex] & \qquad \quad
 $+80 M^4 \, \delta_5 \, (\alpha - 3\beta)^2 + M^2 \alpha^2 \big(11 \delta_0 + 576 \beta \, (\delta_2 - \delta_5)\big)$ \\[1.2ex] & \qquad \quad
 $108 \alpha \beta \big(39 \beta \, (\delta_0 - \delta_3) + 2 M \, (\alpha - 12\beta) \, (\delta_1 - \delta_4)\big)\Big)$ \\
 \multirow{5}{*}{$\bigg. N^f_{4,3}$} & $4 M^2 \Big(4 M^3 \big( 9 \beta^2 ( 39 \delta_1 - 71 \delta_4 ) 
 + 5 \alpha^2 ( 3 \delta_1 + 5 \delta_4 ) 
 + 6 \alpha \beta ( -63 \delta_1 + 41 \delta_4 ) \big)$ \\[1.2ex] & \qquad \quad
 $+48 M^4 \, (\alpha - 3 \beta) \, \big(7 \delta_0 - 6 \delta_3 - 2\delta_2 \, (\alpha - 3\beta) + \delta_4 \, (7 \alpha - 45 \beta) + 2 M \delta_4\big)$\\[1.2ex] & \qquad \quad
 $+M^2 \alpha \big(\alpha \big(47 \delta_3 - 27 \delta_0 - 288 \beta \, (4 \delta_2 - 7 \delta_5)\big) + 348 \beta \, (6 \delta_0 - 5 \delta_3)\big)$\\[1.2ex] & \qquad \quad
 $+12 M \alpha\beta \big(72 M \beta \, (10 \delta_2 - 13 \delta_5)-27 (\alpha + 21\beta) \, (\delta_2 - \delta_5)\big)$\\[1.2ex] & \qquad \quad
 $+216 \alpha\beta \big(\delta_0 \, (2\alpha + 21 \beta) - 3 \delta_3 \, (\alpha + 6 \beta)\big)\Big)$
 \\ \hline \hline
\end{tabular}
\vspace{-0.3cm}
\caption{From top to bottom: polynomials appearing in the coefficients $f_3$ \eqref{f3expansion}, $h_3$ \eqref{h3expansion}, $h_4$ \eqref{h4expansion}, and $f_4$ \eqref{f4expansion}. \label{Tab.QG}}
\end{table}


\bibliography{Instab}

\providecommand{\href}[2]{#2}\begingroup\raggedright\begin{thebibliography}{10}

\bibitem{EventHorizonTelescope:2019dse}
{\scshape Event Horizon Telescope} collaboration, \emph{{First M87 Event Horizon Telescope Results. I. The Shadow of the Supermassive Black Hole}}, \href{https://doi.org/10.3847/2041-8213/ab0ec7}{\emph{Astrophys. J. Lett.} {\bfseries 875} (2019) L1} [\href{https://arxiv.org/abs/1906.11238}{{\ttfamily 1906.11238}}].

\bibitem{EventHorizonTelescope:2022wkp}
{\scshape Event Horizon Telescope} collaboration, \emph{{First Sagittarius A* Event Horizon Telescope Results. I. The Shadow of the Supermassive Black Hole in the Center of the Milky Way}}, \href{https://doi.org/10.3847/2041-8213/ac6674}{\emph{Astrophys. J. Lett.} {\bfseries 930} (2022) L12} [\href{https://arxiv.org/abs/2311.08680}{{\ttfamily 2311.08680}}].

\bibitem{LIGOScientific:2016aoc}
{\scshape LIGO Scientific, Virgo} collaboration, \emph{{Observation of Gravitational Waves from a Binary Black Hole Merger}}, \href{https://doi.org/10.1103/PhysRevLett.116.061102}{\emph{Phys. Rev. Lett.} {\bfseries 116} (2016) 061102} [\href{https://arxiv.org/abs/1602.03837}{{\ttfamily 1602.03837}}].

\bibitem{Cardoso:2019rvt}
V.~Cardoso and P.~Pani, \emph{{Testing the nature of dark compact objects: a status report}}, \href{https://doi.org/10.1007/s41114-019-0020-4}{\emph{Living Rev. Rel.} {\bfseries 22} (2019) 4} [\href{https://arxiv.org/abs/1904.05363}{{\ttfamily 1904.05363}}].

\bibitem{Buoninfante:2024oxl}
N.~Afshordi et~al., \emph{{Black Holes Inside and Out 2024: visions for the future of black hole physics}},  10, 2024, \href{https://arxiv.org/abs/2410.14414}{{\ttfamily 2410.14414}}.

\bibitem{heusler1996black}
M.~Heusler, \emph{Black hole uniqueness theorems}, .

\bibitem{Hawking:2014tga}
S.~W. Hawking, \emph{{Information Preservation and Weather Forecasting for Black Holes}},  \href{https://arxiv.org/abs/1401.5761}{{\ttfamily 1401.5761}}.

\bibitem{Curiel:2018cbt}
E.~Curiel, \emph{{The many definitions of a black hole}}, \href{https://doi.org/10.1038/s41550-018-0602-1}{\emph{Nature Astron.} {\bfseries 3} (2019) 27} [\href{https://arxiv.org/abs/1808.01507}{{\ttfamily 1808.01507}}].

\bibitem{Penrose:1969pc}
R.~Penrose, \emph{{Gravitational collapse: The role of general relativity}}, \href{https://doi.org/10.1023/A:1016578408204}{\emph{Riv. Nuovo Cim.} {\bfseries 1} (1969) 252}.

\bibitem{Penrose:1999vj}
R.~Penrose, \emph{{The question of cosmic censorship}}, \href{https://doi.org/10.1007/BF02702355}{\emph{J. Astrophys. Astron.} {\bfseries 20} (1999) 233}.

\bibitem{Landsman:2021mjt}
K.~Landsman, \emph{{Singularities, black holes, and cosmic censorship: A tribute to Roger Penrose}}, \href{https://doi.org/10.1007/s10701-021-00432-1}{\emph{Found. Phys.} {\bfseries 51} (2021) 42} [\href{https://arxiv.org/abs/2101.02687}{{\ttfamily 2101.02687}}].

\bibitem{Hawking:1975vcx}
S.~W. Hawking, \emph{{Particle Creation by Black Holes}}, \href{https://doi.org/10.1007/BF02345020}{\emph{Commun. Math. Phys.} {\bfseries 43} (1975) 199}.

\bibitem{Bronnikov:2005gm}
K.~A. Bronnikov and J.~C. Fabris, \emph{{Regular phantom black holes}}, \href{https://doi.org/10.1103/PhysRevLett.96.251101}{\emph{Phys. Rev. Lett.} {\bfseries 96} (2006) 251101} [\href{https://arxiv.org/abs/gr-qc/0511109}{{\ttfamily gr-qc/0511109}}].

\bibitem{Bronnikov:2006fu}
K.~A. Bronnikov, V.~N. Melnikov and H.~Dehnen, \emph{{Regular black holes and black universes}}, \href{https://doi.org/10.1007/s10714-007-0430-6}{\emph{Gen. Rel. Grav.} {\bfseries 39} (2007) 973} [\href{https://arxiv.org/abs/gr-qc/0611022}{{\ttfamily gr-qc/0611022}}].

\bibitem{Sotiriou:2011dz}
T.~P. Sotiriou and V.~Faraoni, \emph{{Black holes in scalar-tensor gravity}}, \href{https://doi.org/10.1103/PhysRevLett.108.081103}{\emph{Phys. Rev. Lett.} {\bfseries 108} (2012) 081103} [\href{https://arxiv.org/abs/1109.6324}{{\ttfamily 1109.6324}}].

\bibitem{Herdeiro:2015waa}
C.~A.~R. Herdeiro and E.~Radu, \emph{{Asymptotically flat black holes with scalar hair: a review}}, \href{https://doi.org/10.1142/S0218271815420146}{\emph{Int. J. Mod. Phys. D} {\bfseries 24} (2015) 1542014} [\href{https://arxiv.org/abs/1504.08209}{{\ttfamily 1504.08209}}].

\bibitem{Karakasis:2023hni}
T.~Karakasis, N.~E. Mavromatos and E.~Papantonopoulos, \emph{{Regular compact objects with scalar hair}}, \href{https://doi.org/10.1103/PhysRevD.108.024001}{\emph{Phys. Rev. D} {\bfseries 108} (2023) 024001} [\href{https://arxiv.org/abs/2305.00058}{{\ttfamily 2305.00058}}].

\bibitem{Hui:2012qt}
L.~Hui and A.~Nicolis, \emph{{No-Hair Theorem for the Galileon}}, \href{https://doi.org/10.1103/PhysRevLett.110.241104}{\emph{Phys. Rev. Lett.} {\bfseries 110} (2013) 241104} [\href{https://arxiv.org/abs/1202.1296}{{\ttfamily 1202.1296}}].

\bibitem{Anabalon:2013oea}
A.~Anabalon, A.~Cisterna and J.~Oliva, \emph{{Asymptotically locally AdS and flat black holes in Horndeski theory}}, \href{https://doi.org/10.1103/PhysRevD.89.084050}{\emph{Phys. Rev. D} {\bfseries 89} (2014) 084050} [\href{https://arxiv.org/abs/1312.3597}{{\ttfamily 1312.3597}}].

\bibitem{Babichev:2017guv}
E.~Babichev, C.~Charmousis and A.~Leh\'ebel, \emph{{Asymptotically flat black holes in Horndeski theory and beyond}}, \href{https://doi.org/10.1088/1475-7516/2017/04/027}{\emph{JCAP} {\bfseries 04} (2017) 027} [\href{https://arxiv.org/abs/1702.01938}{{\ttfamily 1702.01938}}].

\bibitem{Ayon-Beato:1998hmi}
E.~Ayon-Beato and A.~Garcia, \emph{{Regular black hole in general relativity coupled to nonlinear electrodynamics}}, \href{https://doi.org/10.1103/PhysRevLett.80.5056}{\emph{Phys. Rev. Lett.} {\bfseries 80} (1998) 5056} [\href{https://arxiv.org/abs/gr-qc/9911046}{{\ttfamily gr-qc/9911046}}].

\bibitem{Bronnikov:2000vy}
K.~A. Bronnikov, \emph{{Regular magnetic black holes and monopoles from nonlinear electrodynamics}}, \href{https://doi.org/10.1103/PhysRevD.63.044005}{\emph{Phys. Rev. D} {\bfseries 63} (2001) 044005} [\href{https://arxiv.org/abs/gr-qc/0006014}{{\ttfamily gr-qc/0006014}}].

\bibitem{Dymnikova:2004zc}
I.~Dymnikova, \emph{{Regular electrically charged structures in nonlinear electrodynamics coupled to general relativity}}, \href{https://doi.org/10.1088/0264-9381/21/18/009}{\emph{Class. Quant. Grav.} {\bfseries 21} (2004) 4417} [\href{https://arxiv.org/abs/gr-qc/0407072}{{\ttfamily gr-qc/0407072}}].

\bibitem{Bronnikov:2017sgg}
K.~A. Bronnikov, \emph{{Nonlinear electrodynamics, regular black holes and wormholes}}, \href{https://doi.org/10.1142/S0218271818410055}{\emph{Int. J. Mod. Phys. D} {\bfseries 27} (2018) 1841005} [\href{https://arxiv.org/abs/1711.00087}{{\ttfamily 1711.00087}}].

\bibitem{Donoghue:2017pgk}
J.~F. Donoghue, M.~M. Ivanov and A.~Shkerin, \emph{{EPFL Lectures on General Relativity as a Quantum Field Theory}},  \href{https://arxiv.org/abs/1702.00319}{{\ttfamily 1702.00319}}.

\bibitem{Bambi:2023try}
C.~Bambi, ed., \emph{{Regular Black Holes. Towards a New Paradigm of Gravitational Collapse}}, Springer Series in Astrophysics and Cosmology. Springer, 2023, \href{https://doi.org/10.1007/978-981-99-1596-5}{10.1007/978-981-99-1596-5}, [\href{https://arxiv.org/abs/2307.13249}{{\ttfamily 2307.13249}}].

\bibitem{BasiBeneito:2022wux}
A.~Bas~i Beneito, G.~Calcagni and L.~Rachwa\l{}, \emph{{Classical and Quantum Nonlocal Gravity}}.
\newblock 2024.
\newblock \href{https://arxiv.org/abs/2211.05606}{{\ttfamily 2211.05606}}.

\bibitem{Edholm:2016hbt}
J.~Edholm, A.~S. Koshelev and A.~Mazumdar, \emph{{Behavior of the Newtonian potential for ghost-free gravity and singularity-free gravity}}, \href{https://doi.org/10.1103/PhysRevD.94.104033}{\emph{Phys. Rev. D} {\bfseries 94} (2016) 104033} [\href{https://arxiv.org/abs/1604.01989}{{\ttfamily 1604.01989}}].

\bibitem{Koshelev:2018hpt}
A.~S. Koshelev, J.~a. Marto and A.~Mazumdar, \emph{{Schwarzschild $1/r$-singularity is not permissible in ghost free quadratic curvature infinite derivative gravity}}, \href{https://doi.org/10.1103/PhysRevD.98.064023}{\emph{Phys. Rev. D} {\bfseries 98} (2018) 064023} [\href{https://arxiv.org/abs/1803.00309}{{\ttfamily 1803.00309}}].

\bibitem{Buoninfante:2018xif}
L.~Buoninfante, A.~S. Cornell, G.~Harmsen, A.~S. Koshelev, G.~Lambiase, J.~a. Marto et~al., \emph{{Towards nonsingular rotating compact object in ghost-free infinite derivative gravity}}, \href{https://doi.org/10.1103/PhysRevD.98.084041}{\emph{Phys. Rev. D} {\bfseries 98} (2018) 084041} [\href{https://arxiv.org/abs/1807.08896}{{\ttfamily 1807.08896}}].

\bibitem{Koshelev:2024wfk}
A.~Koshelev and A.~Tokareva, \emph{{Non-perturbative quantum gravity denounces singular Black Holes}},  \href{https://arxiv.org/abs/2404.07925}{{\ttfamily 2404.07925}}.

\bibitem{Bueno:2024dgm}
P.~Bueno, P.~A. Cano and R.~A. Hennigar, \emph{{Regular Black Holes From Pure Gravity}},  \href{https://arxiv.org/abs/2403.04827}{{\ttfamily 2403.04827}}.

\bibitem{Giacchini:2024exc}
B.~L. Giacchini and I.~Kol\'a\v{r}, \emph{{Toward regular black holes in sixth-derivative gravity}}, \href{https://doi.org/10.1103/PhysRevD.110.104056}{\emph{Phys. Rev. D} {\bfseries 110} (2024) 104056} [\href{https://arxiv.org/abs/2406.00997}{{\ttfamily 2406.00997}}].

\bibitem{DiFilippo:2024mwm}
F.~Di~Filippo, I.~Kol\'a\v{r} and D.~Kubiznak, \emph{{Inner-extremal regular black holes from pure gravity}},  \href{https://arxiv.org/abs/2404.07058}{{\ttfamily 2404.07058}}.

\bibitem{Konoplya:2024hfg}
R.~A. Konoplya and A.~Zhidenko, \emph{{Infinite tower of higher-curvature corrections: Quasinormal modes and late-time behavior of D-dimensional regular black holes}}, \href{https://doi.org/10.1103/PhysRevD.109.104005}{\emph{Phys. Rev. D} {\bfseries 109} (2024) 104005} [\href{https://arxiv.org/abs/2403.07848}{{\ttfamily 2403.07848}}].

\bibitem{Konoplya:2024kih}
R.~A. Konoplya and A.~Zhidenko, \emph{{Dymnikova black hole from an infinite tower of higher-curvature corrections}},  \href{https://arxiv.org/abs/2404.09063}{{\ttfamily 2404.09063}}.

\bibitem{Bena:2022rna}
I.~Bena, E.~J. Martinec, S.~D. Mathur and N.~P. Warner, \emph{{Fuzzballs and Microstate Geometries: Black-Hole Structure in String Theory}},  \href{https://arxiv.org/abs/2204.13113}{{\ttfamily 2204.13113}}.

\bibitem{Rovelli:2014cta}
C.~Rovelli and F.~Vidotto, \emph{{Planck stars}}, \href{https://doi.org/10.1142/S0218271814420267}{\emph{Int. J. Mod. Phys. D} {\bfseries 23} (2014) 1442026} [\href{https://arxiv.org/abs/1401.6562}{{\ttfamily 1401.6562}}].

\bibitem{Saueressig:2015xua}
F.~Saueressig, N.~Alkofer, G.~D'Odorico and F.~Vidotto, \emph{{Black holes in Asymptotically Safe Gravity}}, \href{https://doi.org/10.22323/1.224.0174}{\emph{PoS} {\bfseries FFP14} (2016) 174} [\href{https://arxiv.org/abs/1503.06472}{{\ttfamily 1503.06472}}].

\bibitem{Liebling:2012fv}
S.~L. Liebling and C.~Palenzuela, \emph{{Dynamical boson stars}}, \href{https://doi.org/10.1007/s41114-023-00043-4}{\emph{Living Rev. Rel.} {\bfseries 26} (2023) 1} [\href{https://arxiv.org/abs/1202.5809}{{\ttfamily 1202.5809}}].

\bibitem{Bonanno:2000ep}
A.~Bonanno and M.~Reuter, \emph{{Renormalization group improved black hole space-times}}, \href{https://doi.org/10.1103/PhysRevD.62.043008}{\emph{Phys. Rev. D} {\bfseries 62} (2000) 043008} [\href{https://arxiv.org/abs/hep-th/0002196}{{\ttfamily hep-th/0002196}}].

\bibitem{Eichhorn:2022bgu}
A.~Eichhorn and A.~Held, \emph{{Black holes in asymptotically safe gravity and beyond}},  \href{https://arxiv.org/abs/2212.09495}{{\ttfamily 2212.09495}}.

\bibitem{Platania:2023srt}
A.~Platania, \emph{{Black Holes in Asymptotically Safe Gravity}}.
\newblock 2023.
\newblock \href{https://arxiv.org/abs/2302.04272}{{\ttfamily 2302.04272}}.

\bibitem{Lu:2015cqa}
H.~Lu, A.~Perkins, C.~N. Pope and K.~S. Stelle, \emph{{Black Holes in Higher-Derivative Gravity}}, \href{https://doi.org/10.1103/PhysRevLett.114.171601}{\emph{Phys. Rev. Lett.} {\bfseries 114} (2015) 171601} [\href{https://arxiv.org/abs/1502.01028}{{\ttfamily 1502.01028}}].

\bibitem{Lu:2015psa}
H.~L\"u, A.~Perkins, C.~N. Pope and K.~S. Stelle, \emph{{Spherically Symmetric Solutions in Higher-Derivative Gravity}}, \href{https://doi.org/10.1103/PhysRevD.92.124019}{\emph{Phys. Rev. D} {\bfseries 92} (2015) 124019} [\href{https://arxiv.org/abs/1508.00010}{{\ttfamily 1508.00010}}].

\bibitem{Pravda:2016fue}
V.~Pravda, A.~Pravdova, J.~Podolsky and R.~Svarc, \emph{{Exact solutions to quadratic gravity}}, \href{https://doi.org/10.1103/PhysRevD.95.084025}{\emph{Phys. Rev. D} {\bfseries 95} (2017) 084025} [\href{https://arxiv.org/abs/1606.02646}{{\ttfamily 1606.02646}}].

\bibitem{Bonanno:2019rsq}
A.~Bonanno and S.~Silveravalle, \emph{{Characterizing black hole metrics in quadratic gravity}}, \href{https://doi.org/10.1103/PhysRevD.99.101501}{\emph{Phys. Rev. D} {\bfseries 99} (2019) 101501} [\href{https://arxiv.org/abs/1903.08759}{{\ttfamily 1903.08759}}].

\bibitem{Silveravalle:2022wij}
S.~Silveravalle and A.~Zuccotti, \emph{{Phase diagram of Einstein-Weyl gravity}}, \href{https://doi.org/10.1103/PhysRevD.107.064029}{\emph{Phys. Rev. D} {\bfseries 107} (2023) 064029} [\href{https://arxiv.org/abs/2210.13877}{{\ttfamily 2210.13877}}].

\bibitem{Daas:2022iid}
J.~Daas, K.~Kuijpers, F.~Saueressig, M.~F. Wondrak and H.~Falcke, \emph{{Probing quadratic gravity with the Event Horizon Telescope}}, \href{https://doi.org/10.1051/0004-6361/202244080}{\emph{Astron. Astrophys.} {\bfseries 673} (2023) A53} [\href{https://arxiv.org/abs/2204.08480}{{\ttfamily 2204.08480}}].

\bibitem{Stelle:1976gc}
K.~S. Stelle, \emph{{Renormalization of Higher Derivative Quantum Gravity}}, \href{https://doi.org/10.1103/PhysRevD.16.953}{\emph{Phys. Rev. D} {\bfseries 16} (1977) 953}.

\bibitem{Stelle:1977ry}
K.~S. Stelle, \emph{{Classical Gravity with Higher Derivatives}}, \href{https://doi.org/10.1007/BF00760427}{\emph{Gen. Rel. Grav.} {\bfseries 9} (1978) 353}.

\bibitem{Bueno:2016xff}
P.~Bueno and P.~A. Cano, \emph{{Einsteinian cubic gravity}}, \href{https://doi.org/10.1103/PhysRevD.94.104005}{\emph{Phys. Rev. D} {\bfseries 94} (2016) 104005} [\href{https://arxiv.org/abs/1607.06463}{{\ttfamily 1607.06463}}].

\bibitem{Goroff:1985sz}
M.~H. Goroff and A.~Sagnotti, \emph{{QUANTUM GRAVITY AT TWO LOOPS}}, \href{https://doi.org/10.1016/0370-2693(85)91470-4}{\emph{Phys. Lett. B} {\bfseries 160} (1985) 81}.

\bibitem{Goroff:1985th}
M.~H. Goroff and A.~Sagnotti, \emph{{The Ultraviolet Behavior of Einstein Gravity}}, \href{https://doi.org/10.1016/0550-3213(86)90193-8}{\emph{Nucl. Phys. B} {\bfseries 266} (1986) 709}.

\bibitem{Daas:2024pxs}
J.~Daas, C.~Laporte, F.~Saueressig and T.~van Dijk, \emph{{Rethinking the Effective Field Theory formulation of Gravity}},  \href{https://arxiv.org/abs/2405.12685}{{\ttfamily 2405.12685}}.

\bibitem{Platania:2020knd}
A.~Platania and C.~Wetterich, \emph{{Non-perturbative unitarity and fictitious ghosts in quantum gravity}}, \href{https://doi.org/10.1016/j.physletb.2020.135911}{\emph{Phys. Lett. B} {\bfseries 811} (2020) 135911} [\href{https://arxiv.org/abs/2009.06637}{{\ttfamily 2009.06637}}].

\bibitem{Buoninfante:2025klm}
L.~Buoninfante, \emph{{Remarks on ghost resonances}},  \href{https://arxiv.org/abs/2501.04097}{{\ttfamily 2501.04097}}.

\bibitem{Whitt:1985ki}
B.~Whitt, \emph{{The Stability of Schwarzschild Black Holes in Fourth Order Gravity}}, \href{https://doi.org/10.1103/PhysRevD.32.379}{\emph{Phys. Rev. D} {\bfseries 32} (1985) 379}.

\bibitem{Deffayet:2023wdg}
C.~Deffayet, A.~Held, S.~Mukohyama and A.~Vikman, \emph{{Global and local stability for ghosts coupled to positive energy degrees of freedom}}, \href{https://doi.org/10.1088/1475-7516/2023/11/031}{\emph{JCAP} {\bfseries 11} (2023) 031} [\href{https://arxiv.org/abs/2305.09631}{{\ttfamily 2305.09631}}].

\bibitem{Podolsk__2018}
J.~Podolský, R.~Švarc, V.~Pravda and A.~Pravdová, \emph{Explicit black hole solutions in higher-derivative gravity}, \href{https://doi.org/10.1103/physrevd.98.021502}{\emph{Physical Review D} {\bfseries 98} (2018) }.

\bibitem{Daas:2023axu}
J.~Daas, C.~Laporte and F.~Saueressig, \emph{{Impact of perturbative counterterms on black holes}}, \href{https://doi.org/10.1103/PhysRevD.109.L101504}{\emph{Phys. Rev. D} {\bfseries 109} (2024) L101504} [\href{https://arxiv.org/abs/2311.15739}{{\ttfamily 2311.15739}}].

\bibitem{xActwebpage}
``{xAct: Efficient tensor computer algebra for Mathematica}.'' \url{http://xact.es/index.html}.

\bibitem{Bueno:2016lrh}
P.~Bueno and P.~A. Cano, \emph{{Four-dimensional black holes in Einsteinian cubic gravity}}, \href{https://doi.org/10.1103/PhysRevD.94.124051}{\emph{Phys. Rev. D} {\bfseries 94} (2016) 124051} [\href{https://arxiv.org/abs/1610.08019}{{\ttfamily 1610.08019}}].

\bibitem{Podolsky:2019gro}
J.~Podolsk\'y, R.~\v{S}varc, V.~Pravda and A.~Pravdova, \emph{{Black holes and other exact spherical solutions in Quadratic Gravity}}, \href{https://doi.org/10.1103/PhysRevD.101.024027}{\emph{Phys. Rev. D} {\bfseries 101} (2020) 024027} [\href{https://arxiv.org/abs/1907.00046}{{\ttfamily 1907.00046}}].

\bibitem{Huang:2022urr}
Y.~Huang, D.-J. Liu and H.~Zhang, \emph{{Novel black holes in higher derivative gravity}}, \href{https://doi.org/10.1007/JHEP02(2023)057}{\emph{JHEP} {\bfseries 02} (2023) 057} [\href{https://arxiv.org/abs/2212.13357}{{\ttfamily 2212.13357}}].

\bibitem{Hennigar:2017ego}
R.~A. Hennigar, D.~Kubiz\v{n}\'ak and R.~B. Mann, \emph{{Generalized quasitopological gravity}}, \href{https://doi.org/10.1103/PhysRevD.95.104042}{\emph{Phys. Rev. D} {\bfseries 95} (2017) 104042} [\href{https://arxiv.org/abs/1703.01631}{{\ttfamily 1703.01631}}].

\bibitem{Bueno:2019ltp}
P.~Bueno, P.~A. Cano, J.~Moreno and A.~Murcia, \emph{{All higher-curvature gravities as Generalized quasi-topological gravities}}, \href{https://doi.org/10.1007/JHEP11(2019)062}{\emph{JHEP} {\bfseries 11} (2019) 062} [\href{https://arxiv.org/abs/1906.00987}{{\ttfamily 1906.00987}}].

\bibitem{Bueno:2019ycr}
P.~Bueno, P.~A. Cano and R.~A. Hennigar, \emph{{(Generalized) quasi-topological gravities at all orders}}, \href{https://doi.org/10.1088/1361-6382/ab5410}{\emph{Class. Quant. Grav.} {\bfseries 37} (2020) 015002} [\href{https://arxiv.org/abs/1909.07983}{{\ttfamily 1909.07983}}].

\bibitem{Hennigar:2016gkm}
R.~A. Hennigar and R.~B. Mann, \emph{{Black holes in Einsteinian cubic gravity}}, \href{https://doi.org/10.1103/PhysRevD.95.064055}{\emph{Phys. Rev. D} {\bfseries 95} (2017) 064055} [\href{https://arxiv.org/abs/1610.06675}{{\ttfamily 1610.06675}}].

\bibitem{Hennigar:2018hza}
R.~A. Hennigar, M.~B.~J. Poshteh and R.~B. Mann, \emph{{Shadows, Signals, and Stability in Einsteinian Cubic Gravity}}, \href{https://doi.org/10.1103/PhysRevD.97.064041}{\emph{Phys. Rev. D} {\bfseries 97} (2018) 064041} [\href{https://arxiv.org/abs/1801.03223}{{\ttfamily 1801.03223}}].

\bibitem{Holdom:2002xy}
B.~Holdom, \emph{{On the fate of singularities and horizons in higher derivative gravity}}, \href{https://doi.org/10.1103/PhysRevD.66.084010}{\emph{Phys. Rev. D} {\bfseries 66} (2002) 084010} [\href{https://arxiv.org/abs/hep-th/0206219}{{\ttfamily hep-th/0206219}}].

\bibitem{Saueressig:2021wam}
F.~Saueressig, M.~Galis, J.~Daas and A.~Khosravi, \emph{{Asymptotically flat black hole solutions in quadratic gravity}}, \href{https://doi.org/10.1142/S0218271821420153}{\emph{Int. J. Mod. Phys. D} {\bfseries 30} (2021) 2142015}.

\bibitem{Lu:2024prz}
M.~Lu, J.~Yang and R.~B. Mann, \emph{{Existence of Vacuum Wormholes in Einsteinian Cubic Gravity}},  \href{https://arxiv.org/abs/2410.13996}{{\ttfamily 2410.13996}}.

\bibitem{Alvarez:2023gfg}
E.~\'Alvarez, J.~Anero and E.~Velasco-Aja, \emph{{Variations on the Goroff-Sagnotti operator}}, \href{https://doi.org/10.1103/PhysRevD.108.104043}{\emph{Phys. Rev. D} {\bfseries 108} (2023) 104043} [\href{https://arxiv.org/abs/2301.10773}{{\ttfamily 2301.10773}}].

\bibitem{Anselmi:2013wha}
D.~Anselmi, \emph{{Properties Of The Classical Action Of Quantum Gravity}}, \href{https://doi.org/10.1007/JHEP05(2013)028}{\emph{JHEP} {\bfseries 05} (2013) 028} [\href{https://arxiv.org/abs/1302.7100}{{\ttfamily 1302.7100}}].

\bibitem{deRham:2020ejn}
C.~de~Rham, J.~Francfort and J.~Zhang, \emph{{Black Hole Gravitational Waves in the Effective Field Theory of Gravity}}, \href{https://doi.org/10.1103/PhysRevD.102.024079}{\emph{Phys. Rev. D} {\bfseries 102} (2020) 024079} [\href{https://arxiv.org/abs/2005.13923}{{\ttfamily 2005.13923}}].

\bibitem{Wald:1984rg}
R.~M. Wald, \emph{{General Relativity}}. Chicago Univ. Pr., Chicago, USA, 1984, \href{https://doi.org/10.7208/chicago/9780226870373.001.0001}{10.7208/chicago/9780226870373.001.0001}.

\bibitem{Perkins:2016imn}
A.~Perkins, \emph{{Static spherically symmetric solutions in higher derivative gravity}}, Ph.D. thesis, Imperial Coll., London, 9, 2016.
\newblock 10.25560/44072.

\bibitem{Carballo-Rubio:2022aed}
R.~Carballo-Rubio, V.~Cardoso and Z.~Younsi, \emph{{Toward very large baseline interferometry observations of black hole structure}}, \href{https://doi.org/10.1103/PhysRevD.106.084038}{\emph{Phys. Rev. D} {\bfseries 106} (2022) 084038} [\href{https://arxiv.org/abs/2208.00704}{{\ttfamily 2208.00704}}].

\bibitem{Eichhorn:2022oma}
A.~Eichhorn, A.~Held and P.-V. Johannsen, \emph{{Universal signatures of singularity-resolving physics in photon rings of black holes and horizonless objects}}, \href{https://doi.org/10.1088/1475-7516/2023/01/043}{\emph{JCAP} {\bfseries 01} (2023) 043} [\href{https://arxiv.org/abs/2204.02429}{{\ttfamily 2204.02429}}].

\bibitem{Aktar:2024akk}
S.~Aktar, N.~U. Molla, F.~Rahaman and G.~Mustafa, \emph{{Shadows and Strong Gravitational Lensing Around Black Hole-like Compact Object in Quadratic Gravity}},  \href{https://arxiv.org/abs/2410.18227}{{\ttfamily 2410.18227}}.

\bibitem{Ayzenberg:2023hfw}
D.~Ayzenberg et~al., \emph{{Fundamental Physics Opportunities with the Next-Generation Event Horizon Telescope}},  \href{https://arxiv.org/abs/2312.02130}{{\ttfamily 2312.02130}}.

\end{thebibliography}\endgroup



\providecommand{\href}[2]{#2}\begingroup\raggedright\begin{thebibliography}{100}

\bibitem{tHooft:1974toh}
G.~'t~Hooft and M.~J.~G. Veltman, \emph{{One loop divergencies in the theory of
  gravitation}}, {\emph{Ann. Inst. H. Poincare Phys. Theor. A} {\bfseries 20}
  (1974) 69}.

\bibitem{Christensen:1979iy}
S.~M. Christensen and M.~J. Duff, \emph{{Quantizing Gravity with a Cosmological
  Constant}}, \href{https://doi.org/10.1016/0550-3213(80)90423-X}{\emph{Nucl.
  Phys. B} {\bfseries 170} (1980) 480}.

\bibitem{Goroff:1985th}
M.~H. Goroff and A.~Sagnotti, \emph{{The Ultraviolet Behavior of Einstein
  Gravity}}, \href{https://doi.org/10.1016/0550-3213(86)90193-8}{\emph{Nucl.
  Phys. B} {\bfseries 266} (1986) 709}.

\bibitem{Donoghue:1993eb}
J.~F. Donoghue, \emph{{Leading quantum correction to the Newtonian potential}},
  \href{https://doi.org/10.1103/PhysRevLett.72.2996}{\emph{Phys. Rev. Lett.}
  {\bfseries 72} (1994) 2996}
  [\href{https://arxiv.org/abs/gr-qc/9310024}{{\ttfamily gr-qc/9310024}}].

\bibitem{Donoghue:1994dn}
J.~F. Donoghue, \emph{{General relativity as an effective field theory: The
  leading quantum corrections}},
  \href{https://doi.org/10.1103/PhysRevD.50.3874}{\emph{Phys. Rev. D}
  {\bfseries 50} (1994) 3874}
  [\href{https://arxiv.org/abs/gr-qc/9405057}{{\ttfamily gr-qc/9405057}}].

\bibitem{Burgess:2003jk}
C.~P. Burgess, \emph{{Quantum gravity in everyday life: General relativity as
  an effective field theory}},
  \href{https://doi.org/10.12942/lrr-2004-5}{\emph{Living Rev. Rel.} {\bfseries
  7} (2004) 5} [\href{https://arxiv.org/abs/gr-qc/0311082}{{\ttfamily
  gr-qc/0311082}}].

\bibitem{Stelle:1976gc}
K.~S. Stelle, \emph{{Renormalization of Higher Derivative Quantum Gravity}},
  \href{https://doi.org/10.1103/PhysRevD.16.953}{\emph{Phys. Rev. D} {\bfseries
  16} (1977) 953}.

\bibitem{Modesto:2015ozb}
L.~Modesto and I.~L. Shapiro, \emph{{Superrenormalizable quantum gravity with
  complex ghosts}},
  \href{https://doi.org/10.1016/j.physletb.2016.02.021}{\emph{Phys. Lett. B}
  {\bfseries 755} (2016) 279}
  [\href{https://arxiv.org/abs/1512.07600}{{\ttfamily 1512.07600}}].

\bibitem{Anselmi:2017ygm}
D.~Anselmi, \emph{{On the quantum field theory of the gravitational
  interactions}}, \href{https://doi.org/10.1007/JHEP06(2017)086}{\emph{JHEP}
  {\bfseries 06} (2017) 086}
  [\href{https://arxiv.org/abs/1704.07728}{{\ttfamily 1704.07728}}].

\bibitem{Anselmi:2018ibi}
D.~Anselmi and M.~Piva, \emph{{The Ultraviolet Behavior of Quantum Gravity}},
  \href{https://doi.org/10.1007/JHEP05(2018)027}{\emph{JHEP} {\bfseries 05}
  (2018) 027} [\href{https://arxiv.org/abs/1803.07777}{{\ttfamily
  1803.07777}}].

\bibitem{Anselmi:2018tmf}
D.~Anselmi and M.~Piva, \emph{{Quantum Gravity, Fakeons And Microcausality}},
  \href{https://doi.org/10.1007/JHEP11(2018)021}{\emph{JHEP} {\bfseries 11}
  (2018) 021} [\href{https://arxiv.org/abs/1806.03605}{{\ttfamily
  1806.03605}}].

\bibitem{Donoghue:2019fcb}
J.~F. Donoghue and G.~Menezes, \emph{{Unitarity, stability and loops of
  unstable ghosts}},
  \href{https://doi.org/10.1103/PhysRevD.100.105006}{\emph{Phys. Rev. D}
  {\bfseries 100} (2019) 105006}
  [\href{https://arxiv.org/abs/1908.02416}{{\ttfamily 1908.02416}}].

\bibitem{Donoghue:2019ecz}
J.~F. Donoghue and G.~Menezes, \emph{{Arrow of Causality and Quantum Gravity}},
  \href{https://doi.org/10.1103/PhysRevLett.123.171601}{\emph{Phys. Rev. Lett.}
  {\bfseries 123} (2019) 171601}
  [\href{https://arxiv.org/abs/1908.04170}{{\ttfamily 1908.04170}}].

\bibitem{Wetterich:2019qzx}
C.~Wetterich, \emph{{Quantum scale symmetry}},
  \href{https://arxiv.org/abs/1901.04741}{{\ttfamily 1901.04741}}.

\bibitem{Hawking:1979ig}
S.~W. Hawking and W.~Israel, \emph{{General Relativity}: {An Einstein Centenary
  Survey}}. Univ. Pr., Cambridge, UK, 1979.

\bibitem{Reuter:1996cp}
M.~Reuter, \emph{{Nonperturbative evolution equation for quantum gravity}},
  \href{https://doi.org/10.1103/PhysRevD.57.971}{\emph{Phys. Rev. D} {\bfseries
  57} (1998) 971} [\href{https://arxiv.org/abs/hep-th/9605030}{{\ttfamily
  hep-th/9605030}}].

\bibitem{Wetterich:1992yh}
C.~Wetterich, \emph{{Exact evolution equation for the effective potential}},
  \href{https://doi.org/10.1016/0370-2693(93)90726-X}{\emph{Phys. Lett. B}
  {\bfseries 301} (1993) 90}
  [\href{https://arxiv.org/abs/1710.05815}{{\ttfamily 1710.05815}}].

\bibitem{Morris:1993qb}
T.~R. Morris, \emph{{The Exact renormalization group and approximate
  solutions}}, \href{https://doi.org/10.1142/S0217751X94000972}{\emph{Int. J.
  Mod. Phys. A} {\bfseries 9} (1994) 2411}
  [\href{https://arxiv.org/abs/hep-ph/9308265}{{\ttfamily hep-ph/9308265}}].

\bibitem{Reuter:1993kw}
M.~Reuter and C.~Wetterich, \emph{{Effective average action for gauge theories
  and exact evolution equations}},
  \href{https://doi.org/10.1016/0550-3213(94)90543-6}{\emph{Nucl. Phys. B}
  {\bfseries 417} (1994) 181}.

\bibitem{Berges:2000ew}
J.~Berges, N.~Tetradis and C.~Wetterich, \emph{{Nonperturbative renormalization
  flow in quantum field theory and statistical physics}},
  \href{https://doi.org/10.1016/S0370-1573(01)00098-9}{\emph{Phys. Rept.}
  {\bfseries 363} (2002) 223}
  [\href{https://arxiv.org/abs/hep-ph/0005122}{{\ttfamily hep-ph/0005122}}].

\bibitem{Pawlowski:2005xe}
J.~M. Pawlowski, \emph{{Aspects of the functional renormalisation group}},
  \href{https://doi.org/10.1016/j.aop.2007.01.007}{\emph{Annals Phys.}
  {\bfseries 322} (2007) 2831}
  [\href{https://arxiv.org/abs/hep-th/0512261}{{\ttfamily hep-th/0512261}}].

\bibitem{Gies:2006wv}
H.~Gies, \emph{{Introduction to the functional RG and applications to gauge
  theories}}, \href{https://doi.org/10.1007/978-3-642-27320-9_6}{\emph{Lect.
  Notes Phys.} {\bfseries 852} (2012) 287}
  [\href{https://arxiv.org/abs/hep-ph/0611146}{{\ttfamily hep-ph/0611146}}].

\bibitem{Dupuis:2020fhh}
N.~Dupuis, L.~Canet, A.~Eichhorn, W.~Metzner, J.~M. Pawlowski, M.~Tissier
  et~al., \emph{{The nonperturbative functional renormalization group and its
  applications}},
  \href{https://doi.org/10.1016/j.physrep.2021.01.001}{\emph{Phys. Rept.}
  {\bfseries 910} (2021) 1} [\href{https://arxiv.org/abs/2006.04853}{{\ttfamily
  2006.04853}}].

\bibitem{Souma:1999at}
W.~Souma, \emph{{Nontrivial ultraviolet fixed point in quantum gravity}},
  \href{https://doi.org/10.1143/PTP.102.181}{\emph{Prog. Theor. Phys.}
  {\bfseries 102} (1999) 181}
  [\href{https://arxiv.org/abs/hep-th/9907027}{{\ttfamily hep-th/9907027}}].

\bibitem{Lauscher:2001ya}
O.~Lauscher and M.~Reuter, \emph{{Ultraviolet fixed point and generalized flow
  equation of quantum gravity}},
  \href{https://doi.org/10.1103/PhysRevD.65.025013}{\emph{Phys. Rev. D}
  {\bfseries 65} (2002) 025013}
  [\href{https://arxiv.org/abs/hep-th/0108040}{{\ttfamily hep-th/0108040}}].

\bibitem{Reuter:2001ag}
M.~Reuter and F.~Saueressig, \emph{{Renormalization group flow of quantum
  gravity in the Einstein-Hilbert truncation}},
  \href{https://doi.org/10.1103/PhysRevD.65.065016}{\emph{Phys. Rev. D}
  {\bfseries 65} (2002) 065016}
  [\href{https://arxiv.org/abs/hep-th/0110054}{{\ttfamily hep-th/0110054}}].

\bibitem{Lauscher:2001rz}
O.~Lauscher and M.~Reuter, \emph{{Is quantum Einstein gravity nonperturbatively
  renormalizable?}},
  \href{https://doi.org/10.1088/0264-9381/19/3/304}{\emph{Class. Quant. Grav.}
  {\bfseries 19} (2002) 483}
  [\href{https://arxiv.org/abs/hep-th/0110021}{{\ttfamily hep-th/0110021}}].

\bibitem{Lauscher:2002sq}
O.~Lauscher and M.~Reuter, \emph{{Flow equation of quantum Einstein gravity in
  a higher derivative truncation}},
  \href{https://doi.org/10.1103/PhysRevD.66.025026}{\emph{Phys. Rev. D}
  {\bfseries 66} (2002) 025026}
  [\href{https://arxiv.org/abs/hep-th/0205062}{{\ttfamily hep-th/0205062}}].

\bibitem{Reuter:2002kd}
M.~Reuter and F.~Saueressig, \emph{{A Class of nonlocal truncations in quantum
  Einstein gravity and its renormalization group behavior}},
  \href{https://doi.org/10.1103/PhysRevD.66.125001}{\emph{Phys. Rev. D}
  {\bfseries 66} (2002) 125001}
  [\href{https://arxiv.org/abs/hep-th/0206145}{{\ttfamily hep-th/0206145}}].

\bibitem{Niedermaier:2002eq}
M.~Niedermaier, \emph{{On the renormalization of truncated quantum Einstein
  gravity}}, \href{https://doi.org/10.1088/1126-6708/2002/12/066}{\emph{JHEP}
  {\bfseries 12} (2002) 066}
  [\href{https://arxiv.org/abs/hep-th/0207143}{{\ttfamily hep-th/0207143}}].

\bibitem{Litim:2003vp}
D.~F. Litim, \emph{{Fixed points of quantum gravity}},
  \href{https://doi.org/10.1103/PhysRevLett.92.201301}{\emph{Phys. Rev. Lett.}
  {\bfseries 92} (2004) 201301}
  [\href{https://arxiv.org/abs/hep-th/0312114}{{\ttfamily hep-th/0312114}}].

\bibitem{Codello:2006in}
A.~Codello and R.~Percacci, \emph{{Fixed points of higher derivative gravity}},
  \href{https://doi.org/10.1103/PhysRevLett.97.221301}{\emph{Phys. Rev. Lett.}
  {\bfseries 97} (2006) 221301}
  [\href{https://arxiv.org/abs/hep-th/0607128}{{\ttfamily hep-th/0607128}}].

\bibitem{Machado:2007ea}
P.~F. Machado and F.~Saueressig, \emph{{On the renormalization group flow of
  f(R)-gravity}}, \href{https://doi.org/10.1103/PhysRevD.77.124045}{\emph{Phys.
  Rev. D} {\bfseries 77} (2008) 124045}
  [\href{https://arxiv.org/abs/0712.0445}{{\ttfamily 0712.0445}}].

\bibitem{Codello:2007bd}
A.~Codello, R.~Percacci and C.~Rahmede, \emph{{Ultraviolet properties of
  f(R)-gravity}}, \href{https://doi.org/10.1142/S0217751X08038135}{\emph{Int.
  J. Mod. Phys. A} {\bfseries 23} (2008) 143}
  [\href{https://arxiv.org/abs/0705.1769}{{\ttfamily 0705.1769}}].

\bibitem{Codello:2008vh}
A.~Codello, R.~Percacci and C.~Rahmede, \emph{{Investigating the Ultraviolet
  Properties of Gravity with a Wilsonian Renormalization Group Equation}},
  \href{https://doi.org/10.1016/j.aop.2008.08.008}{\emph{Annals Phys.}
  {\bfseries 324} (2009) 414}
  [\href{https://arxiv.org/abs/0805.2909}{{\ttfamily 0805.2909}}].

\bibitem{Benedetti:2009rx}
D.~Benedetti, P.~F. Machado and F.~Saueressig, \emph{{Asymptotic safety in
  higher-derivative gravity}},
  \href{https://doi.org/10.1142/S0217732309031521}{\emph{Mod. Phys. Lett. A}
  {\bfseries 24} (2009) 2233}
  [\href{https://arxiv.org/abs/0901.2984}{{\ttfamily 0901.2984}}].

\bibitem{Machado:2009ph}
P.~F. Machado and R.~Percacci, \emph{{Conformally reduced quantum gravity
  revisited}}, \href{https://doi.org/10.1103/PhysRevD.80.024020}{\emph{Phys.
  Rev. D} {\bfseries 80} (2009) 024020}
  [\href{https://arxiv.org/abs/0904.2510}{{\ttfamily 0904.2510}}].

\bibitem{Benedetti:2009gn}
D.~Benedetti, P.~F. Machado and F.~Saueressig, \emph{{Taming perturbative
  divergences in asymptotically safe gravity}},
  \href{https://doi.org/10.1016/j.nuclphysb.2009.08.023}{\emph{Nucl. Phys. B}
  {\bfseries 824} (2010) 168}
  [\href{https://arxiv.org/abs/0902.4630}{{\ttfamily 0902.4630}}].

\bibitem{Niedermaier:2009zz}
M.~R. Niedermaier, \emph{{Gravitational Fixed Points from Perturbation
  Theory}}, \href{https://doi.org/10.1103/PhysRevLett.103.101303}{\emph{Phys.
  Rev. Lett.} {\bfseries 103} (2009) 101303}.

\bibitem{Manrique:2009uh}
E.~Manrique and M.~Reuter, \emph{{Bimetric Truncations for Quantum Einstein
  Gravity and Asymptotic Safety}},
  \href{https://doi.org/10.1016/j.aop.2009.11.009}{\emph{Annals Phys.}
  {\bfseries 325} (2010) 785}
  [\href{https://arxiv.org/abs/0907.2617}{{\ttfamily 0907.2617}}].

\bibitem{Manrique:2010am}
E.~Manrique, M.~Reuter and F.~Saueressig, \emph{{Bimetric Renormalization Group
  Flows in Quantum Einstein Gravity}},
  \href{https://doi.org/10.1016/j.aop.2010.11.006}{\emph{Annals Phys.}
  {\bfseries 326} (2011) 463}
  [\href{https://arxiv.org/abs/1006.0099}{{\ttfamily 1006.0099}}].

\bibitem{Groh:2010ta}
K.~Groh and F.~Saueressig, \emph{{Ghost wave-function renormalization in
  Asymptotically Safe Quantum Gravity}},
  \href{https://doi.org/10.1088/1751-8113/43/36/365403}{\emph{J. Phys. A}
  {\bfseries 43} (2010) 365403}
  [\href{https://arxiv.org/abs/1001.5032}{{\ttfamily 1001.5032}}].

\bibitem{Eichhorn:2010tb}
A.~Eichhorn and H.~Gies, \emph{{Ghost anomalous dimension in asymptotically
  safe quantum gravity}},
  \href{https://doi.org/10.1103/PhysRevD.81.104010}{\emph{Phys. Rev. D}
  {\bfseries 81} (2010) 104010}
  [\href{https://arxiv.org/abs/1001.5033}{{\ttfamily 1001.5033}}].

\bibitem{Benedetti:2010nr}
D.~Benedetti, K.~Groh, P.~F. Machado and F.~Saueressig, \emph{{The Universal RG
  Machine}}, \href{https://doi.org/10.1007/JHEP06(2011)079}{\emph{JHEP}
  {\bfseries 06} (2011) 079} [\href{https://arxiv.org/abs/1012.3081}{{\ttfamily
  1012.3081}}].

\bibitem{Manrique:2011jc}
E.~Manrique, S.~Rechenberger and F.~Saueressig, \emph{{Asymptotically Safe
  Lorentzian Gravity}},
  \href{https://doi.org/10.1103/PhysRevLett.106.251302}{\emph{Phys. Rev. Lett.}
  {\bfseries 106} (2011) 251302}
  [\href{https://arxiv.org/abs/1102.5012}{{\ttfamily 1102.5012}}].

\bibitem{Benedetti:2012dx}
D.~Benedetti and F.~Caravelli, \emph{{The Local potential approximation in
  quantum gravity}}, \href{https://doi.org/10.1007/JHEP06(2012)017}{\emph{JHEP}
  {\bfseries 06} (2012) 017} [\href{https://arxiv.org/abs/1204.3541}{{\ttfamily
  1204.3541}}].

\bibitem{Christiansen:2012rx}
N.~Christiansen, D.~F. Litim, J.~M. Pawlowski and A.~Rodigast, \emph{{Fixed
  points and infrared completion of quantum gravity}},
  \href{https://doi.org/10.1016/j.physletb.2013.11.025}{\emph{Phys. Lett. B}
  {\bfseries 728} (2014) 114}
  [\href{https://arxiv.org/abs/1209.4038}{{\ttfamily 1209.4038}}].

\bibitem{Dietz:2012ic}
J.~A. Dietz and T.~R. Morris, \emph{{Asymptotic safety in the f(R)
  approximation}}, \href{https://doi.org/10.1007/JHEP01(2013)108}{\emph{JHEP}
  {\bfseries 01} (2013) 108} [\href{https://arxiv.org/abs/1211.0955}{{\ttfamily
  1211.0955}}].

\bibitem{Becker:2012js}
D.~Becker and M.~Reuter, \emph{{Running boundary actions, Asymptotic Safety,
  and black hole thermodynamics}},
  \href{https://doi.org/10.1007/JHEP07(2012)172}{\emph{JHEP} {\bfseries 07}
  (2012) 172} [\href{https://arxiv.org/abs/1205.3583}{{\ttfamily 1205.3583}}].

\bibitem{Codello:2013fpa}
A.~Codello, G.~D'Odorico and C.~Pagani, \emph{{Consistent closure of
  renormalization group flow equations in quantum gravity}},
  \href{https://doi.org/10.1103/PhysRevD.89.081701}{\emph{Phys. Rev. D}
  {\bfseries 89} (2014) 081701}
  [\href{https://arxiv.org/abs/1304.4777}{{\ttfamily 1304.4777}}].

\bibitem{Ohta:2013uca}
N.~Ohta and R.~Percacci, \emph{{Higher Derivative Gravity and Asymptotic Safety
  in Diverse Dimensions}},
  \href{https://doi.org/10.1088/0264-9381/31/1/015024}{\emph{Class. Quant.
  Grav.} {\bfseries 31} (2014) 015024}
  [\href{https://arxiv.org/abs/1308.3398}{{\ttfamily 1308.3398}}].

\bibitem{Falls:2013bv}
K.~Falls, D.~F. Litim, K.~Nikolakopoulos and C.~Rahmede, \emph{{A bootstrap
  towards asymptotic safety}},
  \href{https://arxiv.org/abs/1301.4191}{{\ttfamily 1301.4191}}.

\bibitem{Becker:2014pea}
D.~Becker and M.~Reuter, \emph{{Towards a $C$-function in 4D quantum gravity}},
  \href{https://doi.org/10.1007/JHEP03(2015)065}{\emph{JHEP} {\bfseries 03}
  (2015) 065} [\href{https://arxiv.org/abs/1412.0468}{{\ttfamily 1412.0468}}].

\bibitem{Falls:2014tra}
K.~Falls, D.~F. Litim, K.~Nikolakopoulos and C.~Rahmede, \emph{{Further
  evidence for asymptotic safety of quantum gravity}},
  \href{https://doi.org/10.1103/PhysRevD.93.104022}{\emph{Phys. Rev. D}
  {\bfseries 93} (2016) 104022}
  [\href{https://arxiv.org/abs/1410.4815}{{\ttfamily 1410.4815}}].

\bibitem{Christiansen:2014raa}
N.~Christiansen, B.~Knorr, J.~M. Pawlowski and A.~Rodigast, \emph{{Global Flows
  in Quantum Gravity}},
  \href{https://doi.org/10.1103/PhysRevD.93.044036}{\emph{Phys. Rev. D}
  {\bfseries 93} (2016) 044036}
  [\href{https://arxiv.org/abs/1403.1232}{{\ttfamily 1403.1232}}].

\bibitem{Becker:2014qya}
D.~Becker and M.~Reuter, \emph{{En route to Background Independence: Broken
  split-symmetry, and how to restore it with bi-metric average actions}},
  \href{https://doi.org/10.1016/j.aop.2014.07.023}{\emph{Annals Phys.}
  {\bfseries 350} (2014) 225}
  [\href{https://arxiv.org/abs/1404.4537}{{\ttfamily 1404.4537}}].

\bibitem{Demmel:2014hla}
M.~Demmel, F.~Saueressig and O.~Zanusso, \emph{{RG flows of Quantum Einstein
  Gravity in the linear-geometric approximation}},
  \href{https://doi.org/10.1016/j.aop.2015.04.018}{\emph{Annals Phys.}
  {\bfseries 359} (2015) 141}
  [\href{https://arxiv.org/abs/1412.7207}{{\ttfamily 1412.7207}}].

\bibitem{Christiansen:2015rva}
N.~Christiansen, B.~Knorr, J.~Meibohm, J.~M. Pawlowski and M.~Reichert,
  \emph{{Local Quantum Gravity}},
  \href{https://doi.org/10.1103/PhysRevD.92.121501}{\emph{Phys. Rev. D}
  {\bfseries 92} (2015) 121501}
  [\href{https://arxiv.org/abs/1506.07016}{{\ttfamily 1506.07016}}].

\bibitem{Morris:2015oca}
T.~R. Morris and Z.~H. Slade, \emph{{Solutions to the reconstruction problem in
  asymptotic safety}},
  \href{https://doi.org/10.1007/JHEP11(2015)094}{\emph{JHEP} {\bfseries 11}
  (2015) 094} [\href{https://arxiv.org/abs/1507.08657}{{\ttfamily
  1507.08657}}].

\bibitem{Ohta:2015efa}
N.~Ohta, R.~Percacci and G.~P. Vacca, \emph{{Flow equation for $f(R)$ gravity
  and some of its exact solutions}},
  \href{https://doi.org/10.1103/PhysRevD.92.061501}{\emph{Phys. Rev. D}
  {\bfseries 92} (2015) 061501}
  [\href{https://arxiv.org/abs/1507.00968}{{\ttfamily 1507.00968}}].

\bibitem{Ohta:2015fcu}
N.~Ohta, R.~Percacci and G.~P. Vacca, \emph{{Renormalization Group Equation and
  scaling solutions for f(R) gravity in exponential parametrization}},
  \href{https://doi.org/10.1140/epjc/s10052-016-3895-1}{\emph{Eur. Phys. J. C}
  {\bfseries 76} (2016) 46} [\href{https://arxiv.org/abs/1511.09393}{{\ttfamily
  1511.09393}}].

\bibitem{Gies:2015tca}
H.~Gies, B.~Knorr and S.~Lippoldt, \emph{{Generalized Parametrization
  Dependence in Quantum Gravity}},
  \href{https://doi.org/10.1103/PhysRevD.92.084020}{\emph{Phys. Rev. D}
  {\bfseries 92} (2015) 084020}
  [\href{https://arxiv.org/abs/1507.08859}{{\ttfamily 1507.08859}}].

\bibitem{Biemans:2016rvp}
J.~Biemans, A.~Platania and F.~Saueressig, \emph{{Quantum gravity on foliated
  spacetimes: Asymptotically safe and sound}},
  \href{https://doi.org/10.1103/PhysRevD.95.086013}{\emph{Phys. Rev. D}
  {\bfseries 95} (2017) 086013}
  [\href{https://arxiv.org/abs/1609.04813}{{\ttfamily 1609.04813}}].

\bibitem{Gies:2016con}
H.~Gies, B.~Knorr, S.~Lippoldt and F.~Saueressig, \emph{{Gravitational Two-Loop
  Counterterm Is Asymptotically Safe}},
  \href{https://doi.org/10.1103/PhysRevLett.116.211302}{\emph{Phys. Rev. Lett.}
  {\bfseries 116} (2016) 211302}
  [\href{https://arxiv.org/abs/1601.01800}{{\ttfamily 1601.01800}}].

\bibitem{Denz:2016qks}
T.~Denz, J.~M. Pawlowski and M.~Reichert, \emph{{Towards apparent convergence
  in asymptotically safe quantum gravity}},
  \href{https://doi.org/10.1140/epjc/s10052-018-5806-0}{\emph{Eur. Phys. J. C}
  {\bfseries 78} (2018) 336}
  [\href{https://arxiv.org/abs/1612.07315}{{\ttfamily 1612.07315}}].

\bibitem{Platania:2017djo}
A.~Platania and F.~Saueressig, \emph{{Functional Renormalization Group Flows on
  Friedman\textendash{}Lema\^\i{}tre\textendash{}Robertson\textendash{}Walker
  backgrounds}}, \href{https://doi.org/10.1007/s10701-018-0181-0}{\emph{Found.
  Phys.} {\bfseries 48} (2018) 1291}
  [\href{https://arxiv.org/abs/1710.01972}{{\ttfamily 1710.01972}}].

\bibitem{Houthoff:2017oam}
W.~B. Houthoff, A.~Kurov and F.~Saueressig, \emph{{Impact of topology in
  foliated Quantum Einstein Gravity}},
  \href{https://doi.org/10.1140/epjc/s10052-017-5046-8}{\emph{Eur. Phys. J. C}
  {\bfseries 77} (2017) 491}
  [\href{https://arxiv.org/abs/1705.01848}{{\ttfamily 1705.01848}}].

\bibitem{Falls:2017lst}
K.~Falls, C.~R. King, D.~F. Litim, K.~Nikolakopoulos and C.~Rahmede,
  \emph{{Asymptotic safety of quantum gravity beyond Ricci scalars}},
  \href{https://doi.org/10.1103/PhysRevD.97.086006}{\emph{Phys. Rev. D}
  {\bfseries 97} (2018) 086006}
  [\href{https://arxiv.org/abs/1801.00162}{{\ttfamily 1801.00162}}].

\bibitem{Knorr:2017fus}
B.~Knorr and S.~Lippoldt, \emph{{Correlation functions on a curved
  background}}, \href{https://doi.org/10.1103/PhysRevD.96.065020}{\emph{Phys.
  Rev. D} {\bfseries 96} (2017) 065020}
  [\href{https://arxiv.org/abs/1707.01397}{{\ttfamily 1707.01397}}].

\bibitem{Christiansen:2017bsy}
N.~Christiansen, K.~Falls, J.~M. Pawlowski and M.~Reichert, \emph{{Curvature
  dependence of quantum gravity}},
  \href{https://doi.org/10.1103/PhysRevD.97.046007}{\emph{Phys. Rev. D}
  {\bfseries 97} (2018) 046007}
  [\href{https://arxiv.org/abs/1711.09259}{{\ttfamily 1711.09259}}].

\bibitem{Becker:2018quq}
M.~Becker and C.~Pagani, \emph{{Geometric operators in the asymptotic safety
  scenario for quantum gravity}},
  \href{https://doi.org/10.1103/PhysRevD.99.066002}{\emph{Phys. Rev. D}
  {\bfseries 99} (2019) 066002}
  [\href{https://arxiv.org/abs/1810.11816}{{\ttfamily 1810.11816}}].

\bibitem{DeBrito:2018hur}
G.~P. De~Brito, N.~Ohta, A.~D. Pereira, A.~A. Tomaz and M.~Yamada,
  \emph{{Asymptotic safety and field parametrization dependence in the $f(R)$
  truncation}}, \href{https://doi.org/10.1103/PhysRevD.98.026027}{\emph{Phys.
  Rev. D} {\bfseries 98} (2018) 026027}
  [\href{https://arxiv.org/abs/1805.09656}{{\ttfamily 1805.09656}}].

\bibitem{Falls:2018ylp}
K.~G. Falls, D.~F. Litim and J.~Schr\"oder, \emph{{Aspects of asymptotic safety
  for quantum gravity}},
  \href{https://doi.org/10.1103/PhysRevD.99.126015}{\emph{Phys. Rev. D}
  {\bfseries 99} (2019) 126015}
  [\href{https://arxiv.org/abs/1810.08550}{{\ttfamily 1810.08550}}].

\bibitem{Knorr:2018kog}
B.~Knorr and F.~Saueressig, \emph{{Towards reconstructing the quantum effective
  action of gravity}},
  \href{https://doi.org/10.1103/PhysRevLett.121.161304}{\emph{Phys. Rev. Lett.}
  {\bfseries 121} (2018) 161304}
  [\href{https://arxiv.org/abs/1804.03846}{{\ttfamily 1804.03846}}].

\bibitem{Becker:2019fhi}
M.~Becker, C.~Pagani and O.~Zanusso, \emph{{Fractal Geometry of Higher
  Derivative Gravity}},
  \href{https://doi.org/10.1103/PhysRevLett.124.151302}{\emph{Phys. Rev. Lett.}
  {\bfseries 124} (2020) 151302}
  [\href{https://arxiv.org/abs/1911.02415}{{\ttfamily 1911.02415}}].

\bibitem{Becker:2019tlf}
M.~Becker and C.~Pagani, \emph{{Geometric Operators in the
  Einstein\textendash{}Hilbert Truncation}},
  \href{https://doi.org/10.3390/universe5030075}{\emph{Universe} {\bfseries 5}
  (2019) 75}.

\bibitem{Becker:2020mjl}
M.~Becker and M.~Reuter, \emph{{Background Independent Field Quantization with
  Sequences of Gravity-Coupled Approximants}},
  \href{https://doi.org/10.1103/PhysRevD.102.125001}{\emph{Phys. Rev. D}
  {\bfseries 102} (2020) 125001}
  [\href{https://arxiv.org/abs/2008.09430}{{\ttfamily 2008.09430}}].

\bibitem{Kluth:2020bdv}
Y.~Kluth and D.~F. Litim, \emph{{Fixed Points of Quantum Gravity and the
  Dimensionality of the UV Critical Surface}},
  \href{https://arxiv.org/abs/2008.09181}{{\ttfamily 2008.09181}}.

\bibitem{Falls:2020qhj}
K.~Falls, N.~Ohta and R.~Percacci, \emph{{Towards the determination of the
  dimension of the critical surface in asymptotically safe gravity}},
  \href{https://doi.org/10.1016/j.physletb.2020.135773}{\emph{Phys. Lett. B}
  {\bfseries 810} (2020) 135773}
  [\href{https://arxiv.org/abs/2004.04126}{{\ttfamily 2004.04126}}].

\bibitem{Knorr:2020ckv}
B.~Knorr, \emph{{Lessons from conformally reduced quantum gravity}},
  \href{https://doi.org/10.1088/1361-6382/abd7c2}{\emph{Class. Quant. Grav.}
  {\bfseries 38} (2021) 065003}
  [\href{https://arxiv.org/abs/2010.00492}{{\ttfamily 2010.00492}}].

\bibitem{Bonanno:2021squ}
A.~Bonanno, T.~Denz, J.~M. Pawlowski and M.~Reichert, \emph{{Reconstructing the
  graviton}},  \href{https://arxiv.org/abs/2102.02217}{{\ttfamily 2102.02217}}.

\bibitem{Martini:2021slj}
R.~Martini, A.~Ugolotti, F.~Del~Porro and O.~Zanusso, \emph{{Gravity in
  $d=2+\epsilon$ dimensions and realizations of the diffeomorphisms group}},
  \href{https://arxiv.org/abs/2103.12421}{{\ttfamily 2103.12421}}.

\bibitem{Knorr:2021niv}
B.~Knorr and M.~Schiffer, \emph{{Non-Perturbative Propagators in Quantum
  Gravity}}, \href{https://doi.org/10.3390/universe7070216}{\emph{Universe}
  {\bfseries 7} (2021) 216} [\href{https://arxiv.org/abs/2105.04566}{{\ttfamily
  2105.04566}}].

\bibitem{Martini:2021lcx}
R.~Martini, A.~Ugolotti and O.~Zanusso, \emph{{The Search for the Universality
  Class of Metric Quantum Gravity}},
  \href{https://doi.org/10.3390/universe7060162}{\emph{Universe} {\bfseries 7}
  (2021) 162} [\href{https://arxiv.org/abs/2105.11870}{{\ttfamily
  2105.11870}}].

\bibitem{Becker:2021pwo}
M.~Becker and M.~Reuter, \emph{{Background Independent Field Quantization with
  Sequences of Gravity-Coupled Approximants II: Metric Fluctuations}},
  \href{https://arxiv.org/abs/2109.09496}{{\ttfamily 2109.09496}}.

\bibitem{Baldazzi:2021orb}
A.~Baldazzi and K.~Falls, \emph{{Essential Quantum Einstein Gravity}},
  \href{https://doi.org/10.3390/universe7080294}{\emph{Universe} {\bfseries 7}
  (2021) 294} [\href{https://arxiv.org/abs/2107.00671}{{\ttfamily
  2107.00671}}].

\bibitem{Daum:2010qt}
J.~E. Daum and M.~Reuter, \emph{{Renormalization Group Flow of the Holst
  Action}}, \href{https://doi.org/10.1016/j.physletb.2012.01.046}{\emph{Phys.
  Lett. B} {\bfseries 710} (2012) 215}
  [\href{https://arxiv.org/abs/1012.4280}{{\ttfamily 1012.4280}}].

\bibitem{Codello:2012sn}
A.~Codello, G.~D'Odorico, C.~Pagani and R.~Percacci, \emph{{The Renormalization
  Group and Weyl-invariance}},
  \href{https://doi.org/10.1088/0264-9381/30/11/115015}{\emph{Class. Quant.
  Grav.} {\bfseries 30} (2013) 115015}
  [\href{https://arxiv.org/abs/1210.3284}{{\ttfamily 1210.3284}}].

\bibitem{Harst:2012ni}
U.~Harst and M.~Reuter, \emph{{The 'Tetrad only' theory space: Nonperturbative
  renormalization flow and Asymptotic Safety}},
  \href{https://doi.org/10.1007/JHEP05(2012)005}{\emph{JHEP} {\bfseries 05}
  (2012) 005} [\href{https://arxiv.org/abs/1203.2158}{{\ttfamily 1203.2158}}].

\bibitem{Daum:2013fu}
J.~E. Daum and M.~Reuter, \emph{{Einstein-Cartan gravity, Asymptotic Safety,
  and the running Immirzi parameter}},
  \href{https://doi.org/10.1016/j.aop.2013.04.002}{\emph{Annals Phys.}
  {\bfseries 334} (2013) 351}
  [\href{https://arxiv.org/abs/1301.5135}{{\ttfamily 1301.5135}}].

\bibitem{Harst:2014vca}
U.~Harst and M.~Reuter, \emph{{A new functional flow equation for
  Einstein\textendash{}Cartan quantum gravity}},
  \href{https://doi.org/10.1016/j.aop.2015.01.006}{\emph{Annals Phys.}
  {\bfseries 354} (2015) 637}
  [\href{https://arxiv.org/abs/1410.7003}{{\ttfamily 1410.7003}}].

\bibitem{Harst:2015eha}
U.~Harst and M.~Reuter, \emph{{On selfdual spin-connections and Asymptotic
  Safety}}, \href{https://doi.org/10.1016/j.physletb.2015.12.016}{\emph{Phys.
  Lett. B} {\bfseries 753} (2016) 395}
  [\href{https://arxiv.org/abs/1509.09122}{{\ttfamily 1509.09122}}].

\bibitem{DeBrito:2019gdd}
G.~P. De~Brito, A.~Eichhorn and A.~D. Pereira, \emph{{A link that matters:
  Towards phenomenological tests of unimodular asymptotic safety}},
  \href{https://doi.org/10.1007/JHEP09(2019)100}{\emph{JHEP} {\bfseries 09}
  (2019) 100} [\href{https://arxiv.org/abs/1907.11173}{{\ttfamily
  1907.11173}}].

\bibitem{deBrito:2020rwu}
G.~P. de~Brito and A.~D. Pereira, \emph{{Unimodular quantum gravity: Steps
  beyond perturbation theory}},
  \href{https://doi.org/10.1007/JHEP09(2020)196}{\emph{JHEP} {\bfseries 09}
  (2020) 196} [\href{https://arxiv.org/abs/2007.05589}{{\ttfamily
  2007.05589}}].

\bibitem{deBrito:2020xhy}
G.~P. de~Brito, A.~D. Pereira and A.~F. Vieira, \emph{{Exploring new corners of
  asymptotically safe unimodular quantum gravity}},
  \href{https://doi.org/10.1103/PhysRevD.103.104023}{\emph{Phys. Rev. D}
  {\bfseries 103} (2021) 104023}
  [\href{https://arxiv.org/abs/2012.08904}{{\ttfamily 2012.08904}}].

\bibitem{deBrito:2021pmw}
G.~P. de~Brito, O.~Melichev, R.~Percacci and A.~D. Pereira, \emph{{Can quantum
  fluctuations differentiate between standard and unimodular gravity?}},
  \href{https://arxiv.org/abs/2105.13886}{{\ttfamily 2105.13886}}.

\bibitem{Ferrero:2021xqg}
R.~Ferrero and M.~Reuter, \emph{{Towards a Geometrization of Renormalization
  Group Histories in Asymptotic Safety}},
  \href{https://doi.org/10.3390/universe7050125}{\emph{Universe} {\bfseries 7}
  (2021) 125} [\href{https://arxiv.org/abs/2103.15709}{{\ttfamily
  2103.15709}}].

\bibitem{Lauscher:2005qz}
O.~Lauscher and M.~Reuter, \emph{{Fractal spacetime structure in asymptotically
  safe gravity}},
  \href{https://doi.org/10.1088/1126-6708/2005/10/050}{\emph{JHEP} {\bfseries
  10} (2005) 050} [\href{https://arxiv.org/abs/hep-th/0508202}{{\ttfamily
  hep-th/0508202}}].

\bibitem{Reuter:2006zq}
M.~Reuter and J.-M. Schwindt, \emph{{Scale-dependent metric and causal
  structures in Quantum Einstein Gravity}},
  \href{https://doi.org/10.1088/1126-6708/2007/01/049}{\emph{JHEP} {\bfseries
  01} (2007) 049} [\href{https://arxiv.org/abs/hep-th/0611294}{{\ttfamily
  hep-th/0611294}}].

\bibitem{Manrique:2008zw}
E.~Manrique and M.~Reuter, \emph{{Bare Action and Regularized Functional
  Integral of Asymptotically Safe Quantum Gravity}},
  \href{https://doi.org/10.1103/PhysRevD.79.025008}{\emph{Phys. Rev. D}
  {\bfseries 79} (2009) 025008}
  [\href{https://arxiv.org/abs/0811.3888}{{\ttfamily 0811.3888}}].

\bibitem{Reuter:2008qx}
M.~Reuter and H.~Weyer, \emph{{Conformal sector of Quantum Einstein Gravity in
  the local potential approximation: Non-Gaussian fixed point and a phase of
  unbroken diffeomorphism invariance}},
  \href{https://doi.org/10.1103/PhysRevD.80.025001}{\emph{Phys. Rev. D}
  {\bfseries 80} (2009) 025001}
  [\href{https://arxiv.org/abs/0804.1475}{{\ttfamily 0804.1475}}].

\bibitem{Reuter:2008wj}
M.~Reuter and H.~Weyer, \emph{{Background Independence and Asymptotic Safety in
  Conformally Reduced Gravity}},
  \href{https://doi.org/10.1103/PhysRevD.79.105005}{\emph{Phys. Rev. D}
  {\bfseries 79} (2009) 105005}
  [\href{https://arxiv.org/abs/0801.3287}{{\ttfamily 0801.3287}}].

\bibitem{Reuter:2011ah}
M.~Reuter and F.~Saueressig, \emph{{Fractal space-times under the microscope: A
  Renormalization Group view on Monte Carlo data}},
  \href{https://doi.org/10.1007/JHEP12(2011)012}{\emph{JHEP} {\bfseries 12}
  (2011) 012} [\href{https://arxiv.org/abs/1110.5224}{{\ttfamily 1110.5224}}].

\bibitem{Nink:2012vd}
A.~Nink and M.~Reuter, \emph{{On the physical mechanism underlying Asymptotic
  Safety}}, \href{https://doi.org/10.1007/JHEP01(2013)062}{\emph{JHEP}
  {\bfseries 01} (2013) 062} [\href{https://arxiv.org/abs/1208.0031}{{\ttfamily
  1208.0031}}].

\bibitem{Nink:2014yya}
A.~Nink, \emph{{Field Parametrization Dependence in Asymptotically Safe Quantum
  Gravity}}, \href{https://doi.org/10.1103/PhysRevD.91.044030}{\emph{Phys. Rev.
  D} {\bfseries 91} (2015) 044030}
  [\href{https://arxiv.org/abs/1410.7816}{{\ttfamily 1410.7816}}].

\bibitem{Reuter:2015rta}
M.~Reuter and G.~M. Schollmeyer, \emph{{The metric on field space, functional
  renormalization, and metric\textendash{}torsion quantum gravity}},
  \href{https://doi.org/10.1016/j.aop.2015.12.004}{\emph{Annals Phys.}
  {\bfseries 367} (2016) 125}
  [\href{https://arxiv.org/abs/1509.05041}{{\ttfamily 1509.05041}}].

\bibitem{Nink:2015lmq}
A.~Nink and M.~Reuter, \emph{{The unitary conformal field theory behind 2D
  Asymptotic Safety}},
  \href{https://doi.org/10.1007/JHEP02(2016)167}{\emph{JHEP} {\bfseries 02}
  (2016) 167} [\href{https://arxiv.org/abs/1512.06805}{{\ttfamily
  1512.06805}}].

\bibitem{Ohta:2015zwa}
N.~Ohta and R.~Percacci, \emph{{Ultraviolet Fixed Points in Conformal Gravity
  and General Quadratic Theories}},
  \href{https://doi.org/10.1088/0264-9381/33/3/035001}{\emph{Class. Quant.
  Grav.} {\bfseries 33} (2016) 035001}
  [\href{https://arxiv.org/abs/1506.05526}{{\ttfamily 1506.05526}}].

\bibitem{Ohta:2016npm}
N.~Ohta, R.~Percacci and A.~D. Pereira, \emph{{Gauges and functional measures
  in quantum gravity I: Einstein theory}},
  \href{https://doi.org/10.1007/JHEP06(2016)115}{\emph{JHEP} {\bfseries 06}
  (2016) 115} [\href{https://arxiv.org/abs/1605.00454}{{\ttfamily
  1605.00454}}].

\bibitem{Pagani:2013fca}
C.~Pagani and R.~Percacci, \emph{{Quantization and fixed points of
  non-integrable Weyl theory}},
  \href{https://doi.org/10.1088/0264-9381/31/11/115005}{\emph{Class. Quant.
  Grav.} {\bfseries 31} (2014) 115005}
  [\href{https://arxiv.org/abs/1312.7767}{{\ttfamily 1312.7767}}].

\bibitem{Pagani:2015ema}
C.~Pagani and R.~Percacci, \emph{{Quantum gravity with torsion and
  non-metricity}},
  \href{https://doi.org/10.1088/0264-9381/32/19/195019}{\emph{Class. Quant.
  Grav.} {\bfseries 32} (2015) 195019}
  [\href{https://arxiv.org/abs/1506.02882}{{\ttfamily 1506.02882}}].

\bibitem{Pagani:2016dof}
C.~Pagani and M.~Reuter, \emph{{Composite Operators in Asymptotic Safety}},
  \href{https://doi.org/10.1103/PhysRevD.95.066002}{\emph{Phys. Rev. D}
  {\bfseries 95} (2017) 066002}
  [\href{https://arxiv.org/abs/1611.06522}{{\ttfamily 1611.06522}}].

\bibitem{Pagani:2019vfm}
C.~Pagani and M.~Reuter, \emph{{Background Independent Quantum Field Theory and
  Gravitating Vacuum Fluctuations}},
  \href{https://doi.org/10.1016/j.aop.2019.167972}{\emph{Annals Phys.}
  {\bfseries 411} (2019) 167972}
  [\href{https://arxiv.org/abs/1906.02507}{{\ttfamily 1906.02507}}].

\bibitem{Pagani:2020say}
C.~Pagani and M.~Reuter, \emph{{Why the Cosmological Constant Seems to Hardly
  Care About Quantum Vacuum Fluctuations: Surprises From Background Independent
  Coarse Graining}},
  \href{https://doi.org/10.3389/fphy.2020.00214}{\emph{Front. in Phys.}
  {\bfseries 8} (2020) 214}.

\bibitem{Donoghue:2019clr}
J.~F. Donoghue, \emph{{A Critique of the Asymptotic Safety Program}},
  \href{https://doi.org/10.3389/fphy.2020.00056}{\emph{Front. in Phys.}
  {\bfseries 8} (2020) 56} [\href{https://arxiv.org/abs/1911.02967}{{\ttfamily
  1911.02967}}].

\bibitem{Bonanno:2020bil}
A.~Bonanno, A.~Eichhorn, H.~Gies, J.~M. Pawlowski, R.~Percacci, M.~Reuter
  et~al., \emph{{Critical reflections on asymptotically safe gravity}},
  \href{https://doi.org/10.3389/fphy.2020.00269}{\emph{Front. in Phys.}
  {\bfseries 8} (2020) 269} [\href{https://arxiv.org/abs/2004.06810}{{\ttfamily
  2004.06810}}].

\bibitem{Bridle:2013sra}
I.~H. Bridle, J.~A. Dietz and T.~R. Morris, \emph{{The local potential
  approximation in the background field formalism}},
  \href{https://doi.org/10.1007/JHEP03(2014)093}{\emph{JHEP} {\bfseries 03}
  (2014) 093} [\href{https://arxiv.org/abs/1312.2846}{{\ttfamily 1312.2846}}].

\bibitem{Demmel:2014sga}
M.~Demmel, F.~Saueressig and O.~Zanusso, \emph{{RG flows of Quantum Einstein
  Gravity on maximally symmetric spaces}},
  \href{https://doi.org/10.1007/JHEP06(2014)026}{\emph{JHEP} {\bfseries 06}
  (2014) 026} [\href{https://arxiv.org/abs/1401.5495}{{\ttfamily 1401.5495}}].

\bibitem{Demmel:2015oqa}
M.~Demmel, F.~Saueressig and O.~Zanusso, \emph{{A proper fixed functional for
  four-dimensional Quantum Einstein Gravity}},
  \href{https://doi.org/10.1007/JHEP08(2015)113}{\emph{JHEP} {\bfseries 08}
  (2015) 113} [\href{https://arxiv.org/abs/1504.07656}{{\ttfamily
  1504.07656}}].

\bibitem{Gonzalez-Martin:2017gza}
S.~Gonzalez-Martin, T.~R. Morris and Z.~H. Slade, \emph{{Asymptotic solutions
  in asymptotic safety}},
  \href{https://doi.org/10.1103/PhysRevD.95.106010}{\emph{Phys. Rev. D}
  {\bfseries 95} (2017) 106010}
  [\href{https://arxiv.org/abs/1704.08873}{{\ttfamily 1704.08873}}].

\bibitem{Biemans:2017zca}
J.~Biemans, A.~Platania and F.~Saueressig, \emph{{Renormalization group fixed
  points of foliated gravity-matter systems}},
  \href{https://doi.org/10.1007/JHEP05(2017)093}{\emph{JHEP} {\bfseries 05}
  (2017) 093} [\href{https://arxiv.org/abs/1702.06539}{{\ttfamily
  1702.06539}}].

\bibitem{Knorr:2018fdu}
B.~Knorr, \emph{{Lorentz symmetry is relevant}},
  \href{https://doi.org/10.1016/j.physletb.2019.01.070}{\emph{Phys. Lett. B}
  {\bfseries 792} (2019) 142}
  [\href{https://arxiv.org/abs/1810.07971}{{\ttfamily 1810.07971}}].

\bibitem{Niedermaier:2006wt}
M.~Niedermaier and M.~Reuter, \emph{{The Asymptotic Safety Scenario in Quantum
  Gravity}}, \href{https://doi.org/10.12942/lrr-2006-5}{\emph{Living Rev. Rel.}
  {\bfseries 9} (2006) 5}.

\bibitem{Reuter:2012id}
M.~Reuter and F.~Saueressig, \emph{{Quantum Einstein Gravity}},
  \href{https://doi.org/10.1088/1367-2630/14/5/055022}{\emph{New J. Phys.}
  {\bfseries 14} (2012) 055022}
  [\href{https://arxiv.org/abs/1202.2274}{{\ttfamily 1202.2274}}].

\bibitem{Eichhorn:2017egq}
A.~Eichhorn, \emph{{Status of the asymptotic safety paradigm for quantum
  gravity and matter}},
  \href{https://doi.org/10.1007/s10701-018-0196-6}{\emph{Found. Phys.}
  {\bfseries 48} (2018) 1407}
  [\href{https://arxiv.org/abs/1709.03696}{{\ttfamily 1709.03696}}].

\bibitem{Percacci:2017fkn}
R.~Percacci, \emph{{An Introduction to Covariant Quantum Gravity and Asymptotic
  Safety}}, vol.~3 of \emph{100 Years of General Relativity}. World Scientific,
  2017, \href{https://doi.org/10.1142/10369}{10.1142/10369}.

\bibitem{Eichhorn:2018yfc}
A.~Eichhorn, \emph{{An asymptotically safe guide to quantum gravity and
  matter}}, \href{https://doi.org/10.3389/fspas.2018.00047}{\emph{Front.
  Astron. Space Sci.} {\bfseries 5} (2019) 47}
  [\href{https://arxiv.org/abs/1810.07615}{{\ttfamily 1810.07615}}].

\bibitem{Reuter:2019byg}
M.~Reuter and F.~Saueressig, \emph{{Quantum Gravity and the Functional
  Renormalization Group}: {The Road towards Asymptotic Safety}}. Cambridge
  University Press, 1, 2019.

\bibitem{Pereira:2019dbn}
A.~D. Pereira, \emph{{Quantum spacetime and the renormalization group: Progress
  and visions}},  in \emph{{Progress and Visions in Quantum Theory in View of
  Gravity}: {Bridging foundations of physics and mathematics}}, 4, 2019,
  \href{https://arxiv.org/abs/1904.07042}{{\ttfamily 1904.07042}}.

\bibitem{Reichert:2020mja}
M.~Reichert, \emph{{Lecture notes: Functional Renormalisation Group and
  Asymptotically Safe Quantum Gravity}},
  \href{https://doi.org/10.22323/1.384.0005}{\emph{PoS} {\bfseries 384} (2020)
  005}.

\bibitem{Pawlowski:2020qer}
J.~M. Pawlowski and M.~Reichert, \emph{{Quantum gravity: a fluctuating point of
  view}},  \href{https://arxiv.org/abs/2007.10353}{{\ttfamily 2007.10353}}.

\bibitem{Griguolo:1995db}
L.~Griguolo and R.~Percacci, \emph{{The Beta functions of a scalar theory
  coupled to gravity}},
  \href{https://doi.org/10.1103/PhysRevD.52.5787}{\emph{Phys. Rev. D}
  {\bfseries 52} (1995) 5787}
  [\href{https://arxiv.org/abs/hep-th/9504092}{{\ttfamily hep-th/9504092}}].

\bibitem{Dou:1997fg}
D.~Dou and R.~Percacci, \emph{{The running gravitational couplings}},
  \href{https://doi.org/10.1088/0264-9381/15/11/011}{\emph{Class. Quant. Grav.}
  {\bfseries 15} (1998) 3449}
  [\href{https://arxiv.org/abs/hep-th/9707239}{{\ttfamily hep-th/9707239}}].

\bibitem{Percacci:2002ie}
R.~Percacci and D.~Perini, \emph{{Constraints on matter from asymptotic
  safety}}, \href{https://doi.org/10.1103/PhysRevD.67.081503}{\emph{Phys. Rev.
  D} {\bfseries 67} (2003) 081503}
  [\href{https://arxiv.org/abs/hep-th/0207033}{{\ttfamily hep-th/0207033}}].

\bibitem{Percacci:2003jz}
R.~Percacci and D.~Perini, \emph{{Asymptotic safety of gravity coupled to
  matter}}, \href{https://doi.org/10.1103/PhysRevD.68.044018}{\emph{Phys. Rev.
  D} {\bfseries 68} (2003) 044018}
  [\href{https://arxiv.org/abs/hep-th/0304222}{{\ttfamily hep-th/0304222}}].

\bibitem{Zanusso:2009bs}
O.~Zanusso, L.~Zambelli, G.~P. Vacca and R.~Percacci, \emph{{Gravitational
  corrections to Yukawa systems}},
  \href{https://doi.org/10.1016/j.physletb.2010.04.043}{\emph{Phys. Lett. B}
  {\bfseries 689} (2010) 90} [\href{https://arxiv.org/abs/0904.0938}{{\ttfamily
  0904.0938}}].

\bibitem{Narain:2009fy}
G.~Narain and R.~Percacci, \emph{{Renormalization Group Flow in Scalar-Tensor
  Theories. I}},
  \href{https://doi.org/10.1088/0264-9381/27/7/075001}{\emph{Class. Quant.
  Grav.} {\bfseries 27} (2010) 075001}
  [\href{https://arxiv.org/abs/0911.0386}{{\ttfamily 0911.0386}}].

\bibitem{Narain:2009gb}
G.~Narain and C.~Rahmede, \emph{{Renormalization Group Flow in Scalar-Tensor
  Theories. II}},
  \href{https://doi.org/10.1088/0264-9381/27/7/075002}{\emph{Class. Quant.
  Grav.} {\bfseries 27} (2010) 075002}
  [\href{https://arxiv.org/abs/0911.0394}{{\ttfamily 0911.0394}}].

\bibitem{Daum:2009dn}
J.-E. Daum, U.~Harst and M.~Reuter, \emph{{Running Gauge Coupling in
  Asymptotically Safe Quantum Gravity}},
  \href{https://doi.org/10.1007/JHEP01(2010)084}{\emph{JHEP} {\bfseries 01}
  (2010) 084} [\href{https://arxiv.org/abs/0910.4938}{{\ttfamily 0910.4938}}].

\bibitem{Shaposhnikov:2009pv}
M.~Shaposhnikov and C.~Wetterich, \emph{{Asymptotic safety of gravity and the
  Higgs boson mass}},
  \href{https://doi.org/10.1016/j.physletb.2009.12.022}{\emph{Phys. Lett. B}
  {\bfseries 683} (2010) 196}
  [\href{https://arxiv.org/abs/0912.0208}{{\ttfamily 0912.0208}}].

\bibitem{Daum:2010bc}
J.~E. Daum, U.~Harst and M.~Reuter, \emph{{Non-perturbative QEG Corrections to
  the Yang-Mills Beta Function}},
  \href{https://doi.org/10.1007/s10714-010-1032-2}{\emph{Gen. Rel. Grav.}
  {\bfseries 43} (2011) 2393}
  [\href{https://arxiv.org/abs/1005.1488}{{\ttfamily 1005.1488}}].

\bibitem{Harst:2011zx}
U.~Harst and M.~Reuter, \emph{{QED coupled to QEG}},
  \href{https://doi.org/10.1007/JHEP05(2011)119}{\emph{JHEP} {\bfseries 05}
  (2011) 119} [\href{https://arxiv.org/abs/1101.6007}{{\ttfamily 1101.6007}}].

\bibitem{Folkerts:2011jz}
S.~Folkerts, D.~F. Litim and J.~M. Pawlowski, \emph{{Asymptotic freedom of
  Yang-Mills theory with gravity}},
  \href{https://doi.org/10.1016/j.physletb.2012.02.002}{\emph{Phys. Lett. B}
  {\bfseries 709} (2012) 234}
  [\href{https://arxiv.org/abs/1101.5552}{{\ttfamily 1101.5552}}].

\bibitem{Eichhorn:2011pc}
A.~Eichhorn and H.~Gies, \emph{{Light fermions in quantum gravity}},
  \href{https://doi.org/10.1088/1367-2630/13/12/125012}{\emph{New J. Phys.}
  {\bfseries 13} (2011) 125012}
  [\href{https://arxiv.org/abs/1104.5366}{{\ttfamily 1104.5366}}].

\bibitem{Eichhorn:2012va}
A.~Eichhorn, \emph{{Quantum-gravity-induced matter self-interactions in the
  asymptotic-safety scenario}},
  \href{https://doi.org/10.1103/PhysRevD.86.105021}{\emph{Phys. Rev. D}
  {\bfseries 86} (2012) 105021}
  [\href{https://arxiv.org/abs/1204.0965}{{\ttfamily 1204.0965}}].

\bibitem{Dona:2012am}
P.~Dona and R.~Percacci, \emph{{Functional renormalization with fermions and
  tetrads}}, \href{https://doi.org/10.1103/PhysRevD.87.045002}{\emph{Phys. Rev.
  D} {\bfseries 87} (2013) 045002}
  [\href{https://arxiv.org/abs/1209.3649}{{\ttfamily 1209.3649}}].

\bibitem{Dona:2013qba}
P.~Don\`a, A.~Eichhorn and R.~Percacci, \emph{{Matter matters in asymptotically
  safe quantum gravity}},
  \href{https://doi.org/10.1103/PhysRevD.89.084035}{\emph{Phys. Rev. D}
  {\bfseries 89} (2014) 084035}
  [\href{https://arxiv.org/abs/1311.2898}{{\ttfamily 1311.2898}}].

\bibitem{Dona:2014pla}
P.~Don\`a, A.~Eichhorn and R.~Percacci, \emph{{Consistency of matter models
  with asymptotically safe quantum gravity}},
  \href{https://doi.org/10.1139/cjp-2014-0574}{\emph{Can. J. Phys.} {\bfseries
  93} (2015) 988} [\href{https://arxiv.org/abs/1410.4411}{{\ttfamily
  1410.4411}}].

\bibitem{Labus:2015ska}
P.~Labus, R.~Percacci and G.~P. Vacca, \emph{{Asymptotic safety in $O(N)$
  scalar models coupled to gravity}},
  \href{https://doi.org/10.1016/j.physletb.2015.12.022}{\emph{Phys. Lett. B}
  {\bfseries 753} (2016) 274}
  [\href{https://arxiv.org/abs/1505.05393}{{\ttfamily 1505.05393}}].

\bibitem{Meibohm:2015twa}
J.~Meibohm, J.~M. Pawlowski and M.~Reichert, \emph{{Asymptotic safety of
  gravity-matter systems}},
  \href{https://doi.org/10.1103/PhysRevD.93.084035}{\emph{Phys. Rev. D}
  {\bfseries 93} (2016) 084035}
  [\href{https://arxiv.org/abs/1510.07018}{{\ttfamily 1510.07018}}].

\bibitem{Oda:2015sma}
K.-y. Oda and M.~Yamada, \emph{{Non-minimal coupling in
  Higgs\textendash{}Yukawa model with asymptotically safe gravity}},
  \href{https://doi.org/10.1088/0264-9381/33/12/125011}{\emph{Class. Quant.
  Grav.} {\bfseries 33} (2016) 125011}
  [\href{https://arxiv.org/abs/1510.03734}{{\ttfamily 1510.03734}}].

\bibitem{Dona:2015tnf}
P.~Don\`a, A.~Eichhorn, P.~Labus and R.~Percacci, \emph{{Asymptotic safety in
  an interacting system of gravity and scalar matter}},
  \href{https://doi.org/10.1103/PhysRevD.93.129904}{\emph{Phys. Rev. D}
  {\bfseries 93} (2016) 044049}
  [\href{https://arxiv.org/abs/1512.01589}{{\ttfamily 1512.01589}}].

\bibitem{Percacci:2015wwa}
R.~Percacci and G.~P. Vacca, \emph{{Search of scaling solutions in
  scalar-tensor gravity}},
  \href{https://doi.org/10.1140/epjc/s10052-015-3410-0}{\emph{Eur. Phys. J. C}
  {\bfseries 75} (2015) 188}
  [\href{https://arxiv.org/abs/1501.00888}{{\ttfamily 1501.00888}}].

\bibitem{Eichhorn:2016esv}
A.~Eichhorn, A.~Held and J.~M. Pawlowski, \emph{{Quantum-gravity effects on a
  Higgs-Yukawa model}},
  \href{https://doi.org/10.1103/PhysRevD.94.104027}{\emph{Phys. Rev. D}
  {\bfseries 94} (2016) 104027}
  [\href{https://arxiv.org/abs/1604.02041}{{\ttfamily 1604.02041}}].

\bibitem{Eichhorn:2016vvy}
A.~Eichhorn and S.~Lippoldt, \emph{{Quantum gravity and Standard-Model-like
  fermions}}, \href{https://doi.org/10.1016/j.physletb.2017.01.064}{\emph{Phys.
  Lett. B} {\bfseries 767} (2017) 142}
  [\href{https://arxiv.org/abs/1611.05878}{{\ttfamily 1611.05878}}].

\bibitem{Wetterich:2016uxm}
C.~Wetterich and M.~Yamada, \emph{{Gauge hierarchy problem in asymptotically
  safe gravity--the resurgence mechanism}},
  \href{https://doi.org/10.1016/j.physletb.2017.04.049}{\emph{Phys. Lett. B}
  {\bfseries 770} (2017) 268}
  [\href{https://arxiv.org/abs/1612.03069}{{\ttfamily 1612.03069}}].

\bibitem{Christiansen:2017gtg}
N.~Christiansen and A.~Eichhorn, \emph{{An asymptotically safe solution to the
  U(1) triviality problem}},
  \href{https://doi.org/10.1016/j.physletb.2017.04.047}{\emph{Phys. Lett. B}
  {\bfseries 770} (2017) 154}
  [\href{https://arxiv.org/abs/1702.07724}{{\ttfamily 1702.07724}}].

\bibitem{Hamada:2017rvn}
Y.~Hamada and M.~Yamada, \emph{{Asymptotic safety of higher derivative quantum
  gravity non-minimally coupled with a matter system}},
  \href{https://doi.org/10.1007/JHEP08(2017)070}{\emph{JHEP} {\bfseries 08}
  (2017) 070} [\href{https://arxiv.org/abs/1703.09033}{{\ttfamily
  1703.09033}}].

\bibitem{Christiansen:2017qca}
N.~Christiansen, A.~Eichhorn and A.~Held, \emph{{Is scale-invariance in
  gauge-Yukawa systems compatible with the graviton?}},
  \href{https://doi.org/10.1103/PhysRevD.96.084021}{\emph{Phys. Rev. D}
  {\bfseries 96} (2017) 084021}
  [\href{https://arxiv.org/abs/1705.01858}{{\ttfamily 1705.01858}}].

\bibitem{Eichhorn:2017eht}
A.~Eichhorn and A.~Held, \emph{{Viability of quantum-gravity induced
  ultraviolet completions for matter}},
  \href{https://doi.org/10.1103/PhysRevD.96.086025}{\emph{Phys. Rev. D}
  {\bfseries 96} (2017) 086025}
  [\href{https://arxiv.org/abs/1705.02342}{{\ttfamily 1705.02342}}].

\bibitem{Eichhorn:2017ylw}
A.~Eichhorn and A.~Held, \emph{{Top mass from asymptotic safety}},
  \href{https://doi.org/10.1016/j.physletb.2017.12.040}{\emph{Phys. Lett. B}
  {\bfseries 777} (2018) 217}
  [\href{https://arxiv.org/abs/1707.01107}{{\ttfamily 1707.01107}}].

\bibitem{Eichhorn:2017lry}
A.~Eichhorn and F.~Versteegen, \emph{{Upper bound on the Abelian gauge coupling
  from asymptotic safety}},
  \href{https://doi.org/10.1007/JHEP01(2018)030}{\emph{JHEP} {\bfseries 01}
  (2018) 030} [\href{https://arxiv.org/abs/1709.07252}{{\ttfamily
  1709.07252}}].

\bibitem{Becker:2017tcx}
D.~Becker, C.~Ripken and F.~Saueressig, \emph{{On avoiding Ostrogradski
  instabilities within Asymptotic Safety}},
  \href{https://doi.org/10.1007/JHEP12(2017)121}{\emph{JHEP} {\bfseries 12}
  (2017) 121} [\href{https://arxiv.org/abs/1709.09098}{{\ttfamily
  1709.09098}}].

\bibitem{Eichhorn:2017sok}
A.~Eichhorn, S.~Lippoldt and V.~Skrinjar, \emph{{Nonminimal hints for
  asymptotic safety}},
  \href{https://doi.org/10.1103/PhysRevD.97.026002}{\emph{Phys. Rev. D}
  {\bfseries 97} (2018) 026002}
  [\href{https://arxiv.org/abs/1710.03005}{{\ttfamily 1710.03005}}].

\bibitem{Christiansen:2017cxa}
N.~Christiansen, D.~F. Litim, J.~M. Pawlowski and M.~Reichert,
  \emph{{Asymptotic safety of gravity with matter}},
  \href{https://doi.org/10.1103/PhysRevD.97.106012}{\emph{Phys. Rev. D}
  {\bfseries 97} (2018) 106012}
  [\href{https://arxiv.org/abs/1710.04669}{{\ttfamily 1710.04669}}].

\bibitem{Eichhorn:2017als}
A.~Eichhorn, Y.~Hamada, J.~Lumma and M.~Yamada, \emph{{Quantum gravity
  fluctuations flatten the Planck-scale Higgs potential}},
  \href{https://doi.org/10.1103/PhysRevD.97.086004}{\emph{Phys. Rev. D}
  {\bfseries 97} (2018) 086004}
  [\href{https://arxiv.org/abs/1712.00319}{{\ttfamily 1712.00319}}].

\bibitem{Alkofer:2018fxj}
N.~Alkofer and F.~Saueressig, \emph{{Asymptotically safe $f(R)$-gravity coupled
  to matter I: the polynomial case}},
  \href{https://doi.org/10.1016/j.aop.2018.07.017}{\emph{Annals Phys.}
  {\bfseries 396} (2018) 173}
  [\href{https://arxiv.org/abs/1802.00498}{{\ttfamily 1802.00498}}].

\bibitem{Gies:2018jnv}
H.~Gies and R.~Martini, \emph{{Curvature bound from gravitational catalysis}},
  \href{https://doi.org/10.1103/PhysRevD.97.085017}{\emph{Phys. Rev. D}
  {\bfseries 97} (2018) 085017}
  [\href{https://arxiv.org/abs/1802.02865}{{\ttfamily 1802.02865}}].

\bibitem{Eichhorn:2018whv}
A.~Eichhorn and A.~Held, \emph{{Mass difference for charged quarks from
  asymptotically safe quantum gravity}},
  \href{https://doi.org/10.1103/PhysRevLett.121.151302}{\emph{Phys. Rev. Lett.}
  {\bfseries 121} (2018) 151302}
  [\href{https://arxiv.org/abs/1803.04027}{{\ttfamily 1803.04027}}].

\bibitem{Eichhorn:2018akn}
A.~Eichhorn, P.~Labus, J.~M. Pawlowski and M.~Reichert, \emph{{Effective
  universality in quantum gravity}},
  \href{https://doi.org/10.21468/SciPostPhys.5.4.031}{\emph{SciPost Phys.}
  {\bfseries 5} (2018) 031} [\href{https://arxiv.org/abs/1804.00012}{{\ttfamily
  1804.00012}}].

\bibitem{Alkofer:2018baq}
N.~Alkofer, \emph{{Asymptotically safe $f(R)$-gravity coupled to matter II:
  Global solutions}},
  \href{https://doi.org/10.1016/j.physletb.2018.12.061}{\emph{Phys. Lett. B}
  {\bfseries 789} (2019) 480}
  [\href{https://arxiv.org/abs/1809.06162}{{\ttfamily 1809.06162}}].

\bibitem{Eichhorn:2018ydy}
A.~Eichhorn, S.~Lippoldt, J.~M. Pawlowski, M.~Reichert and M.~Schiffer,
  \emph{{How perturbative is quantum gravity?}},
  \href{https://doi.org/10.1016/j.physletb.2019.01.071}{\emph{Phys. Lett. B}
  {\bfseries 792} (2019) 310}
  [\href{https://arxiv.org/abs/1810.02828}{{\ttfamily 1810.02828}}].

\bibitem{Pawlowski:2018ixd}
J.~M. Pawlowski, M.~Reichert, C.~Wetterich and M.~Yamada, \emph{{Higgs scalar
  potential in asymptotically safe quantum gravity}},
  \href{https://doi.org/10.1103/PhysRevD.99.086010}{\emph{Phys. Rev. D}
  {\bfseries 99} (2019) 086010}
  [\href{https://arxiv.org/abs/1811.11706}{{\ttfamily 1811.11706}}].

\bibitem{Eichhorn:2018nda}
A.~Eichhorn, S.~Lippoldt and M.~Schiffer, \emph{{Zooming in on fermions and
  quantum gravity}},
  \href{https://doi.org/10.1103/PhysRevD.99.086002}{\emph{Phys. Rev. D}
  {\bfseries 99} (2019) 086002}
  [\href{https://arxiv.org/abs/1812.08782}{{\ttfamily 1812.08782}}].

\bibitem{Eichhorn:2019yzm}
A.~Eichhorn and M.~Schiffer, \emph{{$d=4$ as the critical dimensionality of
  asymptotically safe interactions}},
  \href{https://doi.org/10.1016/j.physletb.2019.05.005}{\emph{Phys. Lett. B}
  {\bfseries 793} (2019) 383}
  [\href{https://arxiv.org/abs/1902.06479}{{\ttfamily 1902.06479}}].

\bibitem{DeBrito:2019rrh}
G.~P. De~Brito, Y.~Hamada, A.~D. Pereira and M.~Yamada, \emph{{On the impact of
  Majorana masses in gravity-matter systems}},
  \href{https://doi.org/10.1007/JHEP08(2019)142}{\emph{JHEP} {\bfseries 08}
  (2019) 142} [\href{https://arxiv.org/abs/1905.11114}{{\ttfamily
  1905.11114}}].

\bibitem{Knorr:2019atm}
B.~Knorr, C.~Ripken and F.~Saueressig, \emph{{Form Factors in Asymptotic
  Safety: conceptual ideas and computational toolbox}},
  \href{https://doi.org/10.1088/1361-6382/ab4a53}{\emph{Class. Quant. Grav.}
  {\bfseries 36} (2019) 234001}
  [\href{https://arxiv.org/abs/1907.02903}{{\ttfamily 1907.02903}}].

\bibitem{Wetterich:2019rsn}
C.~Wetterich, \emph{{Effective scalar potential in asymptotically safe quantum
  gravity}}, \href{https://doi.org/10.3390/universe7020045}{\emph{Universe}
  {\bfseries 7} (2021) 45} [\href{https://arxiv.org/abs/1911.06100}{{\ttfamily
  1911.06100}}].

\bibitem{Reichert:2019car}
M.~Reichert and J.~Smirnov, \emph{{Dark Matter meets Quantum Gravity}},
  \href{https://doi.org/10.1103/PhysRevD.101.063015}{\emph{Phys. Rev. D}
  {\bfseries 101} (2020) 063015}
  [\href{https://arxiv.org/abs/1911.00012}{{\ttfamily 1911.00012}}].

\bibitem{Burger:2019upn}
B.~B\"urger, J.~M. Pawlowski, M.~Reichert and B.-J. Schaefer, \emph{{Curvature
  dependence of quantum gravity with scalars}},
  \href{https://arxiv.org/abs/1912.01624}{{\ttfamily 1912.01624}}.

\bibitem{Platania:2020knd}
A.~Platania and C.~Wetterich, \emph{{Non-perturbative unitarity and fictitious
  ghosts in quantum gravity}},
  \href{https://doi.org/10.1016/j.physletb.2020.135911}{\emph{Phys. Lett. B}
  {\bfseries 811} (2020) 135911}
  [\href{https://arxiv.org/abs/2009.06637}{{\ttfamily 2009.06637}}].

\bibitem{Daas:2020dyo}
J.~Daas, W.~Oosters, F.~Saueressig and J.~Wang, \emph{{Asymptotically safe
  gravity with fermions}},
  \href{https://doi.org/10.1016/j.physletb.2020.135775}{\emph{Phys. Lett. B}
  {\bfseries 809} (2020) 135775}
  [\href{https://arxiv.org/abs/2005.12356}{{\ttfamily 2005.12356}}].

\bibitem{Eichhorn:2020kca}
A.~Eichhorn and M.~Pauly, \emph{{Safety in darkness: Higgs portal to simple
  Yukawa systems}},
  \href{https://doi.org/10.1016/j.physletb.2021.136455}{\emph{Phys. Lett. B}
  {\bfseries 819} (2021) 136455}
  [\href{https://arxiv.org/abs/2005.03661}{{\ttfamily 2005.03661}}].

\bibitem{Eichhorn:2020sbo}
A.~Eichhorn and M.~Pauly, \emph{{Constraining power of asymptotic safety for
  scalar fields}},
  \href{https://doi.org/10.1103/PhysRevD.103.026006}{\emph{Phys. Rev. D}
  {\bfseries 103} (2021) 026006}
  [\href{https://arxiv.org/abs/2009.13543}{{\ttfamily 2009.13543}}].

\bibitem{deBrito:2020dta}
G.~P. de~Brito, A.~Eichhorn and M.~Schiffer, \emph{{Light charged fermions in
  quantum gravity}},
  \href{https://doi.org/10.1016/j.physletb.2021.136128}{\emph{Phys. Lett. B}
  {\bfseries 815} (2021) 136128}
  [\href{https://arxiv.org/abs/2010.00605}{{\ttfamily 2010.00605}}].

\bibitem{Ali:2020znq}
P.~Ali, A.~Eichhorn, M.~Pauly and M.~M. Scherer, \emph{{Constraints on discrete
  global symmetries in quantum gravity}},
  \href{https://doi.org/10.1007/JHEP05(2021)036}{\emph{JHEP} {\bfseries 05}
  (2021) 036} [\href{https://arxiv.org/abs/2012.07868}{{\ttfamily
  2012.07868}}].

\bibitem{Gies:2021upb}
H.~Gies and A.~S. Salek, \emph{{Curvature bound from gravitational catalysis in
  thermal backgrounds}},
  \href{https://doi.org/10.1103/PhysRevD.103.125027}{\emph{Phys. Rev. D}
  {\bfseries 103} (2021) 125027}
  [\href{https://arxiv.org/abs/2103.05542}{{\ttfamily 2103.05542}}].

\bibitem{deBrito:2021pyi}
G.~P. de~Brito, A.~Eichhorn and R.~R. L.~d. Santos, \emph{{The weak-gravity
  bound and the need for spin in asymptotically safe matter-gravity models}},
  \href{https://arxiv.org/abs/2107.03839}{{\ttfamily 2107.03839}}.

\bibitem{Eichhorn:2021tsx}
A.~Eichhorn, M.~Pauly and S.~Ray, \emph{{Towards a Higgs mass determination in
  asymptotically safe gravity with a dark portal}},
  \href{https://arxiv.org/abs/2107.07949}{{\ttfamily 2107.07949}}.

\bibitem{Daas:2021abx}
J.~Daas, W.~Oosters, F.~Saueressig and J.~Wang, \emph{{Asymptotically Safe
  Gravity-Fermion Systems on Curved Backgrounds}},
  \href{https://doi.org/10.3390/universe7080306}{\emph{Universe} {\bfseries 7}
  (2021) 306} [\href{https://arxiv.org/abs/2107.01071}{{\ttfamily
  2107.01071}}].

\bibitem{Houthoff:2020zqy}
W.~Houthoff, A.~Kurov and F.~Saueressig, \emph{{On the scaling of composite
  operators in asymptotic safety}},
  \href{https://doi.org/10.1007/JHEP04(2020)099}{\emph{JHEP} {\bfseries 04}
  (2020) 099} [\href{https://arxiv.org/abs/2002.00256}{{\ttfamily
  2002.00256}}].

\bibitem{Kurov:2020csd}
A.~Kurov and F.~Saueressig, \emph{{On characterizing the Quantum Geometry
  underlying Asymptotic Safety}},
  \href{https://doi.org/10.3389/fphy.2020.00187}{\emph{Front. in Phys.}
  {\bfseries 8} (2020) 187} [\href{https://arxiv.org/abs/2003.07454}{{\ttfamily
  2003.07454}}].

\bibitem{Abbott:1981ke}
L.~F. Abbott, \emph{{Introduction to the Background Field Method}}, {\emph{Acta
  Phys. Polon. B} {\bfseries 13} (1982) 33}.

\bibitem{Litim:2000ci}
D.~F. Litim, \emph{{Optimization of the exact renormalization group}},
  \href{https://doi.org/10.1016/S0370-2693(00)00748-6}{\emph{Phys. Lett. B}
  {\bfseries 486} (2000) 92}
  [\href{https://arxiv.org/abs/hep-th/0005245}{{\ttfamily hep-th/0005245}}].

\bibitem{Knorr:2021slg}
B.~Knorr, \emph{{The derivative expansion in asymptotically safe quantum
  gravity: general setup and quartic order}},
  \href{https://arxiv.org/abs/2104.11336}{{\ttfamily 2104.11336}}.

\bibitem{xActwebpage}
``{xAct: Efficient tensor computer algebra for Mathematica}.''
  \url{http://xact.es/index.html}.

\bibitem{2007CoPhC.177..640M}
J.~M. {Mart{\'{\i}}n-Garc{\'{\i}}a}, R.~{Portugal} and L.~R.~U. {Manssur},
  \emph{{The Invar tensor package}},
  \href{https://doi.org/10.1016/j.cpc.2007.05.015}{\emph{Computer Physics
  Communications} {\bfseries 177} (2007) 640}
  [\href{https://arxiv.org/abs/0704.1756}{{\ttfamily 0704.1756}}].

\bibitem{Brizuela:2008ra}
D.~Brizuela, J.~M. Martin-Garcia and G.~A. Mena~Marugan, \emph{{xPert: Computer
  algebra for metric perturbation theory}},
  \href{https://doi.org/10.1007/s10714-009-0773-2}{\emph{Gen. Rel. Grav.}
  {\bfseries 41} (2009) 2415}
  [\href{https://arxiv.org/abs/0807.0824}{{\ttfamily 0807.0824}}].

\bibitem{2008CoPhC.179..597M}
J.~M. {Mart{\'{\i}}n-Garc{\'{\i}}a}, \emph{{xPerm: fast index canonicalization
  for tensor computer algebra}},
  \href{https://doi.org/10.1016/j.cpc.2008.05.009}{\emph{Computer Physics
  Communications} {\bfseries 179} (2008) 597}
  [\href{https://arxiv.org/abs/0803.0862}{{\ttfamily 0803.0862}}].

\bibitem{2014CoPhC.185.1719N}
T.~{Nutma}, \emph{{xTras: A field-theory inspired xAct package for
  mathematica}},
  \href{https://doi.org/10.1016/j.cpc.2014.02.006}{\emph{Computer Physics
  Communications} {\bfseries 185} (2014) 1719}
  [\href{https://arxiv.org/abs/1308.3493}{{\ttfamily 1308.3493}}].

\bibitem{Alberte:2020jsk}
L.~Alberte, C.~de~Rham, S.~Jaitly and A.~J. Tolley, \emph{{Positivity Bounds
  and the Massless Spin-2 Pole}},
  \href{https://doi.org/10.1103/PhysRevD.102.125023}{\emph{Phys. Rev. D}
  {\bfseries 102} (2020) 125023}
  [\href{https://arxiv.org/abs/2007.12667}{{\ttfamily 2007.12667}}].

\bibitem{Steinwachs:2021jft}
C.~F. Steinwachs, \emph{{Non-perturbative quantum Galileon in the exact
  renormalization group}},
  \href{https://doi.org/10.1088/1475-7516/2021/04/038}{\emph{JCAP} {\bfseries
  04} (2021) 038} [\href{https://arxiv.org/abs/2101.07271}{{\ttfamily
  2101.07271}}].

\bibitem{Groh:2011vn}
K.~Groh, S.~Rechenberger, F.~Saueressig and O.~Zanusso, \emph{{Higher
  Derivative Gravity from the Universal Renormalization Group Machine}},
  \href{https://doi.org/10.22323/1.134.0124}{\emph{PoS} {\bfseries EPS-HEP2011}
  (2011) 124} [\href{https://arxiv.org/abs/1111.1743}{{\ttfamily 1111.1743}}].

\bibitem{Horndeski:1974wa}
G.~W. Horndeski, \emph{{Second-order scalar-tensor field equations in a
  four-dimensional space}},
  \href{https://doi.org/10.1007/BF01807638}{\emph{Int. J. Theor. Phys.}
  {\bfseries 10} (1974) 363}.

\bibitem{Nicolis:2008in}
A.~Nicolis, R.~Rattazzi and E.~Trincherini, \emph{{The Galileon as a local
  modification of gravity}},
  \href{https://doi.org/10.1103/PhysRevD.79.064036}{\emph{Phys. Rev. D}
  {\bfseries 79} (2009) 064036}
  [\href{https://arxiv.org/abs/0811.2197}{{\ttfamily 0811.2197}}].

\bibitem{Deffayet:2009wt}
C.~Deffayet, G.~Esposito-Farese and A.~Vikman, \emph{{Covariant Galileon}},
  \href{https://doi.org/10.1103/PhysRevD.79.084003}{\emph{Phys. Rev. D}
  {\bfseries 79} (2009) 084003}
  [\href{https://arxiv.org/abs/0901.1314}{{\ttfamily 0901.1314}}].

\bibitem{Kobayashi:2019hrl}
T.~Kobayashi, \emph{{Horndeski theory and beyond: a review}},
  \href{https://doi.org/10.1088/1361-6633/ab2429}{\emph{Rept. Prog. Phys.}
  {\bfseries 82} (2019) 086901}
  [\href{https://arxiv.org/abs/1901.07183}{{\ttfamily 1901.07183}}].

\bibitem{Qiu:2011cy}
T.~Qiu, J.~Evslin, Y.-F. Cai, M.~Li and X.~Zhang, \emph{{Bouncing Galileon
  Cosmologies}},
  \href{https://doi.org/10.1088/1475-7516/2011/10/036}{\emph{JCAP} {\bfseries
  10} (2011) 036} [\href{https://arxiv.org/abs/1108.0593}{{\ttfamily
  1108.0593}}].

\bibitem{Kobayashi:2011nu}
T.~Kobayashi, M.~Yamaguchi and J.~Yokoyama, \emph{{Generalized G-inflation:
  Inflation with the most general second-order field equations}},
  \href{https://doi.org/10.1143/PTP.126.511}{\emph{Prog. Theor. Phys.}
  {\bfseries 126} (2011) 511}
  [\href{https://arxiv.org/abs/1105.5723}{{\ttfamily 1105.5723}}].

\bibitem{Sotiriou:2011dz}
T.~P. Sotiriou and V.~Faraoni, \emph{{Black holes in scalar-tensor gravity}},
  \href{https://doi.org/10.1103/PhysRevLett.108.081103}{\emph{Phys. Rev. Lett.}
  {\bfseries 108} (2012) 081103}
  [\href{https://arxiv.org/abs/1109.6324}{{\ttfamily 1109.6324}}].

\bibitem{Sotiriou:2013qea}
T.~P. Sotiriou and S.-Y. Zhou, \emph{{Black hole hair in generalized
  scalar-tensor gravity}},
  \href{https://doi.org/10.1103/PhysRevLett.112.251102}{\emph{Phys. Rev. Lett.}
  {\bfseries 112} (2014) 251102}
  [\href{https://arxiv.org/abs/1312.3622}{{\ttfamily 1312.3622}}].

\bibitem{Herdeiro:2015waa}
C.~A.~R. Herdeiro and E.~Radu, \emph{{Asymptotically flat black holes with
  scalar hair: a review}},
  \href{https://doi.org/10.1142/S0218271815420146}{\emph{Int. J. Mod. Phys. D}
  {\bfseries 24} (2015) 1542014}
  [\href{https://arxiv.org/abs/1504.08209}{{\ttfamily 1504.08209}}].

\bibitem{Ezquiaga:2017ekz}
J.~M. Ezquiaga and M.~Zumalac\'arregui, \emph{{Dark Energy After GW170817: Dead
  Ends and the Road Ahead}},
  \href{https://doi.org/10.1103/PhysRevLett.119.251304}{\emph{Phys. Rev. Lett.}
  {\bfseries 119} (2017) 251304}
  [\href{https://arxiv.org/abs/1710.05901}{{\ttfamily 1710.05901}}].

\bibitem{Vassilevich:2003xt}
D.~V. Vassilevich, \emph{{Heat kernel expansion: User's manual}},
  \href{https://doi.org/10.1016/j.physrep.2003.09.002}{\emph{Phys. Rept.}
  {\bfseries 388} (2003) 279}
  [\href{https://arxiv.org/abs/hep-th/0306138}{{\ttfamily hep-th/0306138}}].

\bibitem{Barvinsky:1985an}
A.~O. Barvinsky and G.~A. Vilkovisky, \emph{{The Generalized Schwinger-Dewitt
  Technique in Gauge Theories and Quantum Gravity}},
  \href{https://doi.org/10.1016/0370-1573(85)90148-6}{\emph{Phys. Rept.}
  {\bfseries 119} (1985) 1}.

\bibitem{Decanini:2005gt}
Y.~Decanini and A.~Folacci, \emph{{Off-diagonal coefficients of the
  Dewitt-Schwinger and Hadamard representations of the Feynman propagator}},
  \href{https://doi.org/10.1103/PhysRevD.73.044027}{\emph{Phys. Rev. D}
  {\bfseries 73} (2006) 044027}
  [\href{https://arxiv.org/abs/gr-qc/0511115}{{\ttfamily gr-qc/0511115}}].

\bibitem{Anselmi:2007eq}
D.~Anselmi and A.~Benini, \emph{{Improved Schwinger-DeWitt techniques for
  higher-derivative corrections to operator determinants}},
  \href{https://doi.org/10.1088/1126-6708/2007/10/099}{\emph{JHEP} {\bfseries
  10} (2007) 099} [\href{https://arxiv.org/abs/0704.2840}{{\ttfamily
  0704.2840}}].

\bibitem{Groh:2011dw}
K.~Groh, F.~Saueressig and O.~Zanusso, \emph{{Off-diagonal heat-kernel
  expansion and its application to fields with differential constraints}},
  \href{https://arxiv.org/abs/1112.4856}{{\ttfamily 1112.4856}}.

\bibitem{Codello:2012kq}
A.~Codello and O.~Zanusso, \emph{{On the non-local heat kernel expansion}},
  \href{https://doi.org/10.1063/1.4776234}{\emph{J. Math. Phys.} {\bfseries 54}
  (2013) 013513} [\href{https://arxiv.org/abs/1203.2034}{{\ttfamily
  1203.2034}}].

\end{thebibliography}\endgroup

\end{document}